\documentclass[10pt,preprint]{aastex}
\usepackage{amsmath,amssymb}
\def\andand{\and}

\def\run#1{{#1}}
\def\z#1{}
\def\AA{A}

\def\pseudofigureone#1#2#3#4{
\begin{figure}
\plotone{#3}
\caption{#4
\label{#1}}
\end{figure}}
\def\pseudofiguretwo#1#2#3#4#5#6{
\begin{figure}
\plottwo{#4}{#5}
\caption{#6
\label{#1}}
\end{figure}}

\usepackage{emulateapj5}
\usepackage{apjfonts}
\journalinfo{Astrophysical Journal {\rm {\bf 612}, 276 (2004)
[e-print~astro-ph/0312046}]}
\def\andand{\and\vskip-4mm}

\def\pseudofigureone#1#2#3#4{{
\refstepcounter{figure}
\label{#1}
\vskip1mm
\plotone{#2}
\vskip1mm
\footnotesize\def\baselinestretch{1.0}
\begin{minipage}{\columnwidth}
{\scshape ~~Fig.}\space\thefigure.--- #4
\end{minipage}
\vskip1mm
}}
\def\pseudofiguretwo#1#2#3#4#5#6{{
\begin{figure*}
\epsscale{2.2}
\plottwo{#2}{#3}
\epsscale{1.0}
\vskip-1mm
\caption{#6
\label{#1}}
\vskip-4mm
\end{figure*}
}}

\slugcomment{\today}
\shorttitle{SIMULATIONS OF THE SMALL-SCALE TURBULENT DYNAMO}
\shortauthors{SCHEKOCHIHIN ET AL.}

\def\ssecref#1{\hbox{\S\,\ref{#1}}}
\def\Secref#1{Section~\ref{#1}}
\def\secref#1{\hbox{\S\,\ref{#1}}}

\def\exref#1{(\ref{#1})}

\def\eqref#1{eq.~(\ref{#1})}
\def\eqsref#1#2{eqs.~(\ref{#1}-\ref{#2})}
\def\eqsandref#1#2{eqs.~(\ref{#1}) and~(\ref{#2})}
\def\Eqref#1{Equation~(\ref{#1})}
\def\Eqsref#1#2{Equations~(\ref{#1}-\ref{#2})}

\def\figref#1{Fig.~\ref{#1}}
\def\Figref#1{Figure~\ref{#1}}
\def\tabref#1{Table~\ref{#1}}

\def\const{{\rm const}}

\def\bea{\begin{eqnarray}}
\def\eea{\end{eqnarray}}

\def\ell{l}
\def\phi{\varphi}

\def\({\left(}
\def\){\right)}
\def\[{\left[}
\def\]{\right]}
\def\<{\langle}
\def\>{\rangle}
\def\l{\left}
\def\r{\right}
\def\bl{\bigl}
\def\br{\bigr}

\def\d{\partial}
\def\diff{d}
\def\Dt{{d\over dt}\,}

\def\unity{{\hat{\mathbb I}}}

\def\vx{{\bf x}}
\def\vy{{\bf y}}
\def\xz{x_0}
\def\vk{{\bf k}}
\def\vu{{\bf u}}
\def\vB{{\bf B}}
\def\vz{{\bf z}}

\def\tvB{{\tilde{\bf B}}}
\def\tB{{\tilde B}}
\def\vJ{{\bf J}}
\def\vF{{\bf F}}
\def\vf{{\bf f}}
\def\vK{{\bf K}}
\def\vb{{\skew{-4}\hat{\bf b}}}
\def\bb{{\hat b}}
\def\vn{{\hat {\bf n}}}

\def\vkk{{\hat {\bf k}}}
\def\kk{{\hat k}}

\def\nut{\tilde\nu}

\def\kapperp{\kappa_\perp}
\def\kappar{\kappa_\parallel}

\def\kapI{\kappa^\text{(i)}}
\def\kapA{\kappa^\text{(a)}}

\def\vKK{{\mathfrak{v}}}
\def\vvKK{{\boldsymbol{\vKK}}}

\def\Brms{B_{\rm rms}}

\def\kperp{k_\perp}
\def\kpar{k_{\parallel}}
\def\kbxj{k_{\vB\times\vJ}}
\def\kbj{k_{\vB\cdot\vJ}}

\def\dpar{\nabla_{\parallel}}
\def\Mbxj{M_{\vB\times\vJ}}
\def\Mbj{M_{\vB\cdot\vJ}}

\def\tbox{\tau_{\rm box}}
\def\Grms{\Gamma_{\rm rms}}
\def\tcorr{\tau_c}

\def\ud{u_\nu}
\def\vth{v_{\rm th}}

\def\kd{k_{\nu}}
\def\kl{k_\lambda}
\def\ld{\ell_{\nu}}
\def\td{\tau_{\nu}}
\def\uf{u_0}
\def\kf{{k_0}}
\def\lf{\ell_0}
\def\tf{\tau_0}
\def\kres{k_{\eta}}
\def\lres{\ell_{\eta}}

\def\ls{\ell_{\rm s}}
\def\ks{k_{\rm s}}
\def\Re{{\rm Re}}
\def\Rel{{\rm Re}_\lambda}
\def\Rm{{\rm Rm}}
\def\Rmc{{\rm Rm}_{\rm c}}
\def\Pr{{\rm Pr}_{\rm m}}
\def\Prc{{\rm Pr}_{\rm m,c}}

\def\usq{{\langle u^2 \rangle}}
\def\Bsq{{\langle B^2 \rangle}}

\def\Bfr{{\langle B^4 \rangle}}

\def\Ksq{{\langle K^2 \rangle}}
\def\Krms{K_{\rm rms}}

\def\kB{k_{\rm rms}}

\def\eps{\epsilon}

\def\gKA{{\bar\gamma}}

\def\eone{\hat {\bf e}_1}
\def\etwo{\hat {\bf e}_2}
\def\ethree{\hat {\bf e}_3}

\def\zetan{\zeta_n}
\def\zetab{\zeta_b}

\def\gperp{\gamma_{\perp}}
\def\sperp{\sigma_\perp}
\def\spar{\sigma_\parallel}

\begin{document}

\title{SIMULATIONS OF THE SMALL-SCALE TURBULENT DYNAMO}
\author{Alexander~A.~Schekochihin,\altaffilmark{1} 
Steven~C.~Cowley,\altaffilmark{2} Samuel~F.~Taylor\altaffilmark{3}}
\affil{Plasma Physics Group, Imperial College, 
Blackett Laboratory, Prince Consort Rd., 
London~SW7~2BW, UK}
\altaffiltext{1}{Present-time address: 
DAMTP/CMS, University of Cambridge, 
Wilberforce Road, Cambridge~CB3~0WA, UK; 
e-mail: as629@damtp.cam.ac.uk.}
\altaffiltext{2}{Also at the Department of Physics and Astronomy, 
UCLA, Los Angeles, CA~90095-1547.}
\altaffiltext{3}{Present-time address:  
Department of Physics, Princeton University, 
Princeton, NJ~08544} 
\author{Jason~L.~Maron\altaffilmark{4}}
\affil{Department of Physics and Astronomy, University of Rochester, 
Rochester, NY 14627 and\\ 
Center for Magnetic Reconnection Studies, 
Department of Physics and Astronomy, University of Iowa, 
Iowa City, IA 52242\\
\null}
\altaffiltext{4}{Present-time address:
Department of Astrophysics, American Museum of Natural History, 
West 79th St., New York, NY 10024-5192} 
\andand
\author{James~C.~McWilliams}
\affil{Department of Atmospheric Sciences, 
UCLA, Los Angeles, CA 90095-1565}

\begin{abstract}

We report the results of an extensive numerical study of 
the small-scale turbulent dynamo. 
The primary focus is on the case of large magnetic Prandtl numbers 
$\Pr$, which is relevant for hot low-density astrophysical plasmas. 
A $\Pr$ parameter scan is given for the model case 
of viscosity-dominated (low-Reynolds-number) turbulence. 
We concentrate on three topics: 
magnetic-energy spectra and saturation levels, 
the structure of the magnetic-field lines, 
and intermittency of the field-strength distribution. 
The main results are as follows: 
(1) the folded structure of the field (direction reversals 
at the resistive scale, field lines curved at the scale of 
the flow) persists from the kinematic to the nonlinear regime; 
(2) the field distribution is self-similar 
and appears to be lognormal during the kinematic regime 
and exponential in the saturated state; and 
(3) the bulk of the magnetic energy is at the resistive 
scale in the kinematic regime and remains there after saturation, 
although the magnetic-energy spectrum becomes much shallower. 
We propose an analytical model of saturation 
based on the idea of partial two-dimensionalization of 
the velocity gradients with respect to the local direction 
of the magnetic folds. The model-predicted saturated spectra 
are in excellent agreement with numerical results. 
Comparisons with large-$\Re$, moderate-$\Pr$ runs are carried out 
to confirm the relevance of these results 
and to test heuristic scenarios of dynamo saturation. 
New features at large $\Re$ 
are elongation of the folds in the nonlinear 
regime from the viscous scale to the box scale and the presence 
of an intermediate nonlinear stage of slower-than-exponential 
magnetic-energy growth accompanied by an increase of the 
resistive scale and partial suppression of the kinetic-energy 
spectrum in the inertial range. 
Numerical results for the saturated state 
do not support scale-by-scale equipartition between 
magnetic and kinetic energies, with a definite excess of magnetic 
energy at small scales. A physical picture of the saturated state 
is proposed. 

\end{abstract}



\section{INTRODUCTION}

\subsection{Large- and Small-Scale Magnetic Fields in Astrophysics}

Magnetic fields are detected everywhere in the universe: 
stars, accretion disks, galaxies, 
and galaxy clusters all carry dynamically 
important magnetic fields \citep[some of the relevant reviews and observations are][]{Kronberg_review,Beck_etal,Minter_Spangler,Zweibel_Heiles,Vallee1,Vallee2,Balbus_Hawley_review,Kulsrud_review,Weiss_Tobias,Title_review,Beck_review1,Beck_review2,Ferriere_review,Kronberg_etal,Kronberg_PToday,Carilli_Taylor,Han_Wielebinski,Widrow_review,Tobias_review,Ossendrijver_review,Han_Ferriere_Manchester,Brandenburg_Subramanian_review}. 
There are two kinds of magnetic fields observed. 
First, there are {\em large-scale fields,} i.e., 
fields spatially coherent at scales comparable 
to the size of the astrophysical object that they inhabit. 
Two examples of such fields are the cyclic dipolar field of 
the Sun and the spiral fields of galaxies. 
Second, there are {\em small-scale fields}: e.g., 
the fluctuating fields in the solar photosphere, 
galaxies, and clusters. 
They are associated with the turbulent motions 
of the constituent plasmas 
and exist at scales below those at which the turbulence is forced. 
Both types of fields are usually strong enough to be 
dynamically important. The challenge is to construct 
a theory of their origin, evolution, and structure consistent 
with observations and to understand the role 
these fields play in the dynamics of astrophysical objects. 

A physically plausible scenario for the origin and maintenance 
of these fields is a turbulent dynamo, which would 
amplify a weak seed field and culminate in a nonlinear 
saturated state. Just as magnetic fields can be 
classified into large- and small-scale varieties, 
there are also two kinds of dynamo responsible for these fields. 
First, three-dimensional velocity fields 
sufficiently random in space and/or time 
will amplify small-scale magnetic fluctuations 
via random stretching of the field 
lines \citep{Batchelor_dynamo,Zeldovich_etal_linear,STF}. 
Since it is a small-scale process, 
we believe that this {\em small-scale dynamo} 
can be studied in a homogeneous and isotropic setting. 
In contrast, the generation and structure of the large-scale 
magnetic fields cannot be understood without going beyond the 
homogeneous picture and taking into account large-scale 
object-specific features: boundary 
conditions, rotation, helicity, mean velocity shear, density 
stratification etc. These effects can combine with turbulence 
to give rise to the second kind of dynamo: 
the {\em large-scale dynamo}. This is usually 
handled in the mean-field framework \citep[e.g.,][]{Moffatt_book,Parker_book,Ruzmaikin_Shukurov_Sokoloff,Brandenburg_Subramanian_review}. 
Mean-field theories tend to assume that all small-scale 
magnetic fluctuations result from the shredding of the 
mean field by the turbulence. Such {\em induced small-scale fields} 
do exist, but they vanish if the mean field vanishes 
and are distinct from the dynamo-generated 
small-scale fields. Neglecting the latter is, in fact, 
not a priori justified 
because the small-scale dynamo is usually much faster 
than the large-scale one: it amplifies magnetic energy 
at the rate of turbulent stretching and produces dynamically 
significant fields before the large-scale fields can grow 
appreciably \citep[e.g.,][]{KA,Kulsrud_review}. 
Procedures that have been devised for incorporating the effect of 
the small-scale fields into the mean-field 
theory \cite[][and references therein]{Vainshtein_Kichatinov,Raedler_Kleeorin_Rogachevskii}
usually do not evolve the small-scale component 
self-consistently, requiring instead certain statistical 
information about the small-scale fields to be 
supplied at the modelling stage. Thus, solving the 
large-scale problem is predicated on making correct assumptions
about the small scales. 

Whether the magnetic fields are dynamo-generated 
or stem from primordial and/or external mechanisms 
\citep{Kulsrud_etal_proto,Kronberg_etal}, 
the key question is how they are maintained in the course 
of their interaction with the ambient turbulence. 
Unless the large-scale field is extremely strong, 
we must always ask 
why it is not quickly churned up by the small-scale turbulence.  
Any large-scale field configuration must be consistent with 
the presence and continued regeneration of considerable amounts 
of small-scale magnetic energy. 
In galaxies and clusters, both large- and small-scale 
fields are in the microgauss range, 
which corresponds to magnetic-energy densities 
in approximate equipartition with the kinetic-energy density 
of the turbulent motions in these objects. 
This suggests that we are observing a 
self-consistent nonlinear state of MHD turbulence. 

While it is surely the interplay between motions and magnetic fields 
at disparate scales that ultimately determines the nature of the 
observed states, 
the above considerations motivate us to focus on the 
homogeneous isotropic small-scale 
MHD turbulence. From the point of view of theoretical physics, 
this is an attempt to understand the universal features before 
tackling the object-specific ones. From the point of view 
of numerical experimentation, this approach is unavoidable 
because simulating 
the full picture requires resolving {\em multiple} 
scale separations that are well beyond the capabilities of current 
computers and are likely to remain so for many years to come. 

\subsection{The Scales in the Problem}
\label{ssec_ranges}

The issue of multiple scale ranges is important, so let us 
examine it in more detail. The hierarchy of scales in an astrophysical 
object can be outlined as follows. The largest spatial scale is 
the system size~$L$. The energy that feeds the turbulence is 
injected at the outer, or forcing, scale~$\lf$. For objects with nontrivial 
geometry such as stars or galactic disks, $\lf\ll L$. 
This is the first scale separation in the problem. 
It is used by mean-field theories to split all fields into 
mean and fluctuating components with averages being done over 
scales of order and below~$\lf$. 

The energy cascades from~$\lf$ down to the 
viscous-dissipation scale~$\ld$. In purely hydrodynamic systems, 
the latter is called the Kolmogorov inner scale. Kolmogorov's 1941 
dimensional theory of turbulence \citep[e.g.,][]{Frisch_book} gives 
$\ld\sim\Re^{-3/4}\lf$. The hydrodynamic Reynolds 
number $\Re\sim\uf\lf/\nu$ is usually fairly large for astrophysical systems, 
so $\ld\ll\lf$ ($\uf$ is the typical velocity at the outer 
scale, $\nu$ is the fluid viscosity). 
This is the second scale separation in the problem. 
It is at scales between the outer and the 
inner scale --- the inertial range --- 
that the universal physics of turbulence is contained. 

If magnetic fields are present, they bring in their own dissipation 
scale~$\lres$ associated with the magnetic diffusivity~$\eta$ of 
the plasma \citep{Spitzer_book}. 
The ratio $\Pr=\nu/\eta$ is called the magnetic Prandtl number. 
Using the Spitzer values for the viscosity and magnetic 
diffusivity of a fully ionized plasma, one finds 
$\Pr\sim 10^{-5} T^4/n$, 
where $T$ is temperature in K and $n$ is the particle 
concentration in cm$^{-3}$. 
In the case of a partially ionized gas with neutral-dominated 
viscosity (e.g., warm interstellar medium), 
we can estimate $\nu\sim\vth/\sigma n$, where $\vth$ is the 
thermal speed and $\sigma$ the atomic cross-section. The formula 
for $\Pr$ is then
$\Pr\sim 10^7 T^2/n$. 
In hot low-density plasmas such as (warm) interstellar and 
intracluster media, as well as in some accretion disks, $\Pr\gg1$
\citep[see, e.g.,][]{Brandenburg_Subramanian_review}. 
Systems with~$\Pr\gg1$ will be the primary focus of this paper. 
If a weak magnetic field is amplified 
by Kolmogorov turbulence, the magnetic-energy 
exponentiation time is the same as the turnover 
time of the viscous-scale eddies, because 
these eddies are the fastest ones. Balancing the viscous-eddy 
turnover time with the magnetic-field diffusion time gives 
an estimate for the resistive scale in the kinematic (weak-field) 
regime: $\lres\sim\Pr^{-1/2}\ld$. As $\Pr\gg1$, $\lres\ll\ld$. 
This is the third scale separation in the problem. 
The scale range in between contains degrees of freedom 
that are accessible to magnetic fluctuations but not 
to velocities (\figref{fig_spectra_cartoon}). It is hardly surprising that 
these degrees of freedom are quickly occupied: 
the small-scale kinematic dynamo 
spreads magnetic energy over the subviscous range and 
piles it up at the resistive scale \citep[e.g.,][]{KA}. 
The key question then is what happens when 
the small-scale fields become sufficiently strong for 
the Lorentz force to react back on the flow. 

\pseudofigureone{fig_spectra_cartoon}{spectra_cartoon.epsf}{f1.ps}{Sketch of 
scale ranges and energy spectra in a large-$\Pr$ medium.}

We have identified three scale ranges, $L\gg\lf\gg\ld\gg\lres$. 
To give a concrete example, let us give order-of-magnitude 
estimates for these scales in our Galaxy. 
The disk diameter is $L\sim10^4$~pc; the supernova scale at which 
the turbulence is stirred is $\lf\sim10^2$~pc; the Reynolds number 
is $\Re\sim10^5$, so the viscous scale is $\ld\sim10^{-2}$~pc; 
the Prandtl number is $\Pr\sim10^{14}$, so the resistive scale is~$10^4$~km, 
a tiny distance by Galactic standards (note \ssecref{ssec_plasma}).
The associated timescales are 
the period of Galactic rotation $T\sim10^8$~yr, the outer-eddy 
turnover time $\tf\sim10^7$~yr, and the viscous-eddy turnover 
time~$\td\sim10^5$~yr. The mean-field dynamo theory 
gives a large-scale Galactic field exponentiating in a 
rotation time $T$, while the timescale for the growth 
of the small-scale magnetic energy is $\td$ \citep{KA}. 
This illustrates our earlier point that the small-scale dynamo 
tends to be faster than the large-scale one.

\subsection{Simulating the Problem}
\label{ssec_numerics}

Simultaneously resolving all of these scales is 
not achievable in numerical simulations. 
Even if the large-scale object-specific features 
are forsaken and only isotropic homogeneous MHD turbulence 
with large $\Re$ and $\Pr$ is considered in a periodic box, 
it is still not possible 
to simulate both the hydrodynamic inertial and the magnetic 
subviscous ranges in the same box. 

In the relevant numerical work so far,
a popular course of action has been to run simulations with~$\Pr=1$ 
\citep[][our runs~\run{a1}, \run{a2}, and~\run{A} in \secref{sec_Re}]{Meneguzzi_Frisch_Pouquet,Kida_Yanase_Mizushima,Kleva_Drake,Miller_etal,Cho_Vishniac,Mininni_Gomez_Mahajan,HBD_apjl}. 
This choice is usually motivated by the view of MHD turbulence as 
a cascade of interacting Alfv\'en-wave packets 
\citep{Kraichnan_IK,GS}. 
In an Alfv\'en wave, velocity and magnetic fluctuations 
are equal, dissipation occurs at either viscous or resistive 
scale, whichever is larger, and whatever 
happens below that scale is usually not expected 
to affect the inertial-range physics. 

In fact, this theory only appears to work 
for {\em anisotropic} MHD turbulence with an externally imposed 
strong mean field 
\citep{Maron_Goldreich,CLV_aniso,Mueller_Biskamp_Grappin}. 
In (forced) simulations 
with zero mean field, one invariably sees that velocity and 
magnetic field are not symmetric, their statistics are not the 
same, they do not dissipate at the same rate or at the same scale, 
and there is no scale-by-scale equipartition of 
kinetic and magnetic energies, as there should have been if 
the turbulence had been purely Alfv\'enic 
\citep[][and see \secref{ssec_facts_sat}]{MCM}. 
The picture that does 
emerge is rather that of a very significant excess of magnetic energy 
at small scales and the magnetic-energy spectra resembling 
the resistively-dominated spectra of the small-scale 
large-$\Pr$ kinematic dynamo.

A natural step beyond $\Pr=1$ is
to look at Prandtl numbers above, but of the order of, 
unity \citep[][our run~\run{B} in \secref{sec_Re}]{Meneguzzi_Frisch_Pouquet,Chou,Brandenburg,Maron_Blackman,MCM,Archontis_etal_kin,Archontis_etal_nlin,HBD_pre} 
more or less in the hope that essential features of the $\Pr\gg1$ 
regime would be captured. This way, 
the inertial range might be resolvable, but the subviscous one (whose 
width is~$\sim\Pr^{1/2}$) is largely sacrificed. 
This approach is, to some extent, justified (see \secref{sec_Re}). 
However, as the asymmetry between magnetic and velocity field becomes 
more pronounced with increasing $\Pr$, one is faced with the 
realization that 
the $\Pr\gtrsim1$ case represents a nonasymptotic mixed 
state about which it is very hard to make any clear 
statements. We have pointed out in 
an earlier paper \citep[][see also \ssecref{ssec_ineff}]{SCHMM_ssim} 
that, in order for the 
effects of the scale separation between $\ld$ and $\lres$ 
on the nonlinear physics 
to be fully present, one must have $\Pr\gg\Re^{1/2}\gg1$. 
Astrophysical plasmas have no problem satisfying this condition, 
but numerically simulating this regime is not, as yet, possible. 

We believe that the key feature of isotropic MHD turbulence 
is the scale separation that arises between velocities and 
magnetic fields. In an atrophied nonasymptotic form, this scale 
separation is discernible even at~$\Pr=1$. 
Therefore, in most of this paper (with the exception of \secref{sec_Re}), 
we follow \citet{Kinney_etal} 
in choosing to resolve the subviscous range while sacrificing 
the inertial one. Namely, we set up our simulation 
in such a way that the forcing and viscous scales are 
comparable, while most of the available resolution is 
spent on the subviscous range.\footnote{Note 
that another approach in which low Reynolds numbers are often considered 
is to study magnetic-field growth and saturation in, 
and nonlinear modifications to, specific deterministic time-independent 
or time-periodic velocity fields with chaotic trajectories 
that are known to be dynamos in the kinematic limit \citep{STF}. 
The best-documented velocity fields of this sort are the ABC flows. 
The strategy then is to 
set up a forcing function that, in the absence of the magnetic 
field, would reproduce a given flow as a solution to the Navier--Stokes 
equation, and to keep $\Re$ below the stability threshold 
of this solution --- or to use some simplified model of the velocity 
equation that accomplishes the same, ---
and to model the fast-dynamo regime by maximizing $\Pr$ 
\citep{Galanti_Sulem_Pouquet,Cattaneo_Hughes_Kim,Maksymczuk_Gilbert}.
This approach differs from ours in that the velocity field is not 
random in time, nor is it homogeneous in space, so such features as 
stagnation points of the flow, transport barriers, etc., 
may acquire a degree of importance 
that cannot be a feature of real turbulence. 
The redeeming advantage here is in being able to deal 
with a velocity field that, unlike in the case of real turbulence, 
does not constitute an unsolved problem by itself, so one 
has more control over the numerical experiment and some of the 
nonlinear effects may be more straightforward to detect. 
In some of these studies, 
$\Re$ is allowed to exceed the hydrodynamic stability threshold 
\citep{Zienicke_Politano_Pouquet,Brummell_Cattaneo_Tobias}, 
in which case the small-scale results are probably 
universal because of the usual physical expectation that 
the small-scale properties of turbulence are insensitive to the 
particular choice of large-scale forcing. Of course, in order for 
such a universality argument to work, $\Re$ must be sufficiently 
large, and resolving large Prandtl numbers is again problematic.} 
At resolutions of $256^3$, 
we can study Prandtl numbers up to 2500, which just 
barely allows us to make some statements about the asymptotic 
$\Pr\gg1$ limit.
Furthermore, this approach allows us to study $\Pr$
dependence of our model, which is the only parameter scan 
that can at present be afforded. Some preliminary results obtained 
in the same setting (at lower resolutions) were published 
in \citet{SMCM_structure}.
 
\subsection{Plan of the Paper}

Our numerical set up, equations, and the limitations of the 
model are discussed in more detail in \secref{sec_model}. 
\Secref{sec_kin} is devoted to a detailed study of 
the kinematic regime, which will be important for understanding 
further developments. \Secref{sec_nlin} treats the nonlinear 
saturated state. Summaries 
are provided at the end of each of these two sections to help the reader 
identify the key points. In \secref{sec_Re}, 
our understanding of the physics of the nonlinear dynamo 
with large Reynolds numbers is outlined 
and some tentative comparisons are made 
with the results of 
simulations with large $\Re$ and $\Pr$ of order unity. 
\Secref{sec_concerns} discusses the scope of applicability 
of the large-$\Pr$-dynamo results 
and discusses directions of future work. 
An itemized summary of the main results and conclusions of the  
paper is given in~\secref{sec_summary}. 

\section{THE MODEL}
\label{sec_model}

\subsection{The Equations and the Code}

We consider the equations of incompressible~MHD:
\bea
\label{NSEq}
\Dt\vu &=& \nu\Delta\vu - \nabla p + \vB\cdot\nabla\vB + \vf,\\
\label{ind_eq}
\Dt\vB &=& \vB\cdot\nabla\vu + \eta\Delta\vB,
\eea
where $d/dt=\d_t+\vu\cdot\nabla$ is the advective time derivative, 
$\vu(t,\vx)$~is the velocity field, 
$\vB(t,\vx)$~is the magnetic field, 
and $\vf(t,\vx)$~is the body force. 
The pressure gradient $\nabla p$ (which includes magnetic pressure)
is determined by the incompressibility condition 
$\nabla\cdot\vu = 0$. We have normalized $p$ and $\vB$ 
by~$\rho$ and~$(4\pi\rho)^{1/2}$, respectively, 
where $\rho=\const$ is plasma density. 

\Eqsref{NSEq}{ind_eq} are solved in a triply periodic box 
by the pseudospectral method 
\citep[see code description in][]{Maron_Goldreich,MCM}. 
The body force $\vf$ is random, 
nonhelical, applied at the largest scales in the box ($\kf/2\pi=1,2$), 
and white in time (i.e., statistically independent at each time step): 
\bea
\label{f_corr}
\<f^i(t,\vx)f^j(t',\vx')\> = \delta(t-t')\,\epsilon^{ij}(\vx-\vx').
\eea
For a white-noise forcing, the average injected power is fixed:
\bea
\label{usq_balance}
{1\over2}\Dt\<u^2\> = -\nu\<|\nabla\vu|^2\> 
- \<\vB\vB:\nabla\vu\> + \eps,
\eea
where
$\eps = \<\vu\cdot\vf\> = (1/2)\epsilon^{ii}(0) = \const$.
The code units are based on box size 1 and injected power 
$\epsilon=1$.  

All runs are listed in \tabref{tab_index}. Runs~\run{S0}-\run{S6} are 
viscosity dominated (\ssecref{ssec_stokes_model}) with \run{S4} 
being the main time-history run. 
Runs~\run{a1}, \run{a2}, \run{A}, and~\run{B} have larger 
$\Re$ but smaller~$\Pr$ (see \secref{sec_Re}).

\subsection{Averaging} 

Mathematically speaking, all fields are random because 
the forcing is random. The system is isotropic 
and homogeneous in space, so all vector fields have zero mean 
and all points in 
the box are equally nonspecial. In theory, the angle brackets 
mean ensemble averages with respect to the forcing realizations 
and to the initial conditions. In practical measurements, 
one can (based on the usual 
ergodic assumption) use volume and time averaging instead. 
In what follows, wherever time histories are plotted, 
only volume averaging is done, while for quantities 
averaged over the kinematic or saturated stages of 
our runs, both volume and time averaging are performed. 
Error bars show the mean square deviations of volume 
averages at particular times from the time-averaged values. 

Averages in the kinematic regime are 
over 20 code time units 
(51 snapshots with time separation $\Delta t=0.4$) for runs S0--S4, 
over 6 time units (31 snapshots, $\Delta t=0.2$) for runs A and B, 
over 35 time units (71 snapshots, $\Delta t=0.5$) for run a1, 
and over 15 time units (31 snapshots, $\Delta t=0.5$) for run a2; 
averages in the saturated state are 
over 20 time units (51 snapshots, $\Delta t=0.4$) for runs S1--S6, 
over 10 time units (51 snapshots, $\Delta t=0.2$) for runs A and B,
and over 30 time units (61 snapshots, $\Delta t=0.5$) for runs a1 and a2. 
We have checked that increasing the averaging intervals 
does not alter the results. 
One code time unit roughly corresponds 
to the box crossing time (roughly the turnover time of the 
outer eddies). 

\subsection{The Viscosity-Dominated Limit}
\label{ssec_stokes_model}

As we can only adequately resolve either the inertial or the 
subviscous range, but not both, we choose to concentrate on 
the latter. In the runs reported in \secref{sec_kin} and 
\secref{sec_nlin} (but not in \secref{sec_Re}), the viscosity is fixed at 
$\nu=5\times10^{-2}$, which effectively leads to $\lf\sim\ld$. 
The velocity field in these runs is not, strictly speaking, turbulent. 
It is still random in time because the external forcing is 
random, but it is smooth in space as a result of very strong viscous 
dissipation. Advection by a random but spatially smooth flow 
is known as the {\em Batchelor regime} \citep{Batchelor_regime}.  
From the point of view of realistic turbulence, 
such a flow is probably a fairly good model of the viscous-scale eddies 
acting on the magnetic fluctuations in the subviscous 
range. In this interpretation, the external forcing $\vf$ 
models the energy supply from the Kolmogorov 
cascade to the motions at the viscous scale. 

We emphasize that,
while $\vf$ is white in time, the resulting velocity 
is not. In order to see this, let us consider \eqref{NSEq} in 
the more drastic limit of $\Re\ll1$, so the hydrodynamic 
nonlinearity can be neglected, and let the Lorentz forces 
be negligible as well. Then the solution of \eqref{NSEq} 
in $\vk$ space~is 
\bea
\label{u_OU}
\vu(t,\vk) = e^{-\nu k^2 t}\vu(t=0,\vk) 
+ \int_0^t\diff\tau\,e^{-\nu k^2\tau}\vf(t-\tau,\vk).
\eea
If $\vf$ is Gaussian, so is $\vu$, but $\vu$ has a finite 
correlation time: 
\bea
\label{u_corr}
\<u^i(t,\vk)\,u^j(t',\vk')\> = (2\pi)^3\delta(\vk+\vk')\, 
{1\over2\nu k^2}\,e^{-\nu k^2|t-t'|}\eps^{ij}(\vk),
\eea
where $\eps^{ij}(\vk)=\int\diff^3 y\,e^{-i\vk\cdot\vy}\eps^{ij}(\vy)$ 
is the Fourier transform of the forcing correlation tensor [\eqref{f_corr}], 
and we took $t,t'\gg(\nu\kf^2)^{-1}$, $\kf$ being the forcing 
wavenumber. The velocity correlation time is then 
$\tcorr\sim(\nu\kf^2)^{-1}$. \Eqsref{u_OU}{u_corr} are
strictly valid only if $\nu k^2\gg ku_k$ for all~$k$. 
In our simulations, we chose $\nu$ so that $\nu\kf^2\sim \kf u_\kf$ 
($\Leftrightarrow \kd\sim\kf$), i.e., just large enough for the 
inertial range to collapse. The obvious estimate for the 
velocity correlation time in this regime is $\tcorr\sim(\kf u_\kf)^{-1}$. 
Given the limitation of not having an inertial 
range, we believe this to be a sensible way 
of modelling turbulent motions. 

Explicit contact can be made between the viscosity-dominated 
model and the analytical results from the kinematic-dynamo theory. 
One of the very few time-dependent random flows for which kinematic dynamo 
is exactly solvable is a Gaussian white velocity field~$\vvKK(t,\vx)$: 
\bea
\label{KK_model}
\<\vKK^i(t,\vx)\,\vKK^j(t',\vx')\> = \delta(t-t')\kappa^{ij}(\vx-\vx').
\eea
This model was introduced by \citet{Kazantsev} and, in the context 
of passive-scalar advection, independently by \citet{Kraichnan_ensemble}. 
For the $\Pr\gg1$ problems, many quantities of interest then turn out 
to depend only on the first few coefficients in the Taylor expansion 
of the velocity correlation tensor: 
\bea
\nonumber
\kappa^{ij}(\vy) =& \kappa_0\delta^{ij} 
- {1\over2}\,\kappa_2\bl(y^2\delta^{ij} - {1\over2}\,y^i y^j\br)\\
& + {1\over4!}\,\kappa_4 y^2\bl(y^2\delta^{ij} - {2\over3}\,y^i y^j\br) + \dots
\label{kappa_coeffs}
\eea
Increasing $\nu$ in \eqsref{u_OU}{u_corr} 
above the value at which $\kd\sim\kf$ 
is equivalent to decreasing the correlation time of the flow. 
Thus, the velocity field in the viscosity-dominated model reduces to the 
Kazantsev velocity in the limit $\nu\to\infty$, 
provided that the force is rescaled to keep the integral 
of the time-correlation function constant: 
$\eps^{ij}(\vk)=\nu^2 k^4\kappa^{ij}(\vk)$
(i.e., $\vf=-\nu\Delta\vvKK$).
Since we owe much of our understanding of the kinematic dynamo 
to the Kazantsev model \citep[e.g.,][]{Kazantsev,Artamonova_Sokoloff,KA,Gruzinov_Cowley_Sudan,Chertkov_etal_dynamo,SCMM_folding,SCHMM_ssim,SCTHMM_aniso}, 
it is convenient to be able to consider the nonlinear problem 
as a well-defined extension of it. 

\subsection{Incompressibility} 

We use incompressible MHD even though 
most astrophysical flows are not incompressible at large scales. 
For example, the supernova forcing of the galactic 
turbulence produces sonic velocities at the outer scale. 
However, motions at smaller scales are subsonic 
and mostly vortical and incompressible 
\citep[e.g.,][]{Boldyrev_Nordlund_Padoan,Porter_Pouquet_Woodward,Balsara_etal}. 
As we are interested in the small-scale dynamo, which 
is driven by the smallest eddies, 
the incompressibility assumption is, therefore, reasonable. 
This view is further supported by 
the satisfactory outcome of a number of comparisons 
(carried out by N.~E.~Haugen 2003, private communication) 
between our simulations (\secref{sec_Re}) 
and the simulations of \citet{HBD_apjl,HBD_pre}, 
who used a compressible grid code.

\subsection{Helicity}

Our simulations are nonhelical in the sense that the net 
kinetic helicity is zero, $\<\vu\cdot(\nabla\times\vu)\>=0$ 
(helicity {\em fluctuations} are allowed). 
The net helicity of the flow can be controlled via the helical 
component of the external forcing \citep{Maron_Blackman}. 
Helicity is usually considered important because it is 
present in some of the astrophysical objects of interest. 
However, net helicity is a large-scale feature 
associated with overall rotation. Its effects should 
only be felt at timescales comparable to the rotation timescale, 
which is usually much longer than the turnover time of the 
turbulent eddies. Furthermore, helicity 
undergoes an inverse cascade from small scales to the largest 
scale in the box \citep{Frisch_etal_hel,PFL}. 
While net helicity is crucial for the generation of the box-scale 
magnetic field (essentially the mean field), the above considerations 
suggest that it probably does not play a significant role in 
the small-scale dynamo. 
In the viscosity-dominated simulations reported in \secref{sec_kin} 
and \secref{sec_nlin}, the box scale is the viscous scale 
and the external forcing models energy influx from 
larger scales, so allowing helical forcing 
is not a physically sensible choice. 
In the large-$\Re$ runs of \secref{sec_Re}, the effect of 
helicity would be interesting to study, but it would introduce 
another scale separation into the problem (between the box size 
and the forcing scale) and, therefore, 
require dramatic increases in numerical resolution. 

\subsection{Plasma Dissipation Processes}
\label{ssec_plasma}

Our choice of Laplacian diffusion operators 
to describe viscous and resistive cutoffs 
is only a very crude model of 
the actual dissipation processes in astrophysical plasmas. 

In a fully-ionized plasma, ions 
become magnetized already at relatively low magnetic-field strengths 
and the hydrodynamic viscosity must be replaced by a more 
general tensor viscosity 
\citep[][see also \citeauthor{Montgomery}~\citeyear{Montgomery}]{Braginskii}, 
which is locally anisotropic and very inefficient at damping 
velocity gradients perpendicular to the magnetic field: 
the perpendicular components of the viscosity tensor 
vanish in the limit of zero Larmor radius  
(the magnetic diffusivity also becomes anisotropic, 
but the difference between its perpendicular and parallel components 
is only a factor of~$\sim2$).
This can alter the physics at subviscous scales 
\citep{Malyshkin_Kulsrud,Balbus_MVI}. 

In partially ionized plasmas (e.g., in the interstellar medium), 
viscous dissipation is controlled by neutral species and can be assumed 
isotropic. However, in the presence of neutrals, ambipolar damping 
might enhance 
magnetic-field diffusion \citep[e.g.,][]{Zweibel_ambi1,KA,Proctor_Zweibel,Brandenburg_Zweibel,Brandenburg_Subramanian,Zweibel_ambi2,Kim_Diamond_ambi}. 

Finally, since the mean free path in astrophysical plasmas is 
often much larger 
than the scale associated with the Spitzer 
resistivity, the magnetic-field diffusion could be superseded 
by kinetic damping effects not contained in the MHD 
description \citep{Foote_Kulsrud,Kulsrud_etal_report}. 

These and other effects outside the MHD paradigm can be important. 
However, in this paper, we have opted to study the problem in the 
simplest available formulation: that provided by \eqsref{NSEq}{ind_eq}. 
We believe that this minimal model already contains 
the essentials of the small-scale dynamo physics.

\section{THE KINEMATIC DYNAMO}
\label{sec_kin}

The kinematic dynamo 
is a natural starting point for a numerical experiment. 
What we learn about magnetic fields in the kinematic 
limit will motivate our investigation of the nonlinear regime. 
This is a rather detailed study, so 
a busy reader anxious to get to the nonlinear matters 
may skip to the short summary in \secref{ssec_kin_sum} 
and on to \secref{sec_nlin}.

\pseudofigureone{fig_eddy}{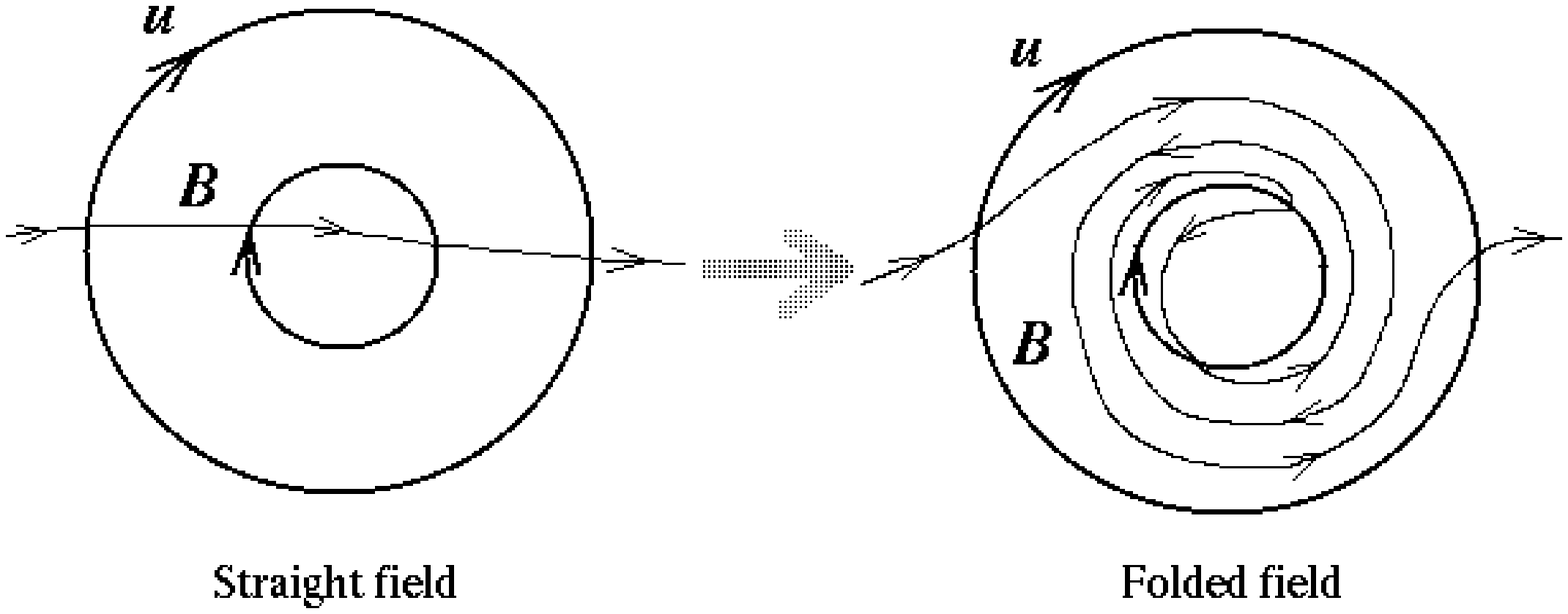}{f2.ps}{Stretching and folding of field lines 
by turbulent eddies.}

The small-scale turbulent dynamo is caused  
by the random stretching of the (nearly) frozen-in 
field lines by the ambient random flow. 
Stretching leads to exponential growth 
of the field strength \citep{Batchelor_dynamo}. 
The growth rate $\gamma\sim\nabla\vu$ 
is then naturally of the order of the inverse eddy-turnover time. 
In the case of Kolmogorov turbulence, it is the turnover time~$\td$ of the 
smallest (viscous-scale) eddies, because they are the fastest 
(see \ssecref{ssec_kin_AB}). 
For the magnetic fields at subviscous scales, the action of these 
eddies is roughly equivalent to the application of a random linear shear. 
Without diffusion ($\eta=0$), stretching is 
not opposed by any dissipation mechanism. 
However, stretching is necessarily accompanied by the refinement of 
the field scale (\figref{fig_eddy}), which 
proceeds exponentially fast in time and also at the eddy-turnover rate. 
Thus, the field scale eventually becomes 
comparable to the diffusion (resistive) scale, bringing an end to the 
diffusion-free evolution. The characteristic time during which 
diffusion-free considerations apply is 
$t\sim\gamma^{-1}\ln(\ld/\lres)\sim\gamma^{-1}\ln\bl(\Pr^{1/2}\br)$, 
assuming the initial field varies at the velocity scales. 

\pseudofigureone{fig_Et}{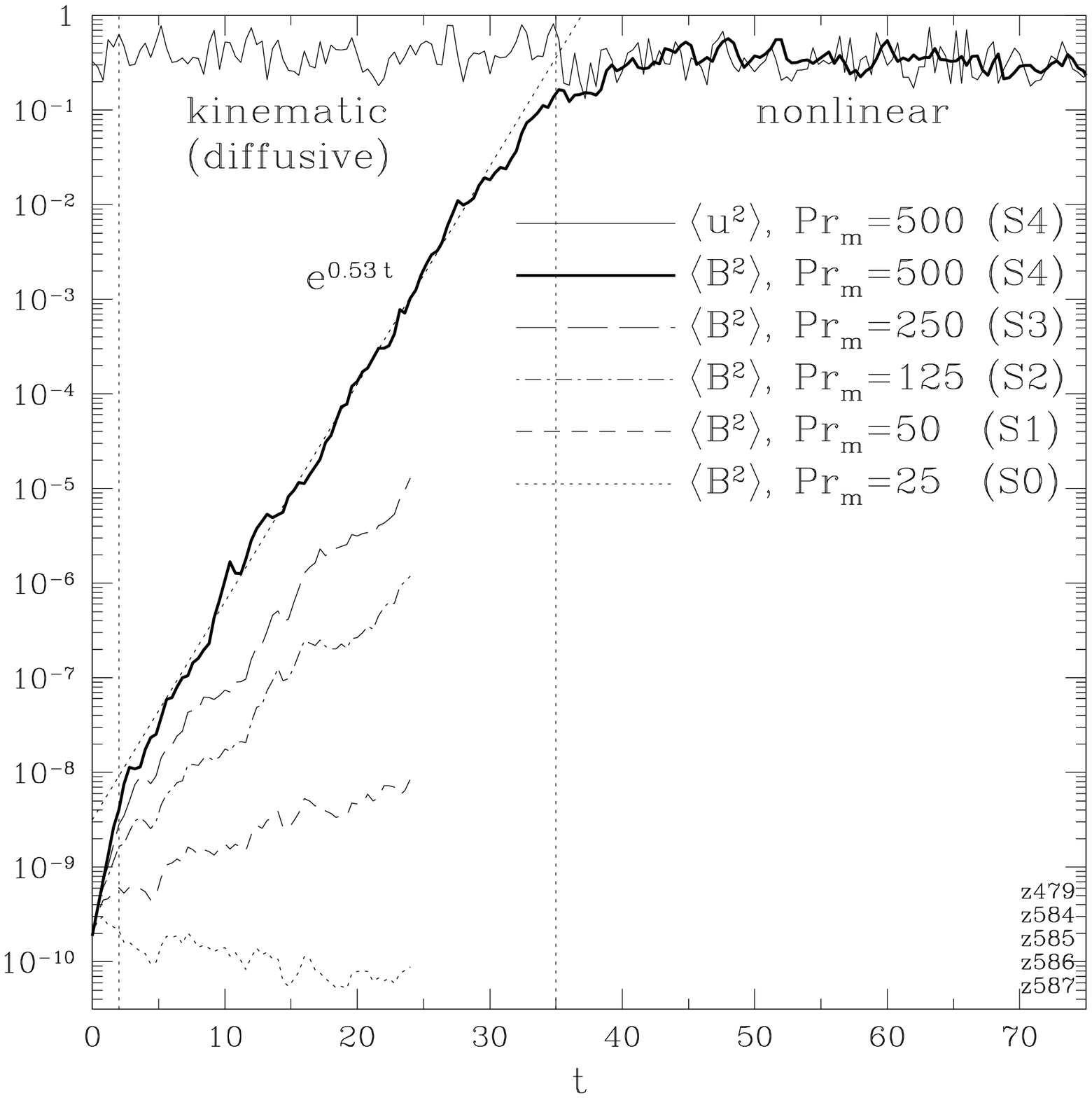}{f3.ps}{Exponential growth and saturation 
of the magnetic energy. Full time history 
is only shown for run~\run{S4}. 
See \tabref{tab_index} for the index of runs.}

\pseudofiguretwo{fig_Mk_kin}{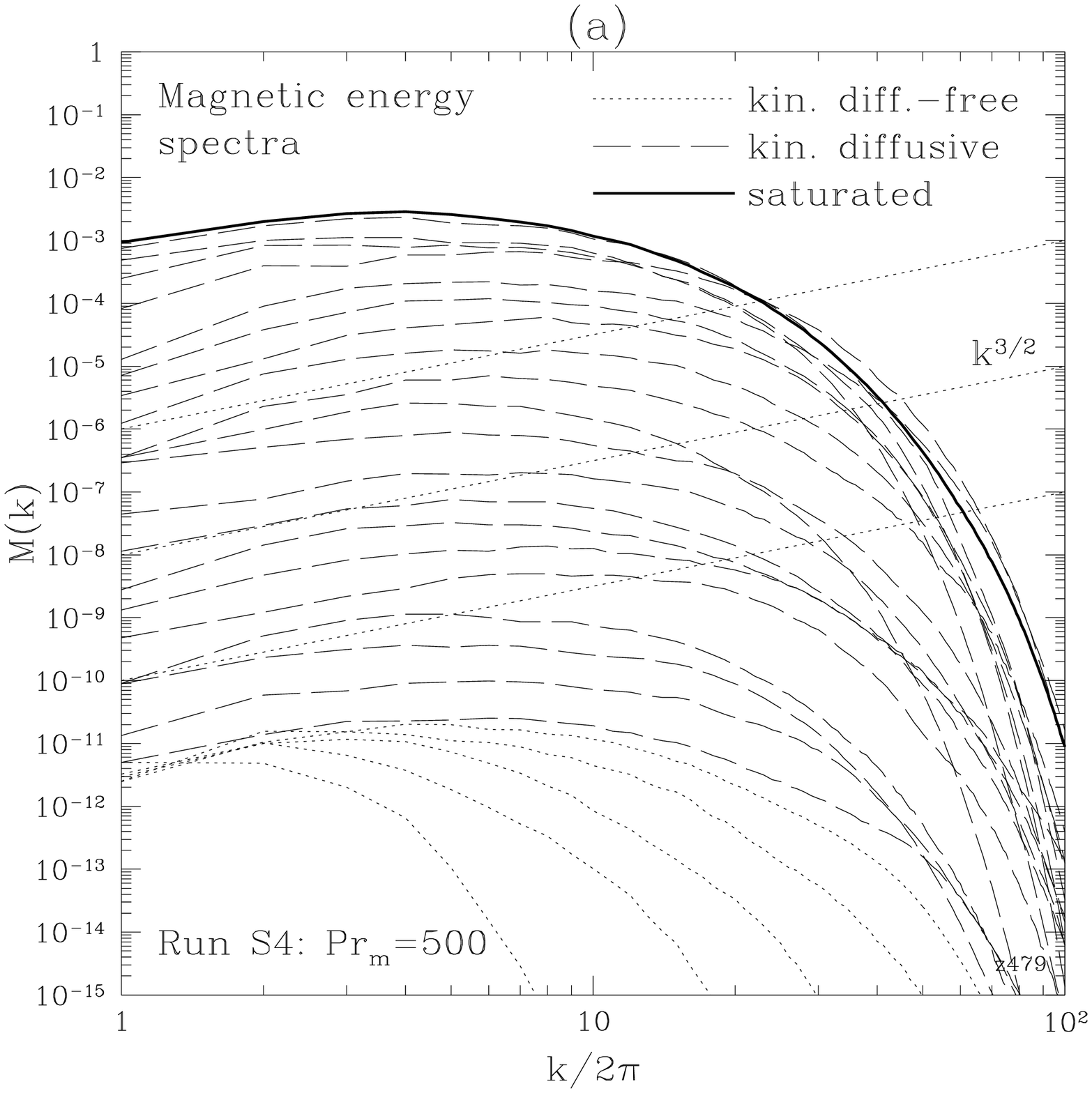}{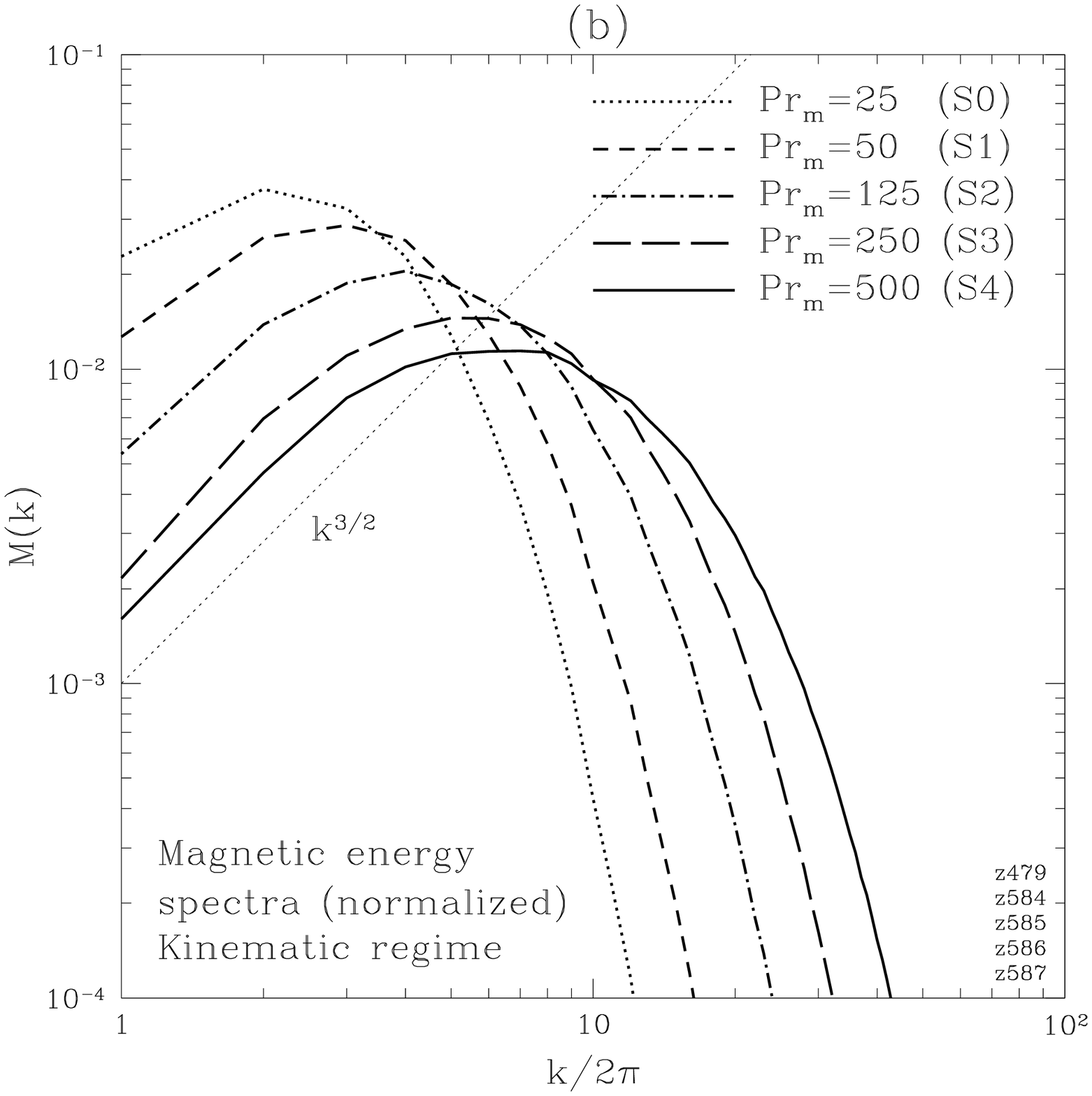}{f4a.ps}{f4b.ps}{(a) Evolution of 
the magnetic-energy spectrum for run \run{S4}. 
The spectra for the diffusion-free regime ($0\le t<2$) are given at 
time intervals of~$\Delta t=0.4$. The subsequent evolution ($2\le t \le 40$) 
is represented by spectra at time intervals~$\Delta t=2$.  
(b) Magnetic-energy spectra 
(normalized by the total magnetic energy and averaged)
during the kinematic stage. 
Note that for run~\run{S0} dynamo is resistively suppressed.}

Once diffusion enters the picture, it will (partially) offset 
the effect of random stretching (see \ssecref{ssec_zeld_mech}). 
If magnetic energy continues to grow exponentially 
in this regime with a growth rate that does not vanish 
as $\eta\to+0$, the dynamo is called {\em fast} 
\citep{Vainshtein_Zeldovich}. 
Whether any given flow is a fast 
dynamo is a problem that very rarely has an analytical 
solution \citep{STF}. 
In practice, since the advent of numerical simulations, 
turbulent flows have more often than not been found to support 
fast dynamo action (see references in \ssecref{ssec_numerics} 
and discussion in \ssecref{ssec_largePr}). 
In what follows, we shall see that magnetic fluctuations that result 
from this dynamo have a distinctive structure 
(see \ssecref{ssec_folding} and \ssecref{ssec_zeld_mech}).

In all our runs, the initial magnetic field had 
a spectrum concentrated at velocity scales so that both diffusion-free 
and diffusive regimes of the kinematic dynamo are realized. 
As a result of resolution constraints, the diffusion-free stage 
is never longer than 2 time units. The diffusive stage is 
quite long, so it is possible to extract 
statistical information by averaging over 20 time 
units. The magnetic energy grows 
exponentially from the initial value of~$\sim10^{-10}$ to 
saturated values of order unity (\figref{fig_Et}). 

In this Section, we study the properties of the growing field.
We concentrate on three main questions. First, 
what is its energy spectrum? 
This is considered in \ssecref{ssec_spectrum}. 
Second, what does the field ``look'' like? 
The field structure is studied in \ssecref{ssec_folding}.
Third, how does it fill the volume? 
This question is treated in \ssecref{ssec_ssim} 
in terms of the field-strength statistics.
\ssecref{ssec_zeld_mech} is a qualitative interlude 
discussing the physical reasons for the existence of 
the small-scale dynamo.

\subsection{The Magnetic-Energy Spectrum and Growth Rate}
\label{ssec_spectrum}

The evolution of the angle-integrated magnetic-energy spectrum, 
$M(k) = (1/2)\int\diff\Omega_\vk k^2\<|\vB(\vk)|^2\>$, 
in the large-$\Pr$ regime is 
the best understood part of the kinematic dynamo 
physics. For the Kazantsev velocity model, $M(k)$ can be shown 
to satisfy a closed integrodifferential equation,  
which, for $k\gg\kd$, reduces to a simple Fokker--Planck equation \citep{Kazantsev,Vainshtein_1,Vainshtein_2,Vainshtein_3,KA,Gruzinov_Cowley_Sudan,SBK_review} 
\bea
\label{M_eq}
\d_t M = {\gKA\over5}\,{\d\over\d k}\(k^2{\d\over\d k}\,M - 4k M\) 
+ 2\gKA M - 2\eta k^2 M,
\eea
where $\gKA=(5/4)\kappa_2\sim\nabla\vu$ 
is comparable to the inverse turnover time of the viscous eddies. 
Integrating \eqref{M_eq}, gives
\bea
\label{Bsq_eq}
\d_t\Bsq = 2\gKA\Bsq - 2\eta\kB^2\Bsq, 
\eea 
so $\gKA$ is the growth rate of the rms magnetic field 
in the absence of diffusion (the stretching rate). 
Without diffusion, \eqref{M_eq} describes a spreading lognormal 
profile with a peak moving exponentially fast toward ever larger 
wavenumbers,  
$k_{\rm peak}\propto\exp[(3/5)\gKA t]$, the amplitude of each 
mode growing exponentially at the rate $(3/4)\gKA$, and the 
dispersion of $\ln k$ increasing linearly in time as $(4/5)\gKA t$. 

Once the resistive scale is reached and diffusion becomes important, 
further increase of $k$ stops. The spectrum is a solution 
of an eigenvalue problem for \eqref{M_eq} with $M(k)\propto\exp(\lambda\gKA t)$. 
Here $\lambda<2$ because diffusion dissipates some of the 
dynamo-generated energy: 
\eqref{Bsq_eq} is an expression of this competition between stretching 
and diffusion. If $\lambda$ tends to a positive $\eta$-independent constant 
in the limit $\eta\to+0$, 
stretching wins and the dynamo is fast. 
In order to solve this eigenvalue problem, 
we must specify the boundary condition at small~$k$. 
In principle, this boundary condition can be determined only by 
solving the original integrodifferential equation of which 
\eqref{M_eq} is an asymptotic form for~$k\gg\kd$. 
This integrodifferential equation contains the velocity 
statistics at all scales and cannot be solved in a general and 
model-independent way. However, the eigenvalues are not 
very sensitive to the boundary condition: 
a natural choice is to impose a zero-flux constraint at some infrared 
cutoff $k_*$ \citep{SCHMM_ssim}, but a number of other reasonable 
options give the same solution\footnote{These include, e.g, requiring that the 
magnetic-field second-order correlation function decay exponentially at large 
distances \citep{Artamonova_Sokoloff,Subramanian} 
or setting $M(k_*)=0$ \citep{SBK_review}. 
\citet{KA} solved the integrodifferential 
equation for the spectrum numerically, using a Kolmogorov spectrum for velocity, 
and also obtained~$\lambda=3/4$. See also \citet{Gruzinov_Cowley_Sudan} for 
yet another argument leading to the same result.}
\bea
\label{Mk_sln}
M(k) & \simeq & (\const)\, e^{\lambda\gKA t}k^{3/2} K_0(k/\kres),\\
\label{lambda_sln}
\lambda & \simeq & {3\over4} - {\pi^2\over5[\ln(\kres/k_*)]^2}, 
\eea
where $\kres=(\gKA/10\eta)^{1/2}$ and $K_0$ is the Macdonald function.  
In the limit $\eta\to+0$, we get $\lambda\to3/4$, a fast dynamo. 
Note that $(3/4)\gKA$ is the rate at which individual $k$-space 
modes grow as a result of random stretching, so the effect of diffusion is 
simply to stop the magnetic energy from spreading to ever larger~$k$ 
\citep{KA,SBK_review}.

The numerical results broadly confirm this theoretical picture. 
In \Figref{fig_Et}, we show the evolution of the magnetic energy 
for several values of~$\Pr$. The growth is exponential in time 
and, for $\Pr=500$,  
the growth rate after resistive scales are reached ($t\gtrsim2$) 
is reduced compared to the diffusion-free 
regime by a factor that is actually quite close 
to $3/8$ predicted by the Kazantsev model.
\footnote{We hesitate to claim that this is 
a corroboration of the specific number that obtains in the Kazantsev 
model rather than a mere coincidence. The Kazantsev velocity is 
$\delta$-correlated in time, which cannot be a good quantitative description 
of the velocity field in our simulations (see \ssecref{ssec_stokes_model}). 
It is well known that finite-correlation-time effects can result in 
order-one corrections to the dynamo growth rate 
\citep{Chandran_tcorr,Kinney_etal,SK_tcorr,Chou,Kleeorin_Rogachevskii_Sokoloff_tcorr}. 
While the scalings in the Kazantsev model may be 
universal (e.g., the power tail of the curvature pdf, \ssecref{ssec_folding}), 
the values of growth rates can only {\em a priori} be considered 
as qualitative predictions.} 
The growth rate at smaller values of $\Pr$ is smaller, as a result of 
nonnegligible reduction by diffusion as suggested by 
the logarithmic correction in \eqref{lambda_sln}. 

The Prandtl number must exceed a certain critical value~$\Prc$ in order for the 
dynamo to be possible. Indeed, in \eqref{lambda_sln}, 
for $k_*\sim\kd$, we have $\kres/k_*\sim\Pr^{1/2}$ and setting $\lambda=0$ 
gives a rough estimate of the threshold 
below which the dynamo is resistively suppressed: $\Prc\sim 26$. 
In our runs, we have found growth at $\Pr=50$ and decay 
at $\Pr=25$, so $\Prc\in(25,50)$. 
Note that both the Kazantsev velocity with smooth correlation tensor~\exref{kappa_coeffs} 
and the velocity field in our simulations are single-scale flows with $\Re\sim1$, 
so $\Prc$ simply corresponds to the critical 
magnetic Reynolds number $\Rmc=\Re\,\Prc$. 
Since for our viscosity-dominated runs $\Re\simeq2$, 
we have $\Rmc\in(50,100)$. 

The diffusion-free spreading of the magnetic energy toward resistive scales 
and the subsequent self-similar growth of the spectrum 
with a peak at the resistive scale 
[\eqref{Mk_sln}] are manifest in \Figref{fig_Mk_kin}a. 
The $k^{3/2}$ scaling appears to be correct (\figref{fig_Mk_kin}b), 
although resolving a broader scaling range is necessary for definite 
corroboration. The occasional strong disruptions of the self-similar growth 
(\figref{fig_Mk_kin}a) are discussed at the end of~\ssecref{ssec_finite}. 

\subsection{The Folded Structure}
\label{ssec_folding}

The small-scale magnetic fields produced by the dynamo are not, in fact, 
completely devoid of large-scale coherence. The smallness of their 
characteristic scale is due primarily to the rapid (in space) field 
reversals transverse to the local field 
direction. Along themselves, 
the fields vary at scales comparable to the size of the eddies. 
This {\em folded structure} is schematically illustrated in \Figref{fig_eddy} 
and is evident in field visualizations (\figref{fig_folding_kin}).

There are many ways to diagnose the folded structure. 
Ott and coworkers studied the field reversals in map dynamos 
in terms of magnetic-flux cancellation: see review by 
\citet{Ott_review} and numerical results 
by \citet{Cattaneo_etal_cancel}. 
\citet{Chertkov_etal_dynamo} considered two-point correlation 
functions of the magnetic field and found large-scale 
correlations along the field and short-scale correlations 
across. The simplest approach is, in fact, to measure 
average characteristic scales at which the field varies  
in the directions perpendicular and parallel to itself. 
In two dimensions, this was done by \citet{Kinney_etal}. 
In three dimensions, the characteristic scales are studied 
in \ssecref{ssec_scales}. A more detailed description of the 
field's ``statistical geometry'' is provided by the distribution 
of the field-line curvature (\ssecref{ssec_curvature}) and by 
its relation to the field strength (\ssecref{ssec_anticorr}). 
The advantage of this approach is that curvature is a local 
quantity, so we only have to look at one-point statistics. 
It is also directly involved in the Lorentz tension 
force [\eqref{F_decomp}] thus providing information 
about the onset of the nonlinear back-reaction (\secref{sec_nlin}). 
An analytical theory based on the Kazantsev model
was developed in an earlier paper~\citep{SCMM_folding}. 
The numerical study below is motivated 
by the quantitative predictions made there. 

\pseudofigureone{fig_folding_kin}{B_z479_zsh100_small.epsf}{f5.ps}{Cross section 
of the magnetic field strength ({\em gray scale}) 
and in-plane field direction ({\em red arrows}) 
at $t=20$ during the kinematic stage of run~\run{S4}. 
Lighter regions correspond to stronger fields. 
Note that what appears to be strong-field ``clumps'' are, in fact, 
cross sections of folds transverse to the page.}

\pseudofiguretwo{fig_kt}{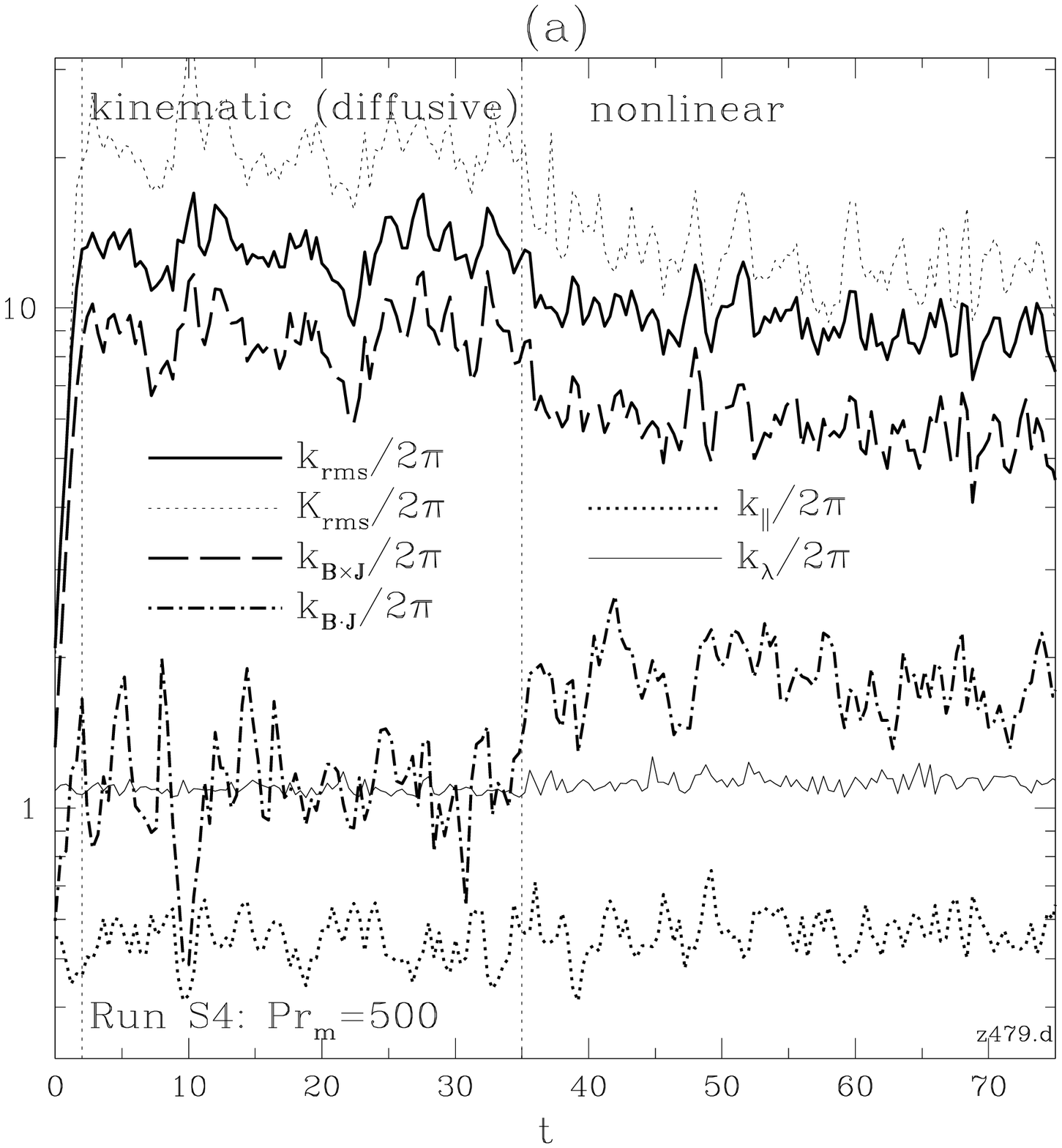}{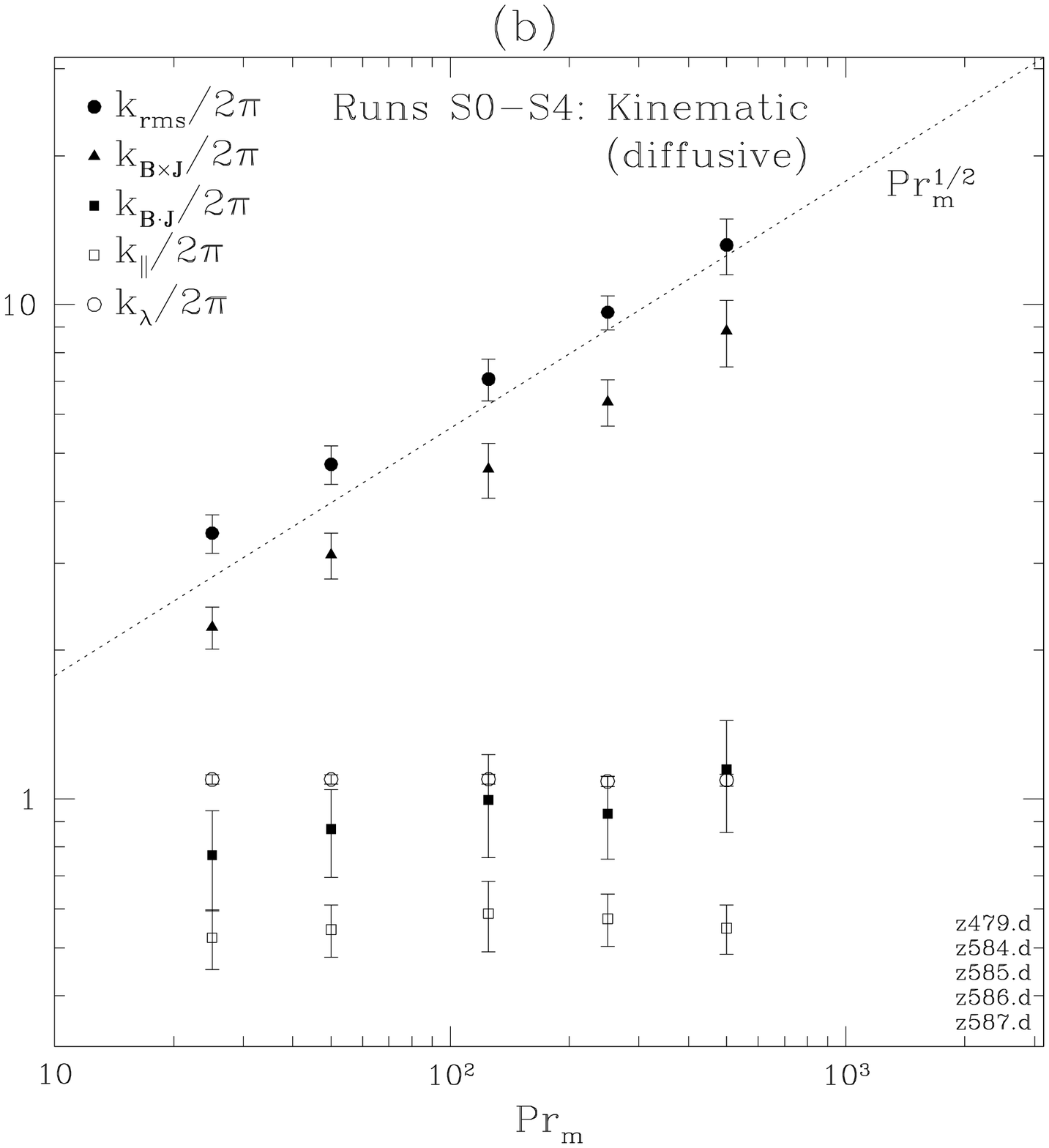}{f6a.ps}{f6b.ps}{(a) 
Evolution of characteristic wavenumbers (defined in \ssecref{ssec_scales})
and of $\Krms=\<|\vb\cdot\nabla\vb|^2\>^{1/2}$ 
for run~\run{S4}. 
(b) Averaged values of the same wavenumbers 
vs.~$\Pr$ during the kinematic stage of runs~\run{S0}-\run{S4}. 
Note that for run~\run{S0} dynamo is resistively suppressed. 
The values plotted are listed in \tabref{tab_index}.}

\subsubsection{Characteristic Scales}
\label{ssec_scales}

We define the characteristic parallel wavenumber of the field 
\bea
\label{kpar_def}
\kpar = \[{\<|\vB\cdot\nabla\vB|^2\>\over\Bfr}\]^{1/2} 
\eea
and its characteristic perpendicular wavenumber
\bea
\label{kbxj_def}
\kbxj = \[{\<|\vB\times\vJ|^2\>\over\Bfr}\]^{1/2},  
\eea
where $\vJ=\nabla\times\vB$. 
The overall field variation is measured 
by the rms wavenumber 
\bea
\label{krms_def}
\kB = \[{1\over\Bsq/2}\int_0^\infty\diff k\,k^2 M(k)\]^{1/2} 
= \[{\<|\nabla\vB|^2\>\over\Bsq}\]^{1/2}. 
\eea 
While both $\kbxj$ and $\kB$ grow exponentially during the 
diffusion-free regime and saturate at 
values that scale as $\kres\sim\kd\Pr^{1/2}$ 
\citep[cf.][]{Brummell_Cattaneo_Tobias}, 
the parallel wavenumber $\kpar$ remains approximately 
constant and comparable to $\kd$ \citep{SCMM_folding}: 
see \figref{fig_kt}. For comparison, we also plot 
(\figref{fig_kt}) the rms wavenumber of the velocity field, 
\bea
\label{kl_def}
\kl = \[{1\over\usq/2}\int_0^\infty\diff k\,k^2 E(k)\]^{1/2} 
= \[{\<|\nabla\vu|^2\>\over\usq}\]^{1/2} = {\sqrt{5}\over\lambda},
\eea
where $E(k)$ is the angle-integrated kinetic-energy spectrum 
and~$\lambda$ is the Taylor microscale of the flow defined in 
the standard way \citep{Frisch_book}. In Kolmogorov turbulence,  
$\lambda\sim\Re^{-1/2}\lf\sim\Re^{1/4}\ld$, so $\kl\sim\Re^{-1/4}\kd$. 

In order to estimate the degree of misalignment 
between the direction-reversing 
fields, it is instructive to look at a characteristic wavenumber of field 
variation in the direction orthogonal to both~$\vB$ and~$\vB\times\vJ$: 
\bea
\label{kbj_def}
\kbj = \[{\<|\vB\cdot\vJ|^2\>\over\Bfr}\]^{1/2}.
\eea
\Figref{fig_kt} and \tabref{tab_index} show that $\kbj\sim\kd$ and does not 
increase with $\Pr$ in the kinematic regime. Thus, the reversing straight 
fields are fairly well aligned and the flux surfaces are sheets.

\subsubsection{Fourth-Order Spectra}
\label{ssec_4order}

One can probe further into the behavior of the quantities $\vB\cdot\nabla\vB$, 
$\vB\times\vJ$, and $\vB\cdot\vJ$ by looking at their spectra. 
All these spectra are, in fact, controlled 
by the spectrum of~$B^2$, which turns out to be flat. 
Define angle-averaged spectra
\bea
\label{def_M4k}
M_4(k)&=&{1\over2}\int\diff\Omega_\vk k^2\<|[B^2](\vk)|^2\>,\\
\Mbxj(k)&=&{1\over2}\int\diff\Omega_\vk k^2\<|[\vB\times\vJ](\vk)|^2\>,\\
\Mbj(k)&=&{1\over2}\int\diff\Omega_\vk k^2\<|[\vB\cdot\vJ](\vk)|^2\>,\\
\label{def_Tk}
T(k)&=&{1\over2}\int\diff\Omega_\vk k^2\<|[\vB\cdot\nabla\vB](\vk)|^2\>.
\eea
We find (\figref{fig_M4})
that all these spectra grow self-similarly and 
\bea
\label{sln_M4k}
& M_4(k) \sim k^0,\qquad\quad
\Mbxj(k) \simeq {1\over4}\,k^2 M_4(k),\\ 
& \Mbj(k) \simeq \kbj^2 M_4(k),\qquad
T(k) \simeq \kpar^2 M_4(k), 
\label{sln_Tk}
\eea
with $\kpar$ and $\kbj$ defined by \eqsandref{kpar_def}{kbj_def}. 

This simple behavior is due to the folded 
structure. If the magnetic-field variation 
is dominated by direction reversals, 
then, for $\kd\ll k\ll\kres$, 
the main contribution to the integral 
\bea
\nonumber
\<|[B^2](\vk)|^2\> = \qquad&\\ 
\int\diff^3 k'\int\diff^3 k''& 
\<\vB(\vk')\cdot\vB(\vk-\vk')\vB(\vk'')\cdot\vB(-\vk-\vk'')\>
\label{convolution_BB}
\eea
is from $k',k''\sim\kres$ with the proviso that $\vk'$, $\vk''$,  
and $\vk$ are all in the direction of the reversal 
(i.e. transverse to the flux sheet). Expanding in~$\vk$, we get 
to zeroth order, $\<|[B^2](\vk)|^2\>\simeq\Bfr\propto k^0$. 
Since the spectrum 
is essentially one-dimensional (peaked for $\vk$ transverse to the 
flux sheet), we get $M_4(k)\sim k^0$. The rest of the fourth-order 
spectra can be worked out in the same way and \eqsref{sln_M4k}{sln_Tk} 
are readily obtained.

These results prove useful in \ssecref{ssec_tension} 
and \ssecref{ssec_tension_AB}.

\pseudofiguretwo{fig_M4}{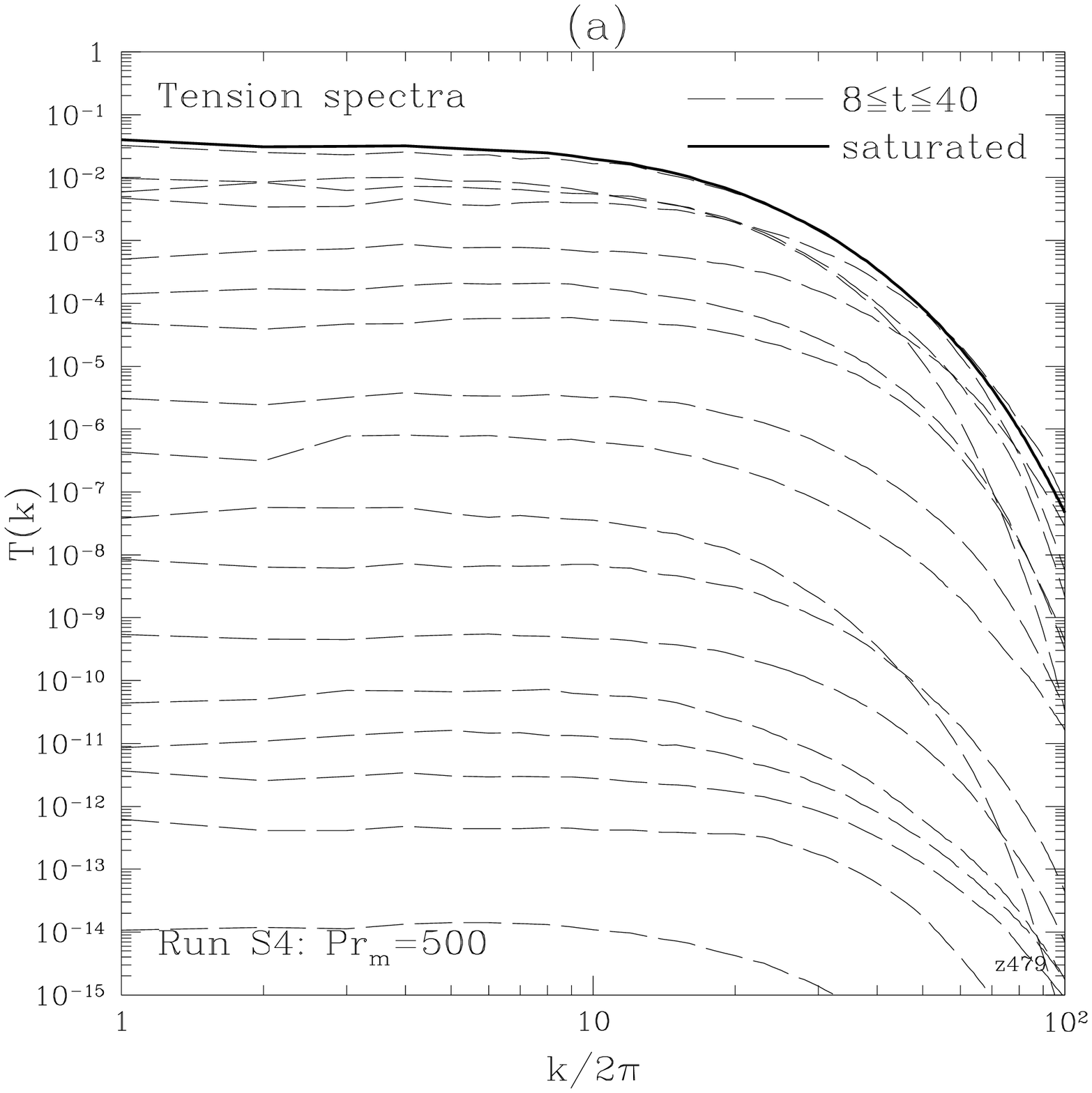}{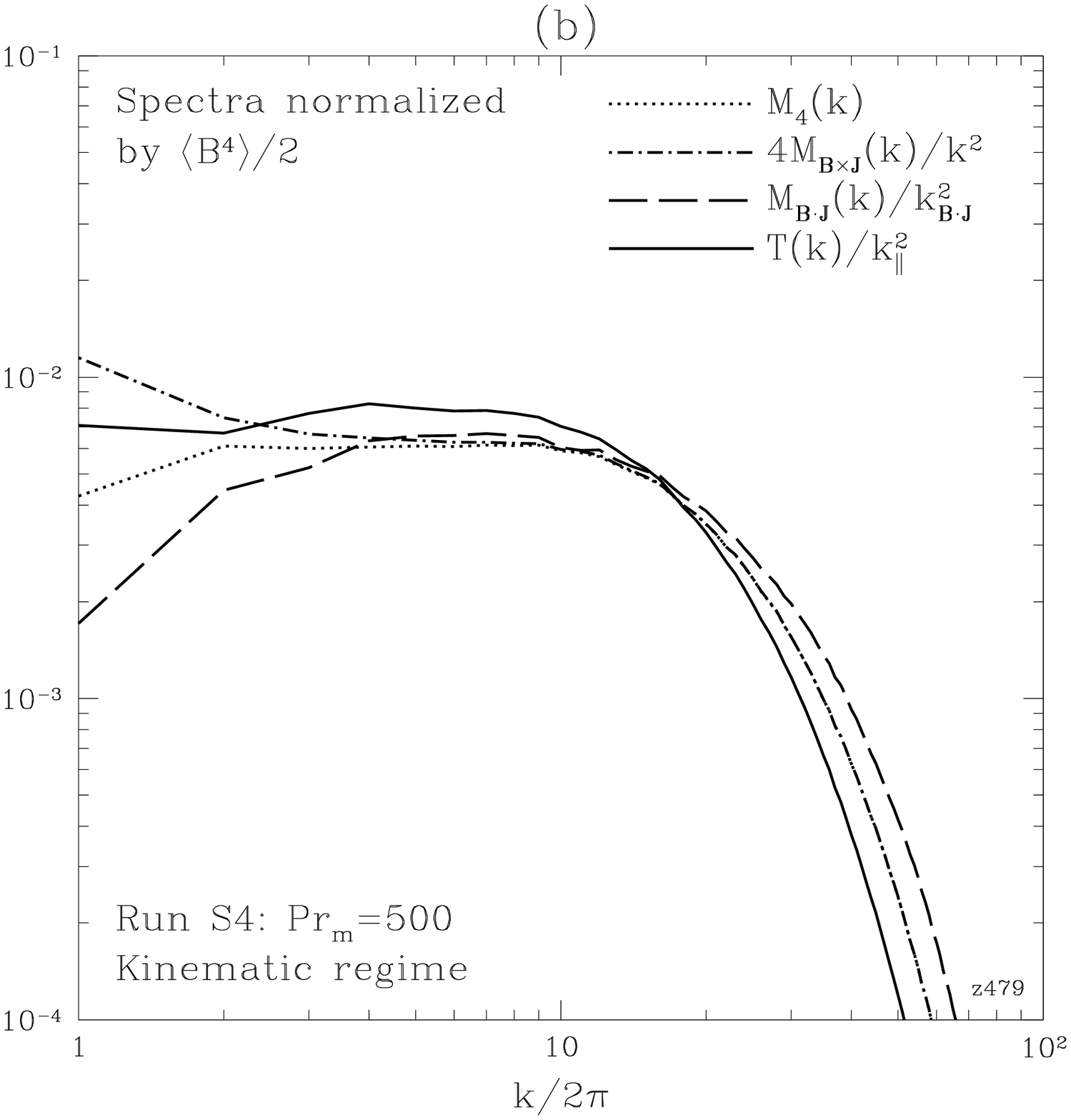}{f7a.ps}{f7b.ps}{(a) Tension 
spectra for run~\run{S4} for~$8\le t\le 40$ at intervals~$\Delta t=2$ 
and in the saturated state. 
(b) Fourth-order spectra defined in \eqsref{def_M4k}{def_Tk} 
(normalized by $\Bfr/2$ and averaged) during the kinematic stage 
of run~\run{S4}. The evolution of all these spectra is analogous 
to that of~$T(k)$ shown in (a).}

\subsubsection{Magnetic-Field Strength and Curvature} 
\label{ssec_anticorr}

The field geometry can be studied in a more detailed way in 
terms of the curvature $\vK=\vb\cdot\nabla\vb$ of the field lines 
($\vb=\vB/B$). 
$K=|\vK|$ satisfies \citep[cf.][]{Drummond_Muench} 
\bea
\label{K_eq}
\Dt K = \bl(\vn\vn:\nabla\vu - 2\vb\vb:\nabla\vu\br) K 
+ \vb\vb:\bl(\nabla\nabla\vu\br)\cdot\vn,
\eea
where $\vn=\vK/K$, and resistive terms have been dropped. 
\Eqref{K_eq} is a straightforward consequence of \eqref{ind_eq}. 
A standard kinematic calculation of the rms curvature, $\Krms=\Ksq^{1/2}$, 
shows that it grows exponentially in the diffusion-free regime 
\citep{Malyshkin_curvature,SCMM_folding}: see \figref{fig_kt}. 
Since curvature is an inverse 
scale, it is clear that in the diffusive regime, $\Krms$ must saturate 
at some $\Pr$-dependent value. 

The growth of~$\Krms$ does not contradict our previous claim that magnetic 
field lines are straight up to the flow scale. 
As the field is stretched and folded, it is organized in long thin 
structures (folds) where it is only significantly curved in the bends 
(turning points). The flow stretches the straight segments 
of the field, while the curved fields in the bends remain weak 
\citep[this is simply a consequence of flux and volume 
conservation: see][]{SCMM_folding,SMCM_structure}. 
Thus, there is anticorrelation between the field-line curvature and 
the field strength. 

The most straightforward statistical measure of this 
anticorrelation is the correlation coefficient defined as follows
\bea
\label{CKB_def}
C_{K,B} = {\<K^2 B^2\> - \Ksq\Bsq\over\Ksq\Bsq}.
\eea 
In all runs, it is found to be within 6\% of its
minimum possible value of~$-1$ (\tabref{tab_index}). 
Furthermore, we notice that the mean square tension force 
[see \eqref{NSEq}] is
\bea
\label{F_decomp}
\<|\vB\cdot\nabla\vB|^2\>=\<B^4 K^2\> + \<B^4|\dpar B/B|^2\>,
\eea
so the fact that $\Krms^2\sim\<|\vB\cdot\nabla\vB|^2/B^4\>$ 
is large (and grows with $\Pr$)
while $\kpar^2=\<|\vB\cdot\nabla\vB|^2\>/\Bfr\sim\kd^2$ 
also indicates that the curvature in the regions of growing field 
remains comparable to the inverse scale of 
the flow.\footnote{The second term in \eqref{F_decomp} represents 
the contribution from the mirror force~$\dpar B/B$, which is also 
large only in the bends, where its rms value grows at the same 
rate as~$\Krms$ \citep{SCMM_folding}. 
The pdf of the mirror force has a power tail with the same 
scaling as the pdf of curvature (\ssecref{ssec_curvature}) and 
is, therefore, also dominated by the outer scales 
(A.~A.~Schekochihin 2003, unpublished).} 

Let us give an illustration of the anticorrelation between 
$B$ and $K$ for a typical snapshot of the field. 
In \citet{SCMM_folding}, we noted that this anticorrelation was manifest  
in cross sections of field-strength and curvature. 
\Figref{fig_folding_kin} does, indeed, show that field lines 
are straight (and direction reversing) in the areas of strong field.
The same point can be made by scatter plots of 
$B$ versus~$K$ during the kinematic 
stage of our run~\run{S4} (\figref{fig_BK_kin}). 
Since the distributions of both magnetic field 
(\ssecref{ssec_ssim}) and curvature (\ssecref{ssec_curvature}) 
are intermittent, the scatter is quite wide.
It is clear, however, that magnetic fields with $B\gtrsim\Brms$ 
have curvatures well below~$\Krms$ while the fields 
with curvatures $K\gtrsim\Krms$ are quite weak. 

If we consider \eqref{K_eq} together 
with the evolution equation for the magnetic-field strength, 
\bea
\label{B_eq}
\Dt B = \bl(\vb\vb:\nabla\vu\br)B + \eta\Delta B - \eta|\nabla\vb|^2 B, 
\eea
and drop both the resistive terms and the second derivatives of 
the velocity field (bending terms), we might observe that 
\bea
\Dt\ln\bl(BK^{1/2}\br) = {1\over2}\bl(\vn\vn:\nabla\vu\br). 
\eea
The formal solution in the comoving frame is 
\bea
\label{BK_sln}
\ln\bl(BK^{1/2}\br)(t) = {1\over2}\int^t\diff t'\,\vn\vn:\nabla\vu\,(t') 
\equiv {1\over4}\,\zetan(t),
\eea
while for the field-strength, we have 
\bea
\label{B_diff_free}
\ln B(t) = \int^t\diff t'\,\vb\vb:\nabla\vu\,(t') 
\equiv {1\over2}\,\zetab(t).
\eea
It can be shown (A.~A.~Schekochihin 2002, unpublished), 
for the Kazantsev model velocity~\exref{KK_model} 
and in neglect of bending and diffusion, 
that the joint pdf of $\zetab(t)/t$ and $\zetan(t)/t$ is 
exactly the same as the joint pdf of $\zeta_1(t)/t$ and 
$\zeta_2(t)/t$, the finite-time Lyapunov exponents corresponding 
to the stretching and the ``null'' direction of the flow 
(see \ssecref{ssec_zeld_mech} for a quick overview of the 
relevant definitions). This means that, while $\zetab(t)$ 
increases linearly with time and is responsible for 
the field stretching, $\zetan(t)$ fluctuates around zero. 
\Eqref{BK_sln} then suggests that $BK^{1/2}$ is a 
special combination in which the effect of stretching 
is cancelled.\footnote{In \citet{SMCM_structure}, we argued 
that $BK\sim\const$ based on a flux-conservation argument, 
which, however, involved some {\em ad hoc} assumptions about the fold 
geometry. \citet{Constantin_Procaccia_Segel} and 
\citet{Brandenburg_Procaccia_Segel} also argued in favor 
of $BK\sim\const$ on the basis of an alternative form of 
\eqref{K_eq}. The numerical evidence presented here 
and in \ssecref{ssec_fold_nlin} 
appears rather to support $BK^{1/2}=\const$. 
It is, however, possible that in the regions of strong 
field and low curvature, the relation between $B$ and $K$ 
is closer to $B\sim1/K$.} 
Least-squares fits performed 
on log-scatter plots of $B$ versus~$K$ such as \figref{fig_BK_kin} 
give $BK^{\alpha}=\const$ with $\alpha\simeq0.47$, which gives 
a measure of both the detailed anticorrelation between $B$ and $K$ 
and the extent to which it can be understood 
via essentially geometrical arguments such as the one we 
have just presented.\footnote{The underlying geometrical nature of 
the curvature-related statistics becomes especially clear in 2D.
The curvature and field-strength in 2D are related by 
$BK^{1/3}=\const$ (again neglecting diffusion and bending). 
This follows from the fact that, 
in 2D, $\vn\perp\vb$, so, $\vb$ being the stretching direction, 
$\vn$ must necessarily be the compressive one: 
incompressibility then requires $\zetan=-\zetab$, 
whence $d\ln(BK^{1/3})/dt=0$. 
A statistical calculation 
for the Kazantsev model shows that the pdf of 
$\ln(BK^{1/3})$ is a $\delta$~function 
(A.~A.~Schekochihin 2002, unpublished). 
A purely geometrical consideration of the field lines also gives 
$BK^{1/3}=\const$ \citep{Thiffeault_curvature}. 
Furthermore, it turns out that the power tail of the curvature pdf 
in 2D derived by \citet{SCMM_folding} 
for the Kazantsev dynamo model, $P_K(K)\sim K^{-5/3}$, can 
be reproduced for the curvature distribution along a generic parabola 
(J.-L.~Thiffeault 2002, private communication). 
It is an open question whether 
similar purely geometric treatment is possible in 3D.\label{fn_curvature}} 
Clearly, these should break down for curvatures close to 
either the inverse flow scale ($K\sim\kd$) or the 
inverse resistive scale ($K\sim\kres$), but it is nontrivial 
that they appear to work quite well away from the cutoffs. 

\pseudofigureone{fig_BK_kin}{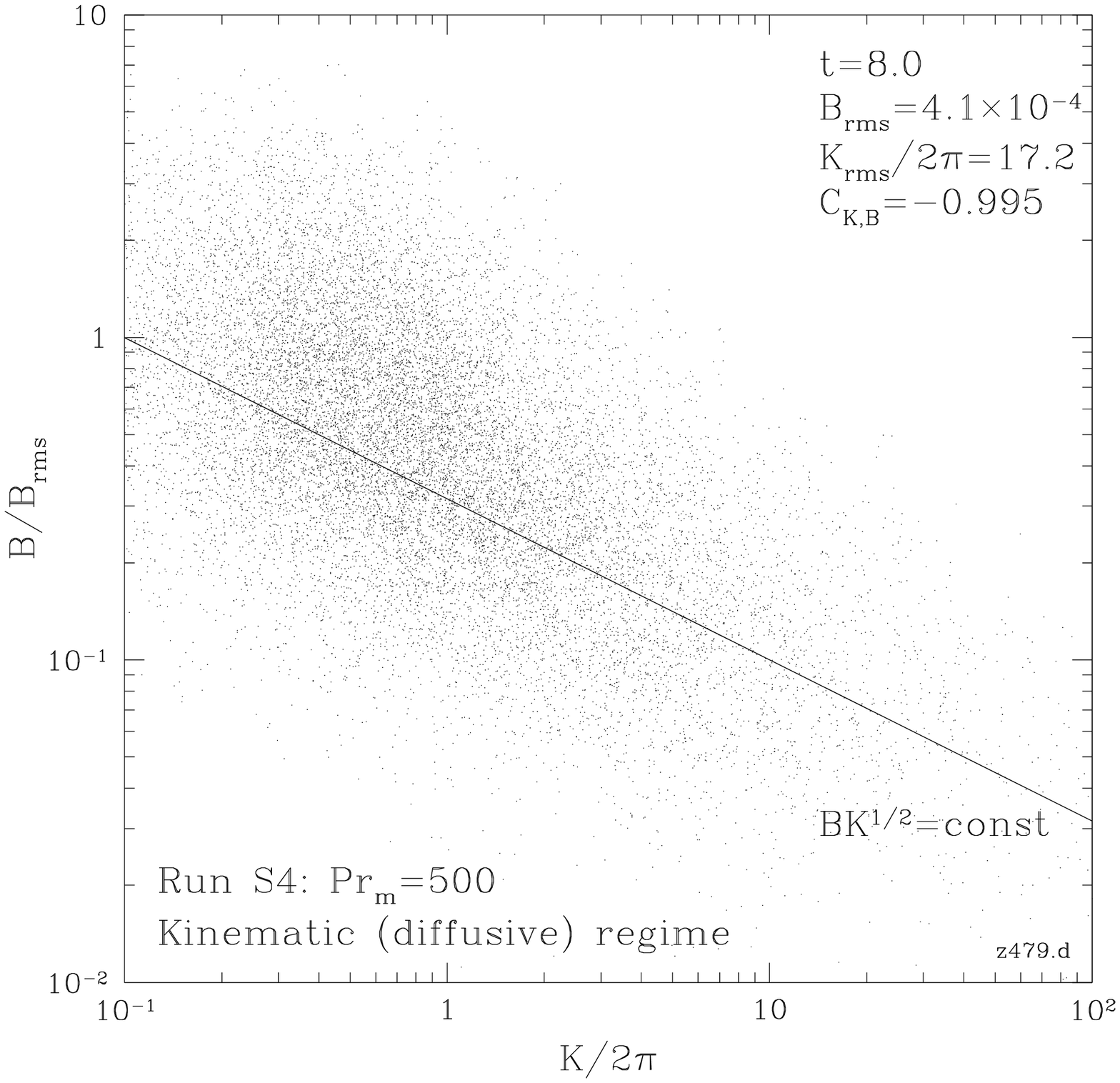}{f8.ps}{Scatter plot of 
$B$ vs.~$K$ at $t=8$ during the kinematic stage of our run~\run{S4} 
(the $256^3$ data were thinned out by a factor of~$1000$).}

\subsubsection{The Curvature Distribution}
\label{ssec_curvature}

We have seen that 
the large values of~$\Krms$ are due to the large values of curvature 
in the bends, where the field itself is weak. 
\Figref{fig_BK_kin} indicates that these only occupy a small fraction 
of the volume (in most places, fields are strong and straight). 
In order to ascertain that this is true, 
as well as to get a more detailed statistical description of the 
field-line geometry, \citet{SCMM_folding} found the pdf of curvature 
analytically: 
\bea
\label{PK_st}
P_K(K)={6\over7}\,{K\over[1+(K/K_*)^2]^{10/7}},
\eea
where $K_*=(2\kappa_4/7\kappa_2)^{1/2}$ [see \eqref{kappa_coeffs} for 
definitions of~$\kappa_2$ and~$\kappa_4$]. This solution means that 
in most of the volume, the field-line curvature is comparable to 
the inverse eddy size ($K\sim K_*\sim\kd$), while the distribution of 
curvatures in the bends is characterized by a power law 
$P_K(K)\sim K^{-13/7}$. This scaling is reproduced very well 
in our simulations (see \figref{fig_PK}). 
Note that we find all curvature-related 
quantities to be quite well converged already after a relatively short 
running time (unlike the quantities containing field-strength, which 
fluctuate very strongly). We believe that this is due to 
statistics of curvature being a geometrical 
property of the field lines independent of the fluctuating 
stretching rates (see footnote~\ref{fn_curvature}).

A stationary power-like pdf of~$K$ is possible because curvature, unlike 
field strength, has an explicit dependence not only on~$\nabla\vu$ 
but also on~$\nabla\nabla\vu$. 
The second derivatives of $\vu$ enter as 
a source term in \eqref{K_eq} and are responsible for bending 
the fields. Thus, the flow scale is explicitly 
present in the curvature equation and, consequently, 
in the curvature statistics. 

Formally speaking, integer moments of the distribution~\exref{PK_st} 
diverge: in the diffusion-free calculation of 
\citet{SCMM_folding}, while the pdf converges 
to the stationary profile~\exref{PK_st}, all moments~$\<K^n\>$ 
grow exponentially 
without bound. In the problem with diffusion, the power tail of the curvature 
pdf is cut off at the resistive scale ($K\sim\kres$) and the moments saturate 
at values that scale with~$\Pr$: e.g., $\Krms\sim\kd\Pr^{2/7}$ asymptotically 
with~$\Pr\to\infty$. 
The numerical results demonstrate the (nontrivial) fact that the curvature 
statistics above 
the resistive scale are not affected by the presence of diffusion. 
In other words, there is no resistive anomaly for the curvature: 
$\eta\to+0$ and $\eta=0$ give the same result. 
Note that the use of hyperdiffusion  does not change the scaling 
of $P_K(K)$ \citep{SMCM_structure} --- another indication that 
curvature statistics do not depend on the dissipation mechanism. 
It can be seen in \ssecref{ssec_ssim} that, unlike the curvature,  
the field strength has statistics that are crucially influenced 
by the resistive cutoff. 

\pseudofigureone{fig_PK}{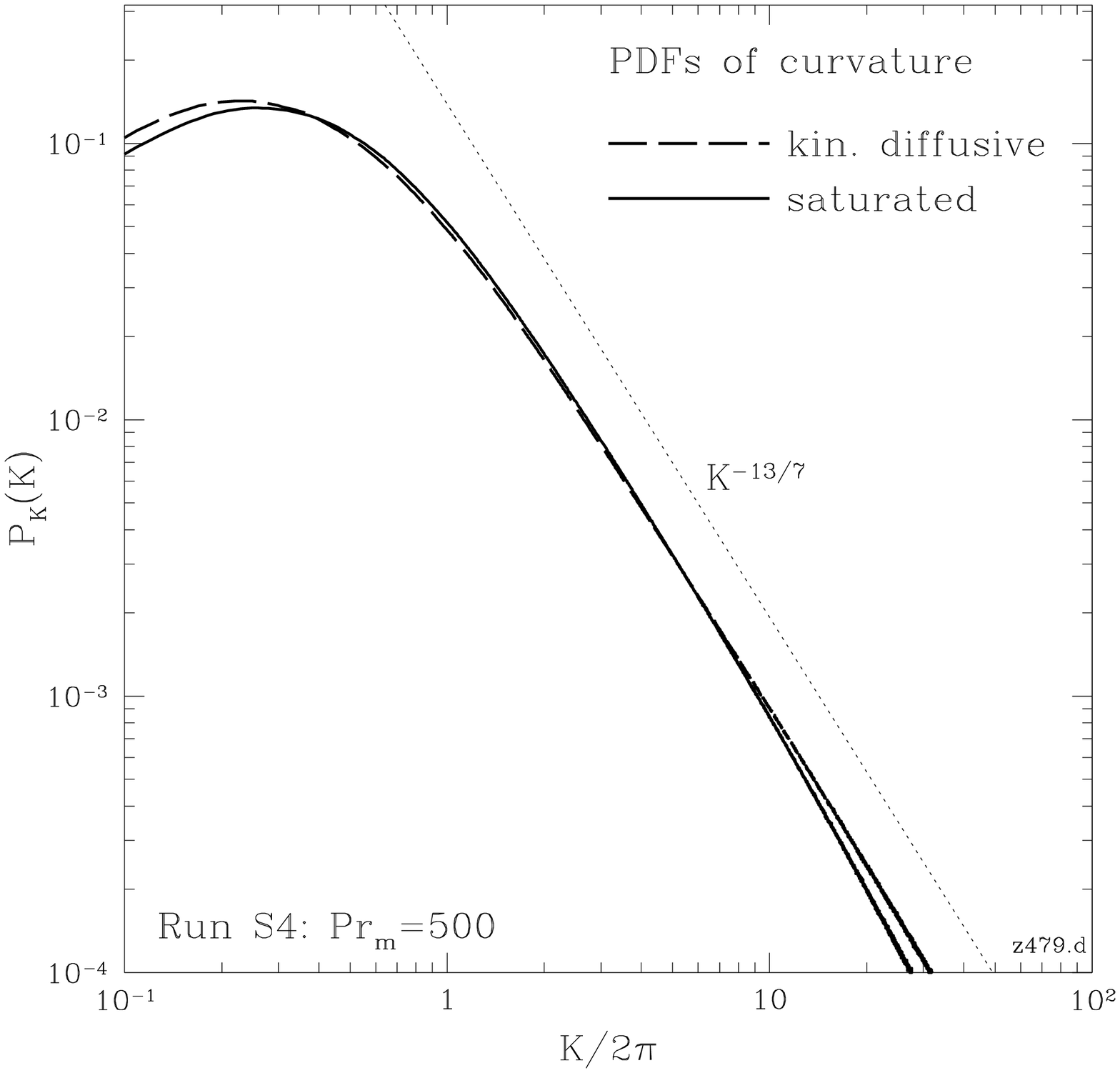}{f9.ps}{Curvature pdf's for run \run{S4} 
during the kinematic regime (with diffusion) 
and in the saturated state. 
Curvature pdf's for all other runs are similar.} 

A note of caution is in order. 
We have found in the course of our numerical 
investigations that the geometric properties of the dynamo-generated 
field, while very well behaved in terms of convergence in time, 
are quite sensitive to spatial resolution. 
Specifically, with our spectral code, if the 
magnetic diffusivity~$\eta$ is not large enough to ensure 
a fairly large separation between the resistive scale and 
the dealiasing cutoff, the folded structure is polluted by 
spurious small-scale ``ringing'' in the field. The method of 
computing curvature is also important. Better-quality results are 
obtained by using the formula 
$\vK=(\vB\cdot\nabla\vB)\cdot\bl(\unity-\vb\vb\br)/B^2$, 
than by directly differentiating the unit 
vector ($\vK=\vb\cdot\nabla\vb$).

\subsection{How Is the Small-Scale Dynamo Possible?}
\label{ssec_zeld_mech}

The folded-structure diagnostics considered in \ssecref{ssec_folding} 
tell us what field configurations are generated by random 
stretching. Diffusion did not figure prominently in the 
theoretical discussion that we have offered in support of our 
numerical results. Its only effect has been to set the 
minimum scale of field variation (reversals and curvature). 
However, if the fields reverse at the resistive scale, 
why are they not destroyed by diffusion? 
In other words, how is the small-scale dynamo possible at all? 

Let us consider the simplest imaginable model of turbulent 
stretching: a velocity field random in time 
and exactly linear in space \citep{Zeldovich_etal_linear}. 
In this model, any physics that depends on the finiteness 
of the flow scale is lost: for example, the stationary curvature 
pdf (\ssecref{ssec_curvature}) cannot be derived for an exactly linear 
velocity (a more serious shortcoming 
will be the incorrect result for the growth rates of $\<B^n\>$: 
see \ssecref{ssec_ssim}). 
The fields produced by a linear flow still 
have a folded structure, but there is no bending, so 
the folds can be arbitrarily long.

If the velocity field is linear, \eqref{ind_eq} is solved (in the comoving frame) 
by the Ansatz 
\bea
\label{Zeld_ansatz}
\vB(t,\vx) = \int{\diff k_0^3\over(2\pi)^3}\,\tvB(t,\vk_0) 
e^{i\vk(t,\vk_0)\cdot\vx},
\eea
where $\vk(0,\vk_0)=\vk_0$, and $\tvB(0,\vk_0)=\vB_0(\vk_0)$ is the Fourier 
transform of the initial field. After direct substitution 
into \eqref{ind_eq}, we find the following solution
\bea
\label{B_sln}
\tB^i(t,\vk_0) &=& {\d x^i\over\d\xz^m}(t)\, B_0^m(\vk_0)\,
\exp\[-\eta\int_0^t\diff t' k^2(t,\vk_0)\],\\ 
\label{k_sln}
k_i &=& {\d\xz^m\over\d x^i}(t)\, k_{0m},
\eea
where $\vx_0\to\vx(t)$ is the transformation of variables 
induced by the flow, i.e., 
$\vx(t) = \vx_0(t) + \int_0^t\diff t'\vu(t',\vx(t'))$.
For a linear velocity field, 
$\d x^i/\d\xz^m$ depends on time only.

Solution~\exref{B_sln} is a generalization of the 
Cauchy solution of the ideal induction equation.
It can be used to express the statistics of the magnetic field 
in terms of its initial statistics and the statistics of the 
(contravariant) metric associated with the flow
\bea
\label{g_def}
g^{ij}(t) = {\d x^i\over\d\xz^m}{\d x^j\over\d\xz^m}.
\eea
The metric contains all the statistical information necessary to describe 
random advection of arbitrary tensor fields by the ambient velocity field 
\citep[e.g.][]{BS_metric}. Since $g^{ij}$ is symmetric and positive definite, 
it has a set of real 
positive eigenvalues $\{e^{\zeta_1},e^{\zeta_2},e^{\zeta_3}\}$ 
and an orthonormal basis $\{\eone,\etwo,\ethree\}$.
This is the Lyapunov basis of the flow, and $\zeta_i(t)/t$ are 
the finite-time Lyapunov exponents. The quantities $\zeta_i$ can be 
considered ordered so that $\zeta_1\ge\zeta_2\ge\zeta_3$. 
Incompressibility requires that $\zeta_1+\zeta_2+\zeta_3=0$ at all times. 
Therefore, $\zeta_1>0$ corresponds to the stretching direction, 
$\zeta_3<0$ to the compressive 
direction, and $\zeta_2$ to the ``null'' direction. 
The exponent $\zeta_2$ can be either 
positive or negative. In time-reversible flows 
(e.g., in the Kazantsev model), 
$\<\zeta_2\>=0$ (in real turbulent flows, 
it is usually positive). 
Note that Lyapunov eigenvectors stabilize exponentially fast 
in time, while the finite-time Lyapunov exponents $\zeta_i/t$ 
converge to constant values as $\sim1/t$ \citep[][]{Goldhirsch_Sulem_Orszag}. 

\Eqref{B_sln} implies that $\vB$ aligns with the stretching direction 
$\eone$. In an ideal fluid, 
\bea
\label{BB_ideal}
B^2\sim\exp\,(\zeta_1) 
\eea
for a typical realization.\footnote{In other words, $\zeta_1=\zetab$, where 
$\zetab$ is defined in \eqref{B_diff_free}. As we noted in 
\ssecref{ssec_anticorr}, it is also possible to show that 
the curvature $\vK=\vb\cdot\nabla\vb$ aligns with 
the ``null'' direction and $\zeta_2=\zetan$ [\eqref{BK_sln}] 
for a linear velocity field.} 
While $\tvB$ transforms as a vector [times the exponential factor due to 
diffusion: see \eqref{B_sln}], $\vk$ transforms as a covector 
[\eqref{k_sln}].  
Therefore, it aligns with the stretching direction of the {\em covariant} 
metric, 
which is the inverse of~$g^{ij}$. Its stretching direction 
is then the same as the compressive direction of~$g^{ij}$, 
with exponent $-\zeta_3>0$, 
so $\vk$ aligns with the compressive direction 
$\ethree$ and grows exponentially in time. 
Therefore, most modes $\tvB(t,\vk_0)$ [\eqref{B_sln}] 
decay superexponentially.  
The only modes that are not thus suppressed are those for which 
the angle between the initial wavenumber~$\vk_0$ and the compressive 
direction 
is very close to~$90^\circ$, with the window of allowed angles narrowing 
exponentially fast in time \citep{Zeldovich_etal_linear}. 
Since $\vk\perp\tvB$ and $\tvB\parallel\eone$, 
we have $\vk\parallel\etwo$, i.e., 
the surviving fields can have reversals 
only in the ``null'' direction. In contrast,  
in two dimensions, the field aligned with the stretching direction must 
necessarily reverse in the compressive direction, so the folds 
are destroyed by diffusion \citep[see simulations by][]{Kinney_etal}. 
These statements (illustrated by \figref{fig_zeldovich}) 
explain qualitatively why small-scale dynamo works in three dimensions 
but not in two dimensions \citep{Zeldovich_antidynamo}. 
Even in three dimensions, diffusion leads to suppression of most modes. 
\citet{Zeldovich_etal_linear} showed that
the remaining ones, with the correct alignment of 
$\vk$ and $\vB$, are 
sufficient to make the total magnetic energy 
grow as 
\bea
\label{BB_resistive}
B^2\sim\exp\,[(\zeta_1-\zeta_2)/2].
\eea  

\pseudofigureone{fig_zeldovich}{zeldovich.epsf}{f10.ps}{Fold alignment in three 
and two dimensions.}

In the linear-velocity model, 
a system of infinite size can contain 
an infinite number of initial wavenumbers~$\vk_0$ (no new modes 
are produced by a linear velocity field). Only an exponentially 
small fraction of $\vk_0$ values contribute to the growing field 
at any given time. 
In a finite system, only a finite number of modes can exist, 
limited by the system-size and resistive cutoffs. 
In a linear velocity field, they would all eventually be suppressed.
In a finite-scale flow, 
besides stretching, there is also bending of folds, 
which effectively means that new modes are made. 
Thus, the effect of the finiteness of the flow scale 
is that modes are continually resupplied. 

One might wonder how 
the \citet{Zeldovich_etal_linear} small-scale dynamo 
mechanism relates to the classic 
``stretch-twist-fold'' (STF) picture of the fast dynamo 
\citep{Vainshtein_Zeldovich,STF}. The mechanism outlined 
above makes the small-scale dynamo possible as a {\em statistical} 
effect: the random flow produces certain typical structures (folds), 
then diffusion selects a subset of these structures that
turns out to be sufficiently large {\em on the average} 
to give rise to energy growth. No specific form of 
the flow is prescribed: three-dimensionality, 
randomness in time, and spatial smoothness are all that 
is required. In contrast, the STF picture 
shows what might happen to a flux loop in three dimensions 
if a particular set of transformations is applied to it. 
\citet{Vainshtein_etal_kin} studied numerically 
the evolution of a single circular field line in a chaotic flow 
that was designed to favor STF 
transformations. They observed that only at the very beginning 
of the evolution could the line be seen to actually stretch, 
twist, and fold. It then quickly became randomly tangled 
and, while the field did grow exponentially, it was not 
possible to claim that the mechanism of this growth was 
successive stretching, twisting, and folding. 
Note also that in order to twist the field lines, 
the flow must possess pointwise helicity (net helicity 
is not required because the sense of the twist can be 
arbitrary). However, numerical experiments by 
\citet{Hughes_Cattaneo_Kim} suggest that flows with 
zero pointwise helicity can still generate small-scale fields. 
Thus, while the STF mechanism is, in principle, a viable dynamo, 
it remains unclear whether a generic turbulent flow  
contains enough STF realizations 
to amplify magnetic energy this way, whether this is the 
statistically prevalent kind of dynamo, and what its 
relation to the \citet{Zeldovich_etal_linear} mechanism~is.

It would be worthwhile to make careful measurements of the fold alignment 
with respect to the local Lyapunov basis in numerical simulations. 
We leave such a study outside the scope of this paper. 

\subsection{Self-Similar Growth and Intermittency}
\label{ssec_ssim}

The subviscous range is bounded on the IR side by the 
viscous scale $\kd^{-1}$ (the flow scale) and on the UV side 
by the resistive scale~$\kres^{-1}$. Taking $\kd\to0$ 
amounts to assuming a linear velocity field; 
$\kres\to\infty$ is the diffusion-free limit. 
Both limits are singular for at least some 
of the statistical quantities of interest. 
For the magnetic-energy spectrum, the resistive 
cutoff~$\kres$ is explicitly present, but the 
IR cutoff $k_*\sim\kd$ only 
enters via a small correction to the growth rate 
[\eqref{lambda_sln}]. On the other hand, the flow scale 
is present explicitly in the quantities related to 
the field structure: thus, e.g., 
the field-line curvature is allowed to have a stationary 
pdf only because of the broken scale invariance on the 
IR side [$K_*\sim\kd$, \eqref{PK_st}], while 
diffusion simply cuts off the pdf at large $K$.
In both cases, analytical theory is facilitated 
because one of the cutoffs does not matter. 

For the field-strength distribution, both cutoffs turn out to be 
important. Analytical theory exists only for the diffusion-free 
case ($\kres=\infty$, \ssecref{ssec_diff_free}) and for a 
purely linear velocity field with diffusion ($\kd=0$, $\kres$ finite, 
\ssecref{ssec_linear}). Neither of the two limits 
correctly describes intermittency of the magnetic field 
generated by a finite-scale flow (\ssecref{ssec_finite}). 
The linear-velocity model (reviewed in \ssecref{ssec_zeld_mech}) 
is very helpful in understanding the way the small-scale 
dynamo works in the presence of diffusion. The inadequacy 
of this model as a theory of intermittency has not until now been 
properly appreciated, so we would like to emphasize the importance 
of the results reported in \ssecref{ssec_finite}.

\subsubsection{The Diffusion-Free Case}
\label{ssec_diff_free}

The situation in the diffusion-free regime is straightforward. 
\Eqref{ind_eq} with $\eta=0$ admits 
the formal solution~\exref{B_diff_free} 
for $B(t)$ in the comoving frame, 
so, on the basis of the Central Limit Theorem, 
$B$~should have a lognormal distribution. 
As usual, the Kazantsev model~\exref{KK_model} 
allows for an exact solution. The resulting 
pdf of~$B$ is, indeed, lognormal, with a growing mean 
$\<\ln[B(t)]\>=(3/4)\,\kappa_2 t$ and dispersion~$D=\kappa_2 t$ 
[see~\eqref{lognormal_PDF}].
As the width of this lognormal profile grows with time, the field 
becomes increasingly more intermittent: the moments of the field 
strength are 
\bea
\label{Bn_ideal}
\<B^n\>\propto\exp[(1/4)\kappa_2 n(n+3)\,t].
\eea
This formula can be obtained both from direct averaging 
of the induction equation \citep[e.g.,][]{SK_tcorr} 
and from averaging \eqref{BB_ideal} over the statistics 
of the finite-time Lyapunov exponent $\zeta_1(t)/t$ 
\citep{Chertkov_etal_dynamo}. 

A good conventional measure 
of intermittency is the kurtosis $\Bfr/\Bsq^2\propto\exp(2\kappa_2 t)$, 
which can be roughly interpreted as the inverse volume-filling fraction. 
The kurtosis and all other normalized moments 
$\<B^{mn}\>/\<B^m\>^n$ 
grow exponentially. This means that the dominant contribution to each 
moment~$\<B^n\>$ comes from a different substructure, which occupies 
an exponentially decreasing fraction of the volume compared with those 
contributing to lower moments. This is because 
the fields everywhere are stretched exponentially, but with  
a fluctuating stretching rate $\zeta_1(t)/t$ [\eqref{BB_ideal}], 
so any occasional difference between 
substructures tends to be amplified exponentially.  

\pseudofiguretwo{fig_Bn}{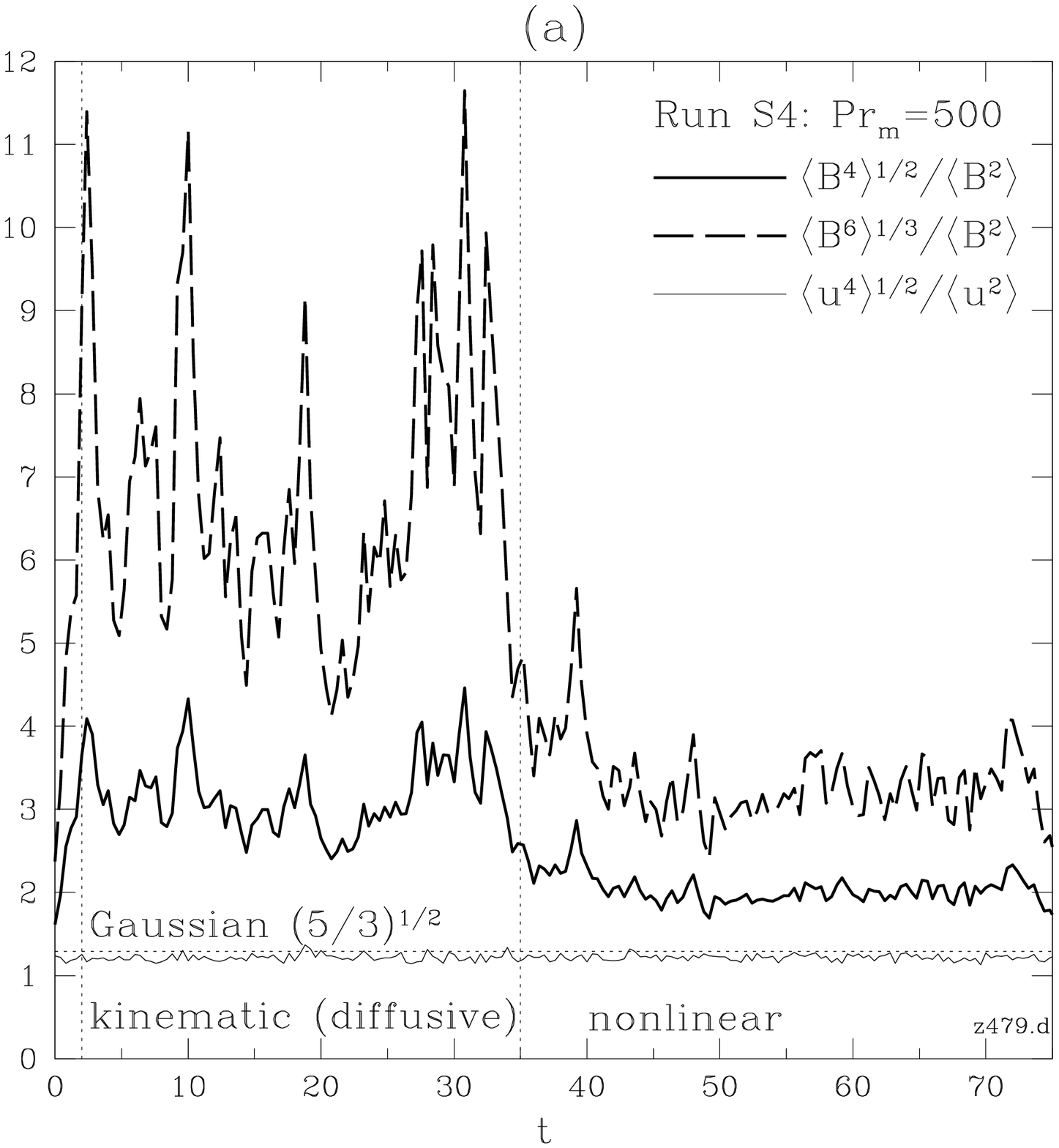}{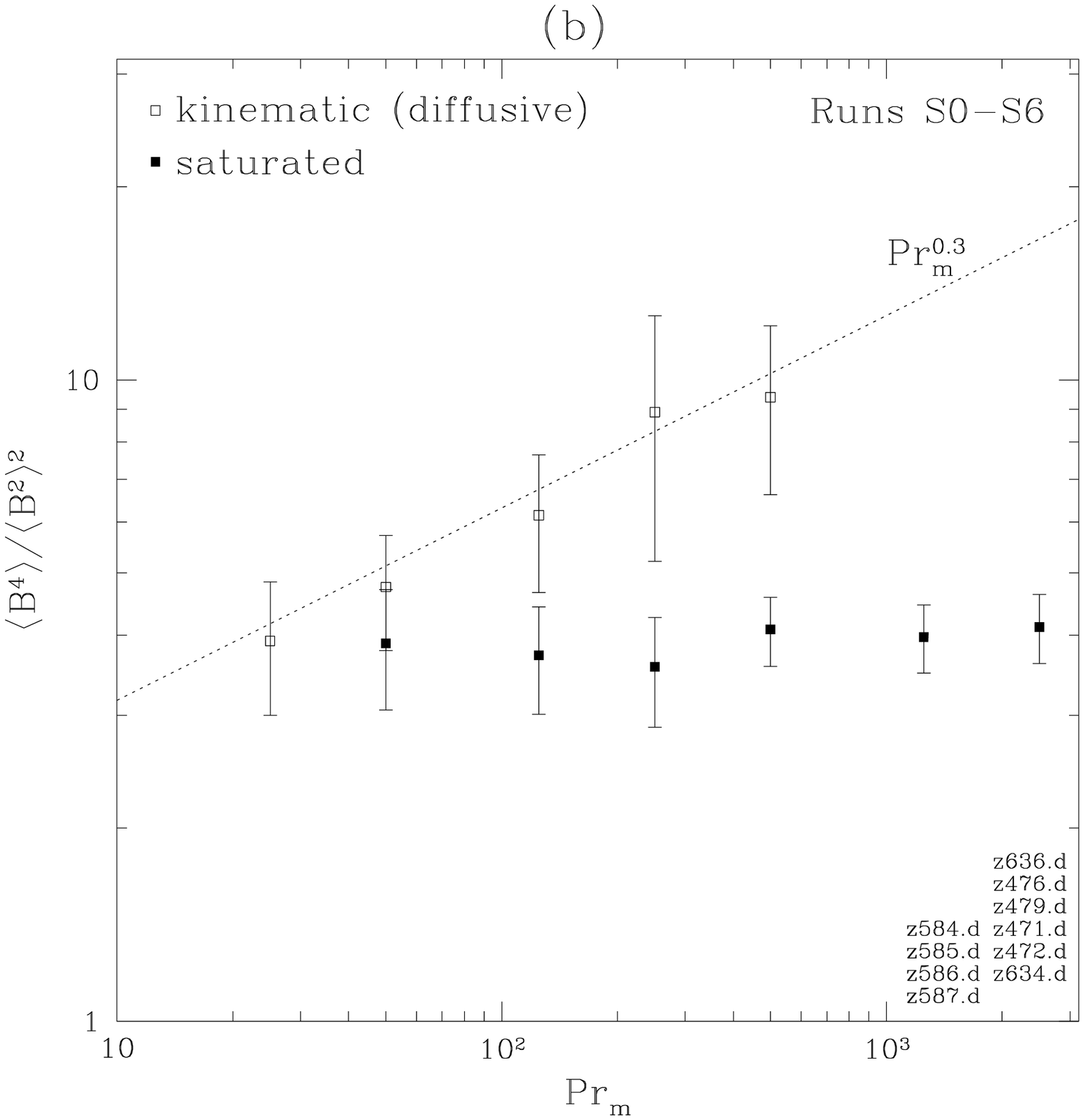}{f11a.ps}{f11b.ps}{(a) Evolution of normalized 
moments of the magnetic-field strength for run~\run{S4}. 
The (square root of) kurtosis of 
the velocity field is given for comparison. 
(b) Kurtosis $\Bfr/\Bsq^2$
of the magnetic-field strength during the kinematic 
stage of runs \run{S0}-\run{S4} and 
in the saturated state of runs \run{S1}-\run{S6}. 
The values plotted are listed in \tabref{tab_index}.}

\subsubsection{A Model with Diffusion and a Linear Velocity Field}
\label{ssec_linear}

The situation changes dramatically when diffusion is introduced. 
Exact theory for one-point statistics in this case is problematic 
even for the Kazantsev velocity, because 
of the closure problem associated with the diffusion term. 
The problem is solvable, however, for the particular case when  
the velocity field is exactly linear and the system domain is infinite \citep{Zeldovich_etal_linear,Chertkov_etal_dynamo,Nazarenko_West_Zaboronski,West_etal}. 
The general outline of the solution was given 
in \ssecref{ssec_zeld_mech} leading to the growth 
of a typical field realization 
according to \eqref{BB_resistive}. 
\citet{Chertkov_etal_dynamo} worked out the moments 
$\<B^{2n}\>$ averaged over the (Gaussian) distribution of~$\zeta_i$ 
for the Kazantsev model and found 
\bea
\label{Bn_resistive}
\<B^{2n}\>\propto\Pr^{5n/4}\exp[(3/16)\kappa_2 n(n+4)\,t].
\eea
Thus, it appeared that, in the diffusive regime, while the specific expression 
for the moments changed, the intermittency continued to increase exponentially 
in time just as it did in the diffusion-free case.

\subsubsection{The Finite-Scale Flow: Self-Similarity}
\label{ssec_finite}

We already saw in \ssecref{ssec_folding} 
that the linear-velocity model is not sufficient
to describe the field structure. 
The intermittency properties of the field strength are also affected --- 
and, indeed, drastically changed, --- by the presence 
of the finite system scale.

In an infinite system, 
intermittency can grow with time because 
for ever higher moments $\<B^n\>$, ever smaller sets of substructures 
can always be found in which the field has exponentially outgrown the rest 
of the system and that, therefore, dominantly contribute to~$\<B^n\>$. 
By contrast, in a finite system, 
only a finite number of exponentially growing substructures can exist, 
so the contribution to all moments must eventually come from the 
same fastest-growing one. The statistics of the field should, therefore, 
be {\em self-similar}, with $\<B^n\>$ growing at rates 
proportional to~$n$, not~$n^2$, and all normalized moments 
$\<B^{mn}\>/\<B^m\>\<B^n\>$ saturating.\footnote{This effect was 
not properly appreciated in our 
discussion of the kinematic dynamo in \citet{SMCM_structure}. 
Note that for the problem of passive-scalar decay, self-similar behavior 
was found in 2D maps \citep[the so-called ``strange 
mode'': see][]{Pierrehumbert_strange_mode,Pierrehumbert_lattice,Sukhatme_Pierrehumbert} 
and even in scalar-mixing experiments \citep{Rothstein_Henry_Gollub}. 
Numerical simulations of the advection-diffusion equation in two dimensions 
have shown the same property \citep{Fereday_Haynes}. 
In map dynamos studied by Ott and coworkers \citep{Ott_review}, 
all moments of the field also grew at the same rate.}

In our simulations, there is unambiguous evidence for the self-similar 
evolution. 
After initial diffusion-free growth, the normalized moments saturate 
(\figref{fig_Bn}a). The pdf of the magnetic-field 
strength becomes self-similar: namely, 
the pdf's of $B/\Brms$ collapse onto a single 
stationary profile throughout the kinematic stage of the 
dynamo (\figref{fig_PB_kin}). 
The large-$B$ tail of the pdf of~$B/\Brms$ 
is reasonably 
well fitted by a lognormal distribution. Specifically, suppose 
that the pdf of $z=\ln B$ is 
\bea
\label{lognormal_PDF}
P_z(z)=(\pi D)^{-1/2}\exp\bl[-\bl(z-\<\ln B\>\br)^2/D\br].
\eea 
Then $\<B^n\>\propto\exp\[\<\ln B\> n + D n^2/4\]$, 
so $D=\ln\bl(\Bfr^{1/2}/\Bsq\br)$. 
In the diffusive regime, $D=\const$ and 
the field-strength statistics become self-similar: 
the pdf of $\zeta=\ln(B/\Brms)$ is stationary, 
\bea
\label{lognormal_fit}
P_\zeta(\zeta)=(\pi D)^{-1/2}\exp\bl[-\bl(\zeta+D/2\br)^2/D\br].
\eea 
The lognormal fit in \Figref{fig_PB_tail} is obtained by calculating 
$D$ from the numerical data and comparing the profile~\exref{lognormal_fit} 
with the numerically calculated pdf. 
The fit is qualitative but 
decent, considering the simplicity of the chosen 
profile~\exref{lognormal_fit} and large statistical errors in determining 
$\Bfr/\Bsq^2$ (\figref{fig_Bn}b). 

The pdf at small $B$ appears to be power-like,  
$BP(B)\sim B^{2.7...2.8}$ in all regimes (\figref{fig_PB_kin}b). 
This tail extends to the limit of our diagnostic's resolution 
[$BP(B)\sim10^{-5}$ at $B\sim 10^{-3}\Brms$, not shown], 
but there may be an unresolved lognormal tail at even smaller~$B$. 
It is clear that, while the large-$B$ tail describes the straight segments 
of the folds, the small-$B$ tail gives the field-strength 
distribution for the weak fields in the bends. 

\pseudofiguretwo{fig_PB_kin}{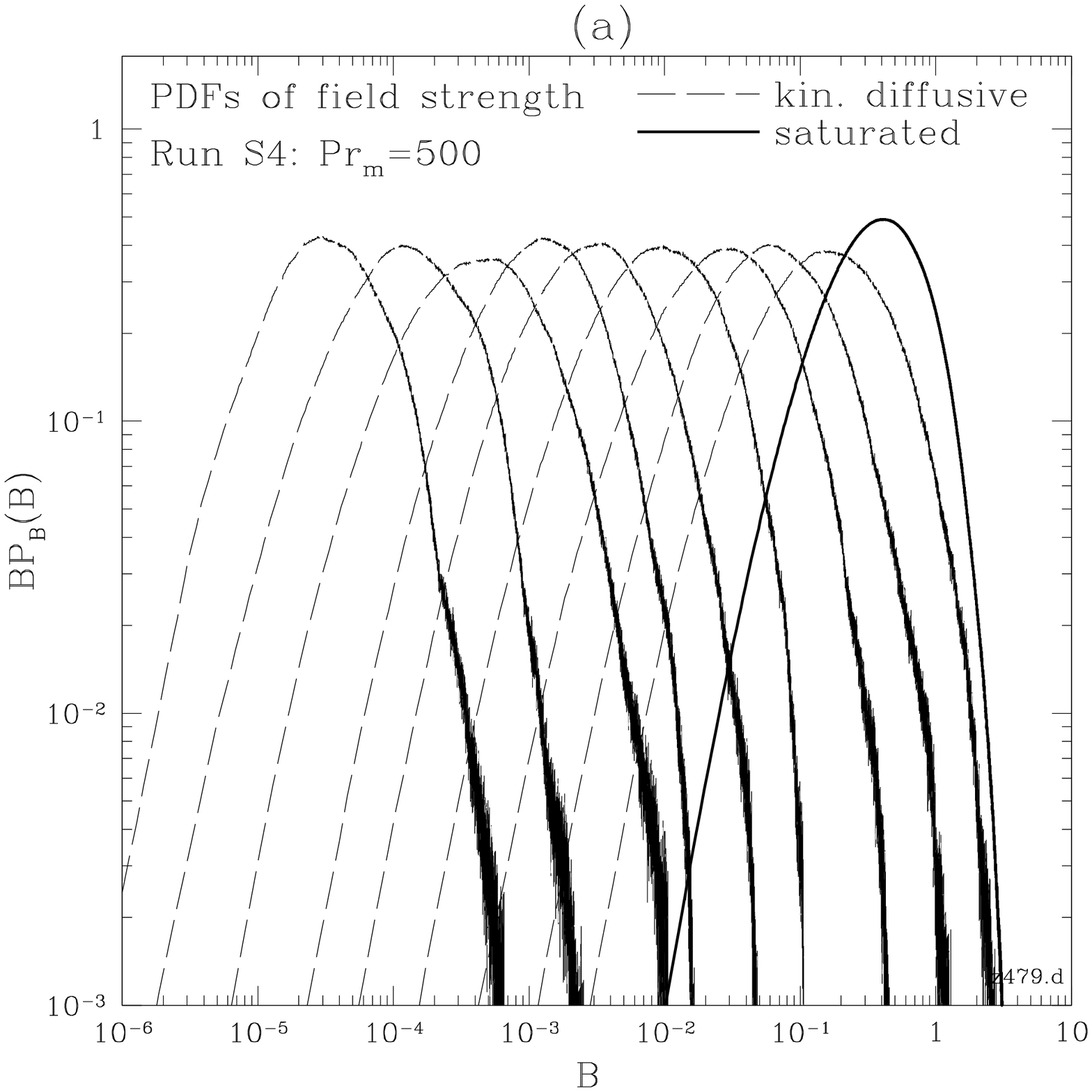}{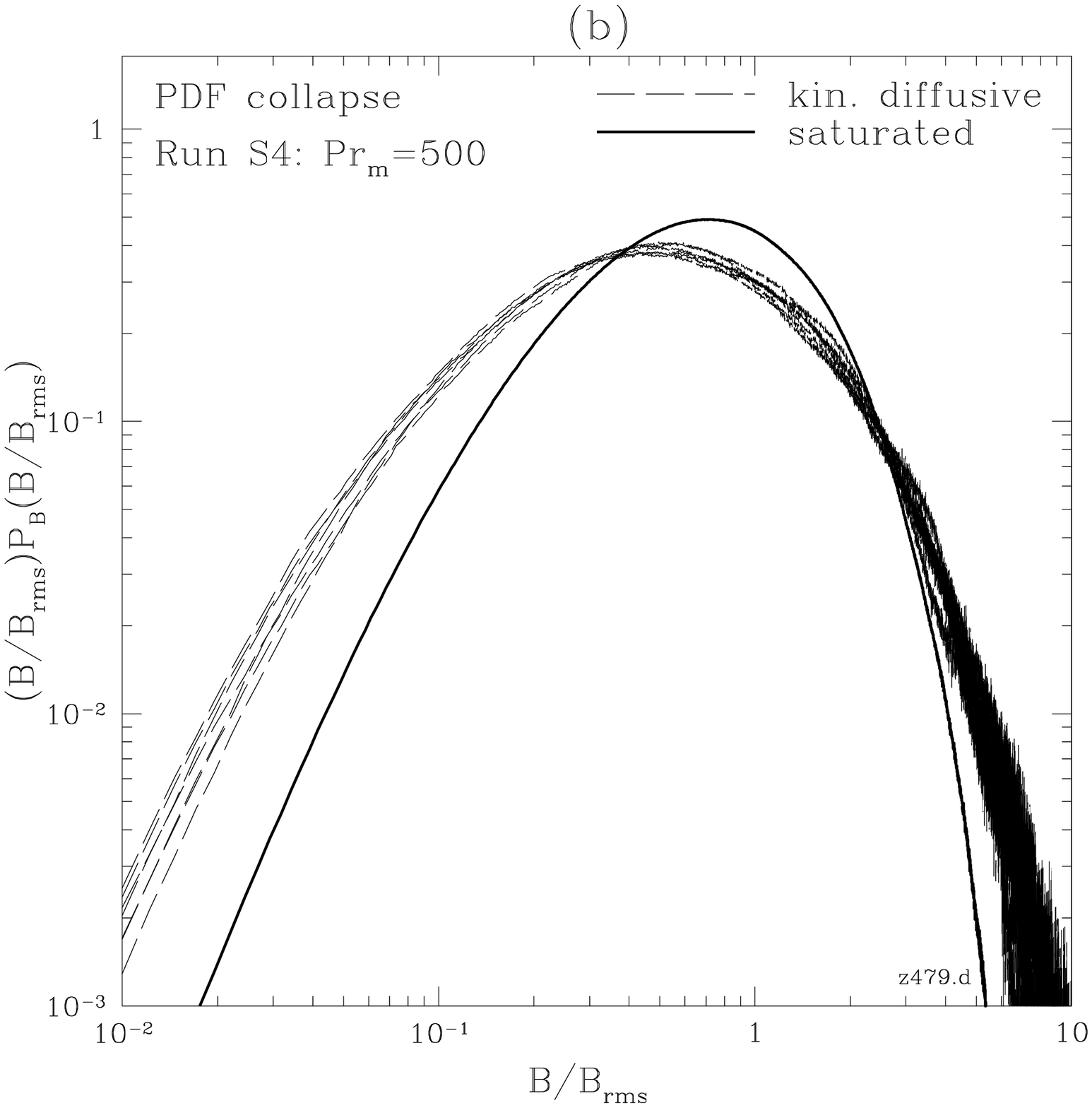}{f12a.ps}{f12b.ps}{(a) 
Evolution of the pdf of the magnetic-field strength for run~\run{S4}. 
The pdf's for the diffusion-free regime ($0\le t<2$) are given at 
time intervals of~$\Delta t=0.4$. The subsequent evolution ($2\le t \le 40$) 
is represented by pdf's at time intervals~$\Delta t=4$.  
(b) Collapse of the field-strength pdf's in the kinematic 
stage of run~\run{S4} onto a self-similar profile. 
These are pdf's of~$B/\Brms$. 
The pdf in the saturated state is shown for comparison.}

Note that, in space, intermittency 
of the field strength means that the growing 
fields do not uniformly fill the volume. More specifically, 
intermittency is often associated with the presence of 
``coherent structures'' that have disparate spatial dimensions 
For the dynamo-generated magnetic fields, they are the folds, for 
which the disparate dimensions are their length 
and the field-reversal scale. 
A linear velocity field allows the folds to be elongated indefinitely, 
thus giving rise to arbitrarily large aspect ratios between the two 
scales. In reality, the length of the folds cannot be larger than 
the scale of the flow because of the bending of the folds. 
The aspect ratio is then bounded from above by the maximum allowed 
scale separation~$\sim\Pr^{1/2}$.

The hypothetical lognormal pdf~\exref{lognormal_PDF} is 
self-similar only if its dispersion~$D$ does not depend on time. 
In contrast, in the diffusion-free regime, we had $D\sim\gamma t$, 
where $\gamma\sim\nabla\vu$ is the stretching rate. 
In the case of $\eta>0$ and linear velocity field, 
the formula for $\<B^n\>$ derived by \citet{Chertkov_etal_dynamo} 
[\eqref{Bn_resistive}] is also consistent 
with a lognormal distribution for which $D\sim\gamma t$. 
Both results are only valid transiently, during the time 
that it takes magnetic fluctuations to reach the resistive 
scale (in the former case) or the system (flow) scale (in the latter 
case). Since the scale separation is~$\sim\Pr^{1/2}$ and 
the spreading over scales proceeds exponentially fast 
at the rate~$\sim\gamma$ (\ssecref{ssec_spectrum}), the time during 
which intermittency increases is $t_*\sim\gamma^{-1}\ln\Pr^{1/2}$. 
Physically, this is the time necessary to form a typical fold 
with length of the order of the flow scale and 
field reversals at the resistive scale --- starting 
from either a flow-scale or a resistive-scale fluctuation. 
We might conjecture that this time determines the 
characteristic magnitude of the dispersion in the self-similar 
regime: $D\sim\gamma t_*\sim\ln\Pr^{1/2}$, which implies that 
the kurtosis~$\Bfr/\Bsq^2=\exp(2D)$ increases with~$\Pr$ in a power-like 
fashion. The specific power law depends on prefactors that 
may be nonuniversal. The same holds for other normalized moments. 
\Figref{fig_Bn}b is an attempt to test 
this hypothesis for a sequence of simulations with increasing~$\Pr$. 
The fluctuations in the kinematic regime are very large 
(see \figref{fig_Bn}a), so 
the error bars are too wide to allow us to claim 
definite confirmation of the power-like behavior, but our results 
are consistent with a $\Bfr/\Bsq^2\sim\Pr^{0.3}$. 

A caveat is in order. At our resolutions, the lognormal fit is, 
in fact, not 
the only one that is compatible with the numerical evidence. 
Stretched-exponential and even steep-power-tail 
[$P(B)\sim B^{-4.5}$] fits of comparable quality can be achieved. 
Note, however, that 
a stretched exponential would not be compatible with a 
$\Pr$-dependent kurtosis. 

We emphasize that the self-similarity reported here is statistical, 
not exact. Namely, it does not imply that the magnetic field 
is simply a growing eigenmode of the induction 
equation~\exref{ind_eq}. Such an eigenmode does exist for some 
finite-scale nonrandom (and time-independent) flows and maps \citep{STF}.
The self-similar pdf we have found here is a natural generalization 
of this eigenmode dynamo to random flows. 

A testimony to the statistical nature of the self-similarity 
is the tendency of the growing field configuration to undergo 
sudden disruptions when the self-similar profiles are rapidly 
destroyed and then regenerated by the dynamo. These disruptions 
are manifested, e.g., in sudden destruction of the large-$k$ 
tails of magnetic spectra (\figref{fig_Mk_kin}a). 
These events are correlated with large downward fluctuations of 
the velocity field. Indeed, when such a fluctuation occurs, 
the stretching rate 
$\gamma\sim\nabla\vu$ drops, the resistive scale 
$\lres\sim(\eta/\gamma)^{1/2}$ increases, and the whole 
field structure with direction reversals at the previously 
established smaller resistive scale is quickly dissipated. 
Since it is the stronger straight direction-reversing fields 
that are destroyed, the tail of the field-strength pdf is 
suppressed. The self-similar statistics are quickly 
reestablished as the magnetic field adjusts to the new 
stretching rate. 
Of course, the disruptions are nothing more than 
extreme events in what is a continuous 
series of fluctuations. Their presence shows the degree of disorder 
behind the self-similar statistics. 

\pseudofigureone{fig_PB_tail}{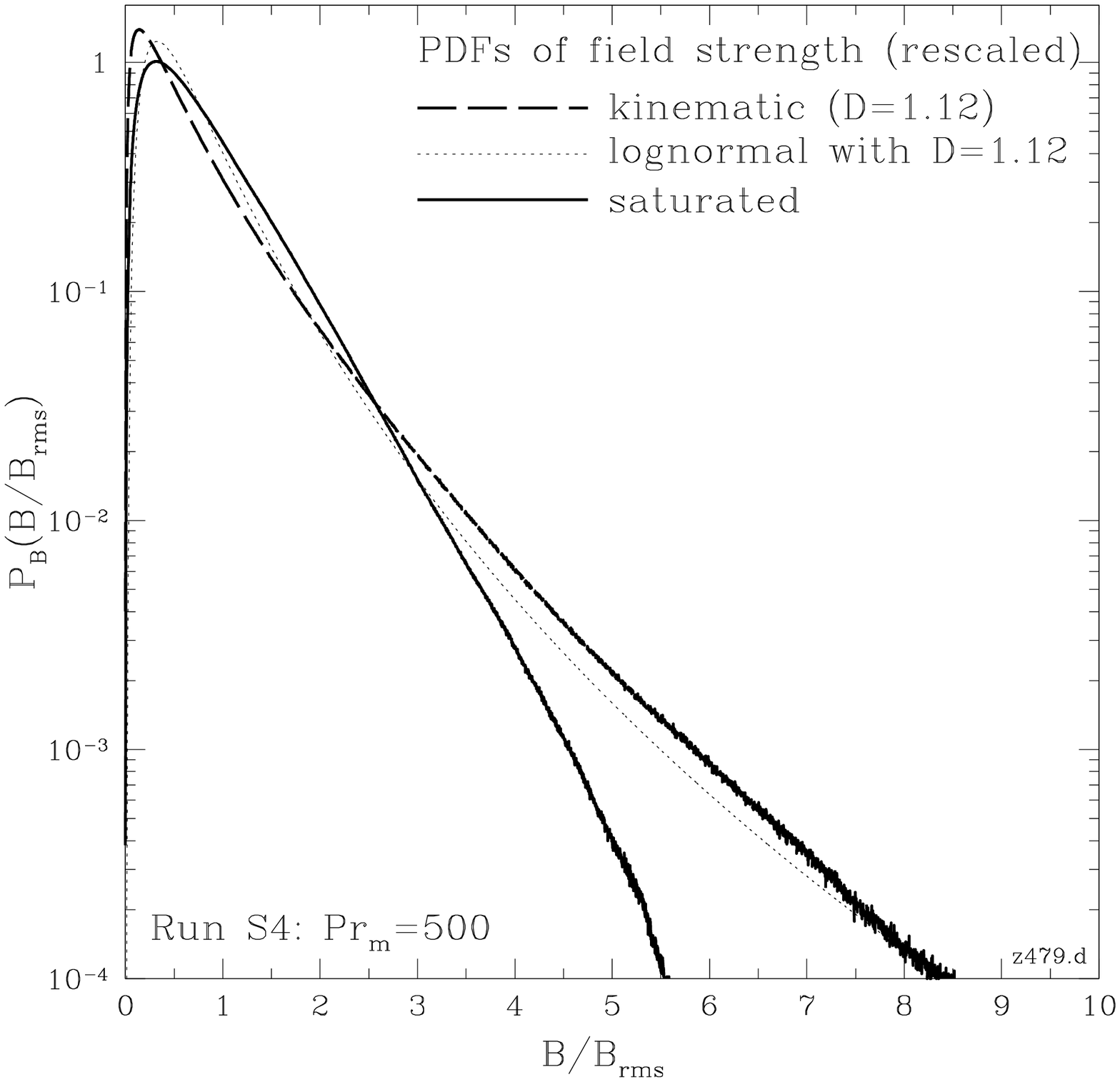}{f13.ps}{
The pdf of $B/\Brms$ for run~\run{S4} 
(averaged over the kinematic diffusive stage) 
and the lognormal profile~\exref{lognormal_fit} with the same~$D$.
Also given is the pdf in the saturated state. 
The pdf's for all other runs are similar.}

\subsection{Summary of the Kinematic Dynamo}
\label{ssec_kin_sum}

We have established the following properties of 
the kinematic regime. 

\begin{enumerate}

\item The bulk of the exponentially growing magnetic energy is
concentrated at the resistive scale~$\lres\sim\Pr^{-1/2}\ld$. 
The magnetic-energy spectrum peaks at $\kres\sim\lres^{-1}$ 
and in the subviscous range $\kd\ll k\ll\kres$, 
it has a positive spectral exponent consistent with 
the theoretical prediction $k^{3/2}$ \citep{Kazantsev,KA}. 

\item The fields generated by the dynamo have folded structure: 
they reverse direction at the resistive scale with the field 
lines remaining straight (and fairly well aligned) up to the 
scale of the flow. It is the direction reversals that are 
responsible for the magnetic-energy spectrum peaking 
at the resistive scale. 
In the regions of curved field (bends), 
the field is weak. The field's strength and its curvature 
are, thus, anticorrelated. The theory of 
\citet{SCMM_folding} is broadly confirmed. 

\item The growing direction-reversing fields are intermittent 
in space. The field-strength distribution is consistent with 
a lognormal profile and evolves self-similarly with time. 
The volume fraction occupied by the growing fields decreases 
with increasing~$\Pr$ in a power-like fashion. 

\end{enumerate}

The \citet{Zeldovich_etal_linear} mechanism of 
fold alignment (\ssecref{ssec_zeld_mech}) explains 
the existence of the dynamo. 

\section{THE NONLINEAR SATURATED STATE}
\label{sec_nlin}

The kinematic growth of the magnetic energy must culminate in a 
saturated nonlinear state. When does the nonlinearity 
[the Lorentz tension force in \eqref{NSEq}]
become important? We showed in \ssecref{ssec_folding} that the kinematic 
dynamo produced folded fields for which $\vB\cdot\nabla\vB\sim\kd B^2$. 
This means that the tension force can start to oppose the purely hydrodynamic 
terms in \eqref{NSEq} once 
$\vB\cdot\nabla\vB\sim\vu\cdot\nabla\vu~\Leftrightarrow~B^2\sim\ud^2$, i.e., 
when the magnetic energy becomes comparable to the energy of the viscous-scale 
eddies. Note that this is a nontrivial statement that only holds for 
the folded field: in an unstructured field with the 
same~$\kB\sim\kres$, we would have had $\vB\cdot\nabla\vB\sim\kres B^2$ 
and the nonlinearity would have set in at much lower field 
energies $B^2/\ud^2\sim\kd/\kres\sim\Pr^{-1/2}$.

\pseudofigureone{fig_E_eta}{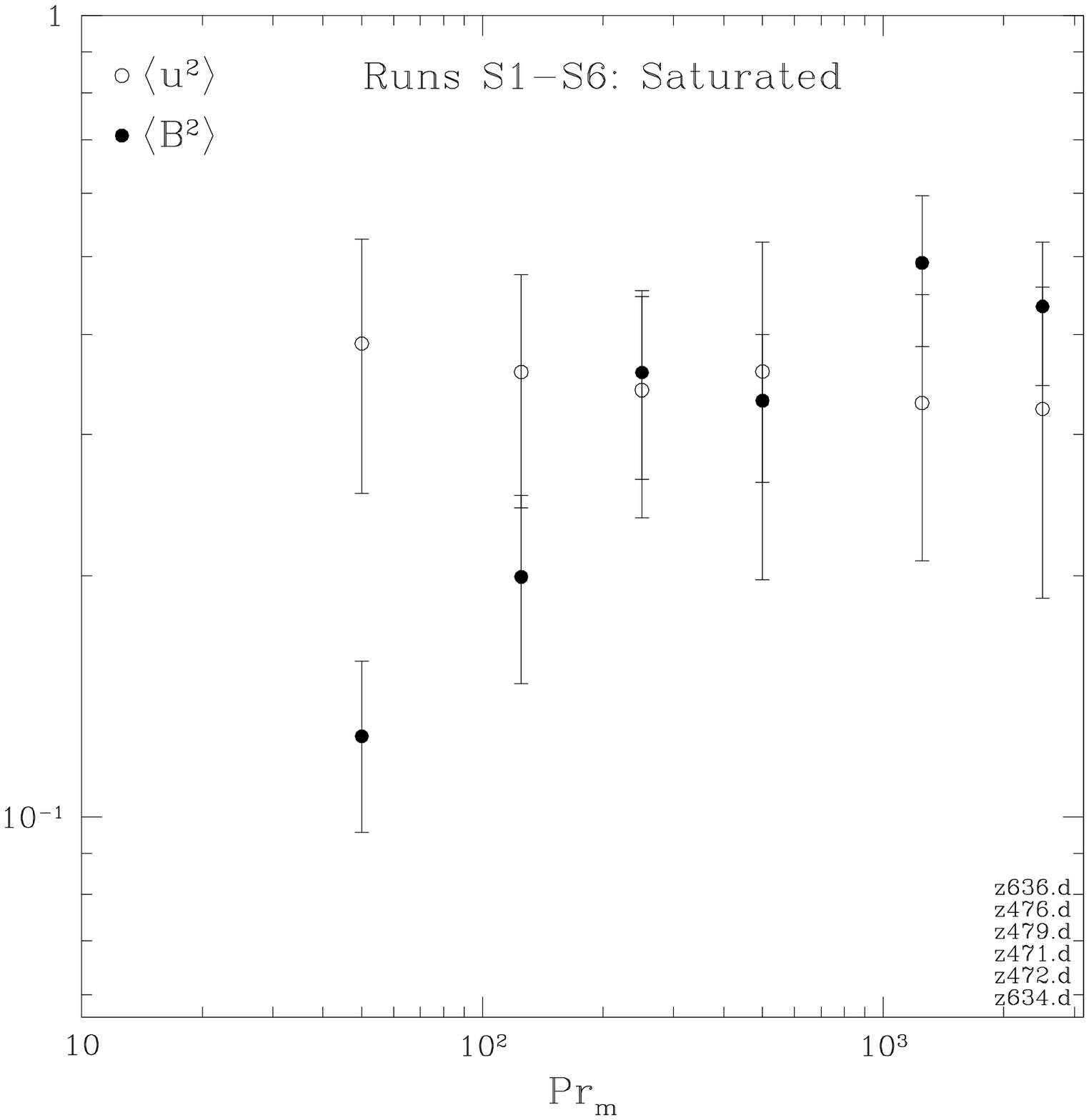}{f14.ps}{Saturated values of the 
kinetic and magnetic energies vs.~$\Pr$. The values plotted are 
listed in \tabref{tab_index}.}

In runs \run{S1}-\run{S6}, the viscous scale is the same as 
the forcing scale, so the above arguments imply saturation with 
roughly equal kinetic and magnetic energies. We find that the 
saturated values of~$\Bsq$ grow with~$\Pr$ toward an apparent 
constant asymptotic value somewhat above~$\usq$ 
\citep[\figref{fig_E_eta}, cf.][]{Brummell_Cattaneo_Tobias}. 

Although a number of heuristic models of back-reaction have 
recently been 
proposed \citep[][and \ssecref{ssec_sat}]{Subramanian_diff,Subramanian_nlin,Kim_nlin1,Kim_nlin2,Boldyrev_nlin,Nazarenko_Falkovich_Galtier,SMCM_structure,SCHMM_ssim,SCTHMM_aniso}, 
the exact mechanism of the nonlinear saturation in isotropic MHD turbulence 
is not currently well understood even on the qualitative level. 
One must, therefore, probe the nonlinear physics by careful numerical 
analysis. 
As always in turbulence problems, only statistical 
statements can be expected to have 
any degree of universal applicability. 
The statistical averages are replaced by time averages, so, in order 
to collect statistics, simulations must be run for a long time {\em after} 
the statistical steady state has been achieved. 
We present statistical results averaged over 20 time units, 
after all discernible systematic change of the nonlinear state ceased. 

We first establish the survival of the folded field 
structure in the saturated state (\ssecref{ssec_fold_nlin}), 
then look at the magnetically induced subviscous velocity 
fluctuations (\ssecref{ssec_tension}), and, finally, 
study the magnetic-energy spectrum, for which a 
nonlinear saturation model is described and 
tested (\ssecref{ssec_sat}). 
Intermittency in the saturated state is considered 
in \ssecref{ssec_intermittency_nlin}.
A summary is given in \ssecref{ssec_sum_visc}. 

\pseudofigureone{fig_folding_sat}{B_z479_zsh270_small.epsf}{f15.ps}{Field 
structure at $t=76$ in the saturated state of run~\run{S4}.
This is the nonlinear counterpart of \figref{fig_folding_kin}.}

\subsection{Persistence of the Folded Structure}
\label{ssec_fold_nlin}

A key fact that we learn from numerical simulations is that
the field stays folded after its 
energy saturates (\figref{fig_folding_sat}). 
All main features of the folded structure 
observed in the kinematic regime (\ssecref{ssec_folding}) carry over 
to the saturated state:

\begin{enumerate}

\item Field varies along itself at the scale of the flow. 
Its characteristic parallel wavenumber [\eqref{kpar_def}] 
is $\kpar\sim\kd$ and does not depend on~$\Pr$, while 
$\kbxj$ and $\kB$ [\eqsref{kbxj_def}{krms_def}] increase with~$\Pr$ 
\citep[\figref{fig_k_nlin}, cf.][]{Brandenburg_Procaccia_Segel,Brummell_Cattaneo_Tobias}. 
The scaling $\kB\sim\kbxj\sim\Pr^{1/2}$ appears to be recovered 
asymptotically with $\Pr$, but is much less well satisfied 
at our finite $\Pr$ than it was in the kinematic case 
(cf.~\figref{fig_kt}b). 
This is consistent with the flattening of the magnetic-energy 
spectrum discussed in \ssecref{ssec_sat}. 

\item The magnetic-field strength and curvature remain anticorrelated 
with the correlation coefficient~$C_{K,B}$ 
[\eqref{CKB_def}] within 4\% of~$-1$ (\tabref{tab_index})
and the log-scatter plots of $B$ versus~$K$ least squares fitted 
by $BK^\alpha=\const$ with $\alpha\simeq0.45$ 
(\figref{fig_BK_nlin}; 
cf.~\ssecref{ssec_anticorr}).

\item The curvature pdf in the saturated state has a power tail 
with scaling very close to the kinematic~$K^{-13/7}$ 
(the scaling in the saturated case may be 
slightly steeper than in the kinematic case: see \figref{fig_PK}). 
The bulk of the pdf is concentrated at the flow scales
(cf.~\ssecref{ssec_curvature}).

\end{enumerate}

\pseudofigureone{fig_k_nlin}{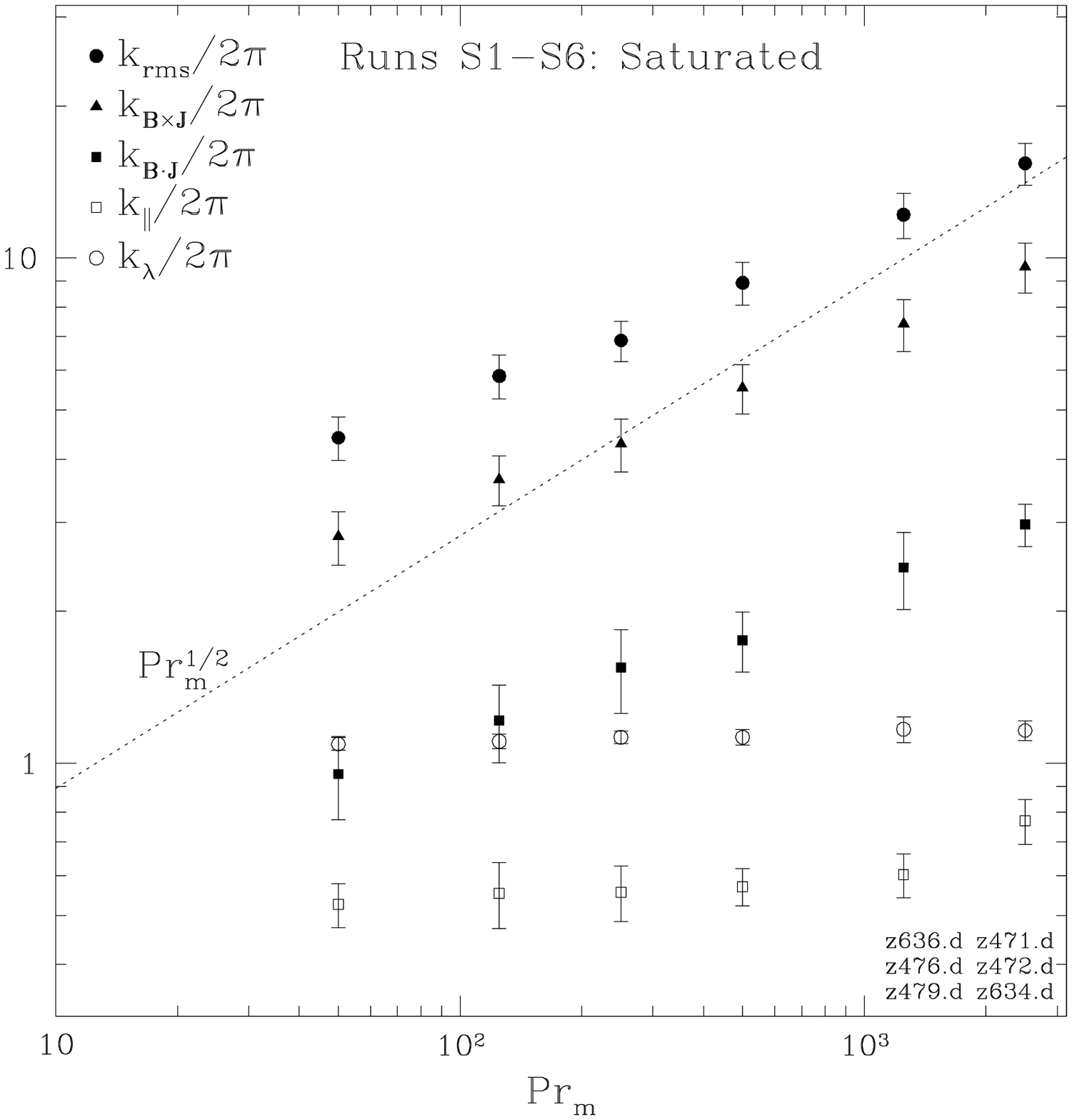}{f16.ps}{Averaged values of various 
characteristic wavenumbers (defined in \ssecref{ssec_folding})
vs.~$\Pr$ in the saturated state of runs~\run{S1}-\run{S6}. 
The values plotted are listed in \tabref{tab_index}.
Note that the value of $\kpar$ for run~\run{S6} is too large 
as a result of this run being somewhat underresolved.}

Physically, it is clear that the preservation of the folded structure 
is inevitable. The large viscosity of the medium does not allow 
the setting up of the kind of detailed small-scale flows that would 
be required to unfold the field without destroying it. 
Furthermore, a wholesale evacuation of the small-scale degrees 
of freedom would lead to an unstable situation in which any significant 
velocity fluctuation would set in motion small-scale field generation 
and refolding. The latter point 
was checked numerically by \citet{MCM}. 

The persistence of the kinematic scaling of the tail of the curvature 
pdf also makes sense. The tail describes the curvature 
distribution in the bends of the folds, where the magnetic field remains 
weak even when field energies in the straight segments of the folds approach 
equipartition levels. Therefore, the situation in the bends is quasikinematic, 
so the curvature statistics are almost the same as in the kinematic limit.

We note one feature of the field structure 
in the saturated state that differs from the kinematic regime:
$\kbj$ [\eqref{kbj_def}] increases somewhat (\figref{fig_kt}a) 
and appears to scale with $\Pr$ in the same way as 
$\kB$ and $\kbxj$: we find $\kB:\kbxj:\kbj\sim5:3:1$ (\figref{fig_k_nlin}). 
This indicates that the direction-reversing fields are 
less well aligned than in the kinematic case. 
Geometrically, one can interpret this effect as a transition 
from flux sheets to flux ribbons. 
While intuitively such a transition 
might be attributable to increased mixing efficiency of 
the velocity field in the saturated state (see \ssecref{ssec_loc_aniso}), 
we do not currently have 
a satisfactory theoretical picture of this nonlinear effect. 

Saturation requires some 
form of suppression of stretching by the tension force in \eqref{NSEq}. 
Since the stretching motions are large-scale, the back-reaction 
must have a large-scale coherence. The folded fields, while 
formally small-scale, possess such a large-scale coherence in the 
form of slow (in space) parallel variation. 
The persistence of the 
folded structure in the saturated state 
is crucial for the saturation model in \ssecref{ssec_sat}.

\pseudofigureone{fig_BK_nlin}{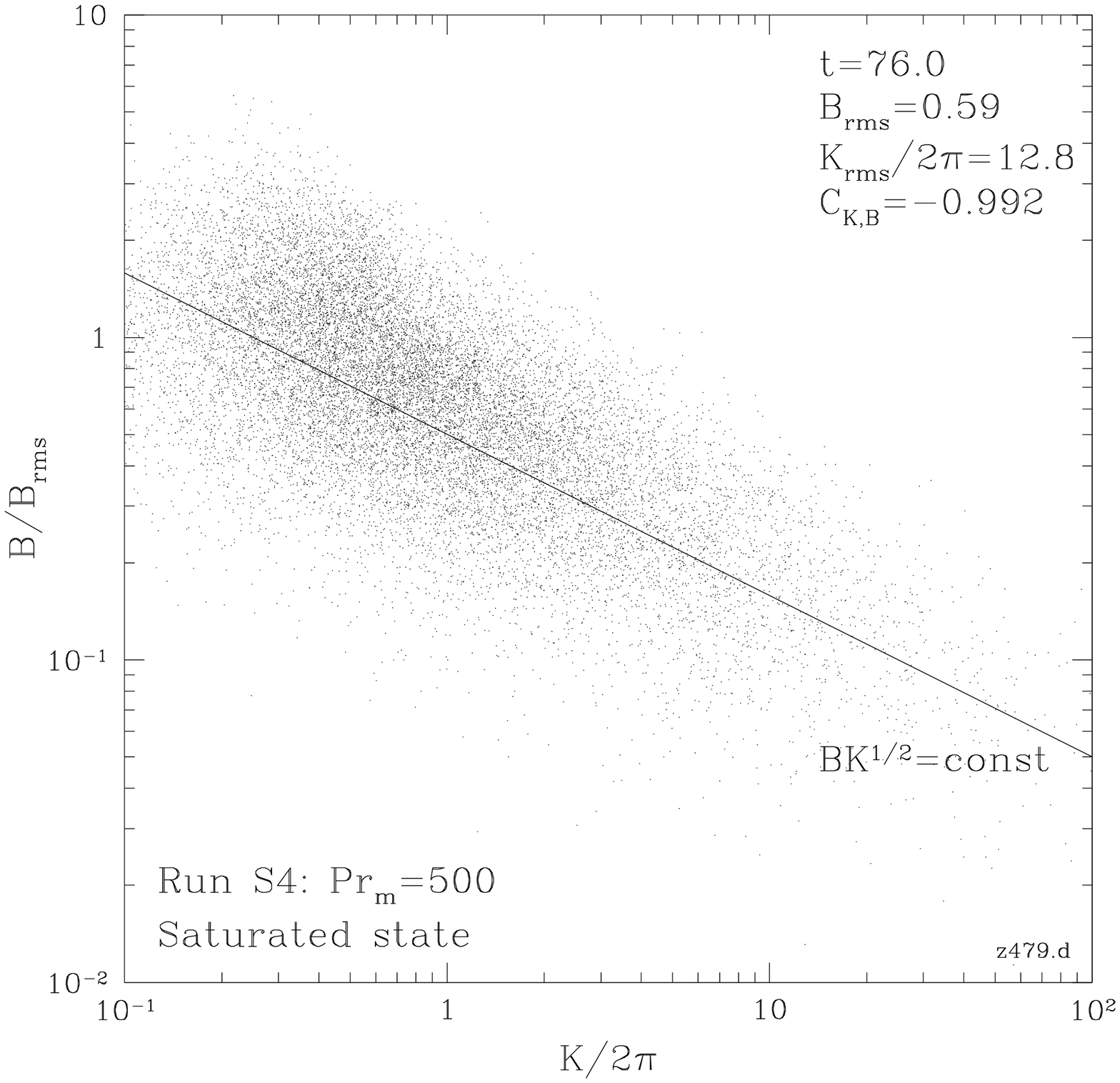}{f17.ps}{Scatter plot of 
$B$ vs.~$K$ at $t=76$ in the saturated state of our run~\run{S4} 
(the $256^3$ data were thinned out by a factor of~$1000$).} 

\subsection{The Tension Force and the Subviscous Velocity Spectrum}
\label{ssec_tension}

The first traces of nonlinearity appear already 
in the kinematic stage in the form of the small-scale modification 
of the velocity spectrum. For $k\gg\kd$, the Navier--Stokes 
equation~\exref{NSEq} reduces to the following force balance 
(in $\vk$ space):
\bea
\label{visc_bal_eq}
\nu k^2\vu(\vk) \simeq -i\vk p(\vk) + \vF(\vk),
\eea
where $\vF=\vB\cdot\nabla\vB$ is the tension force.
Pressure is determined from the incompressibility constraint, 
$\vk\cdot\vu(\vk)=0$, and we get for the velocity spectrum 
at $k\gg\kd$
\bea
\nonumber
E(k) &=& {1\over2}\int\diff\Omega_\vk\, k^2 \<|\vu(\vk)|^2\>\\ 
&\simeq& {1\over2\nu^2 k^4}\int\diff\Omega_\vk\, k^2 
\l\<\vF(\vk)\cdot\(\unity-{\vk\vk\over k^2}\)\cdot\vF(-\vk)\r\>.
\label{Ek_tail}
\eea
By the same argument as in \ssecref{ssec_4order}, 
the angle integral on the right-hand side is~$\sim k^0$, 
so $E(k)\sim k^{-4}$. 
\Figref{fig_Ek} shows that the velocity spectrum does indeed develop 
such a tail at subviscous scales already during the kinematic-dynamo 
stage. The tail thickens as the tension force grows ($F\propto B^2$). 

\pseudofigureone{fig_Ek}{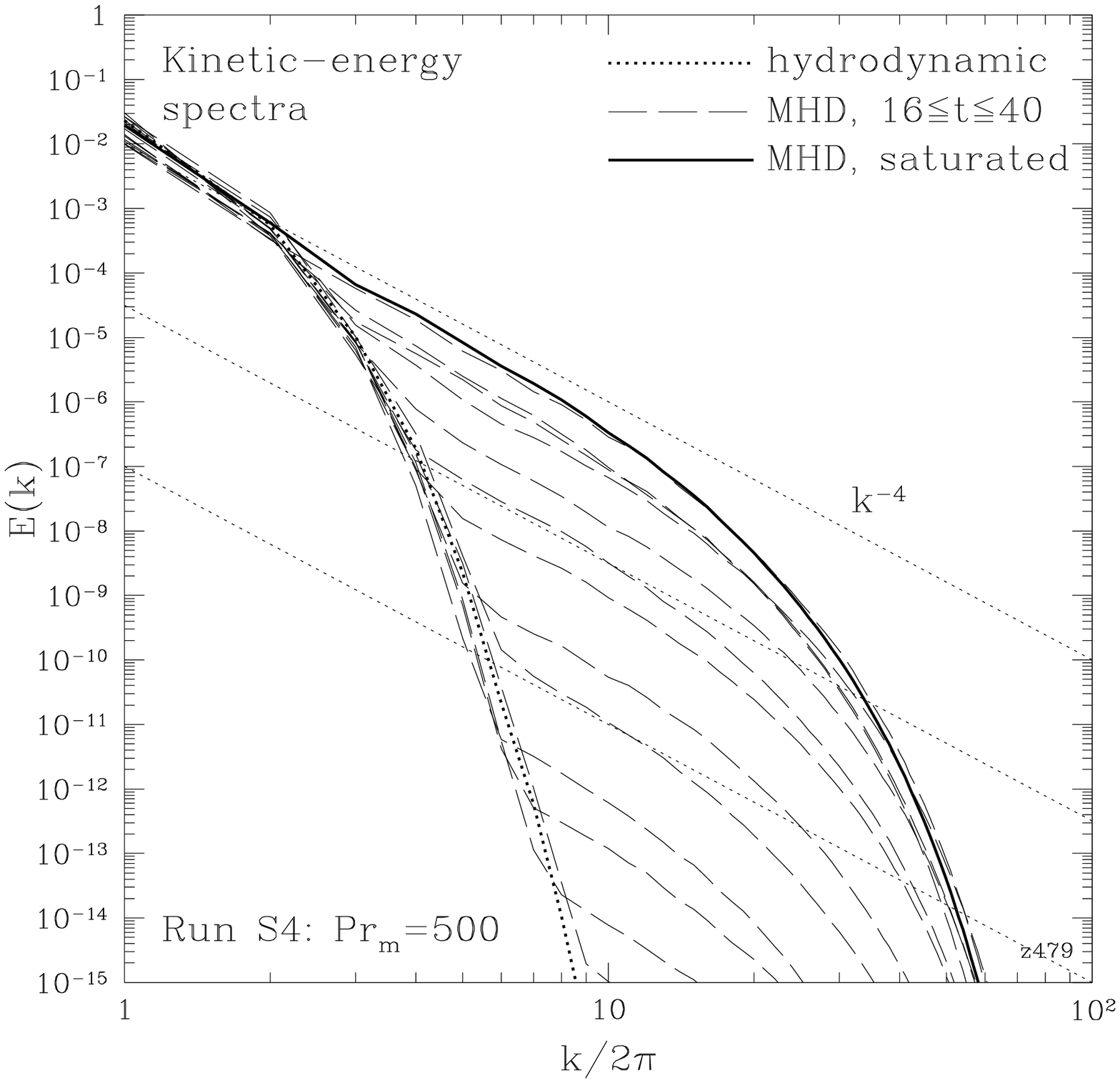}{f18.ps}{Evolution of 
the kinetic-energy spectrum for run~\run{S4}. 
We plot the spectrum in the purely hydrodynamic regime ($\vB=0$), the spectrum 
in the saturated state, 
and the evolving spectra for~$16\le t\le 40$ at time 
intervals of~$\Delta t=2$. 
The corresponding evolution of the tension spectra is given 
in \Figref{fig_M4}a.}

In the saturated state, the tension spectrum remains flat, as 
does $M_4(k)$, and the relation $T(k)\simeq\kpar^2 M_4(k)$ 
persists: indeed, since the folded 
structure is preserved, the argument given 
in \ssecref{ssec_4order} should continue to apply. 
The velocity spectrum keeps its $k^{-4}$ tail, 
and we find that 
\bea
\label{visc_bal}
E(k)\simeq {2\over3}\,{1\over\nu^2 k^4}\,T(k)
\eea
is quite well satisfied (\figref{fig_TandE}). 

\pseudofigureone{fig_TandE}{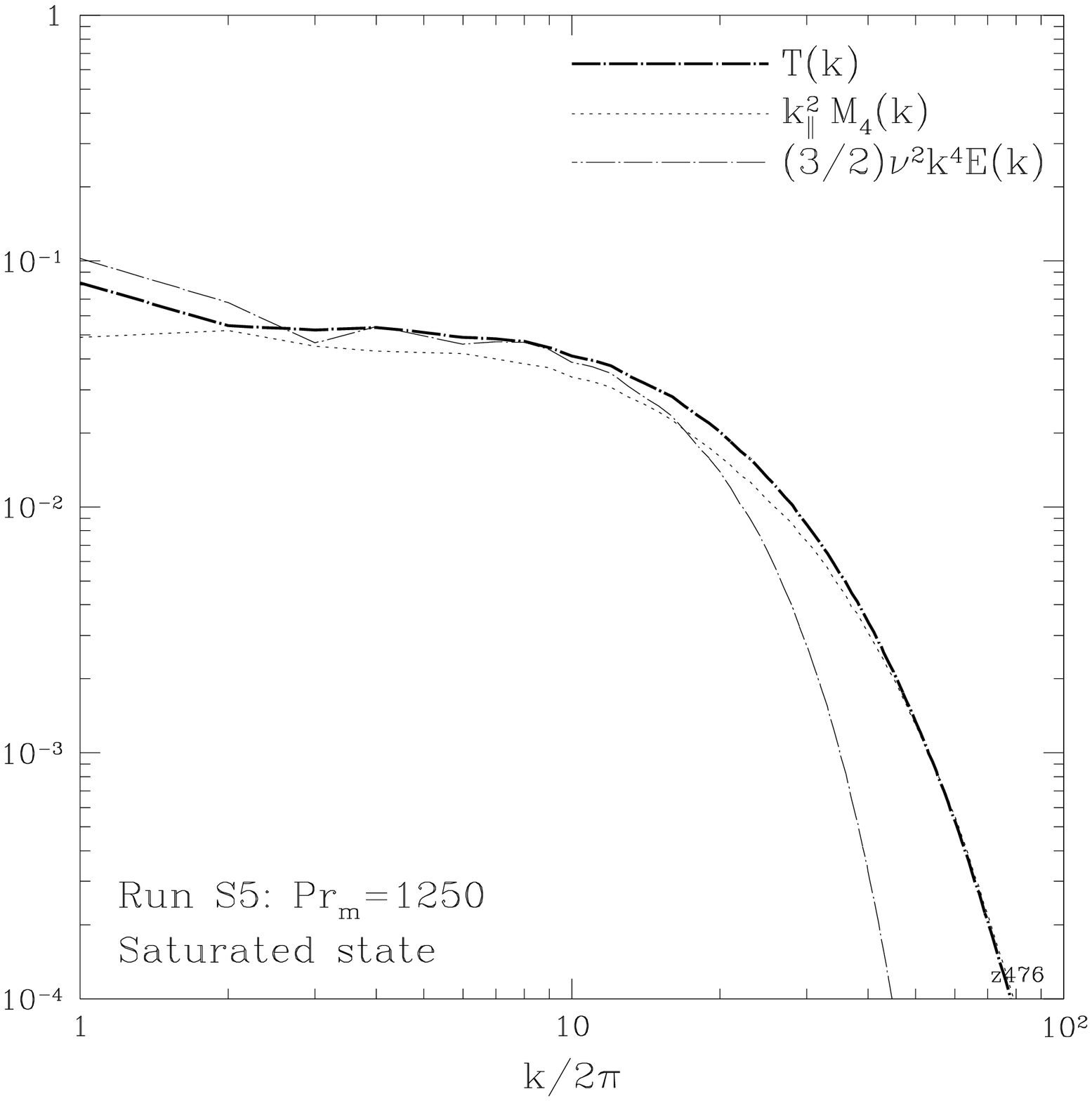}{f19.ps}{Tension 
spectrum for run~\run{S5} (saturated state). 
For comparison, we also plot the kinetic-energy spectrum 
$E(k)$ compensated by~$(3/2)\nu^2 k^4$ 
and the spectrum of~$B^2$, 
multiplied by~$\kpar^2$ [as defined by \eqref{kpar_def}]. 
The results for runs~\run{S1}-\run{S4} and \run{S6} are analogous.}

Thus, the subviscous $k^{-4}$ tail of the velocity spectrum 
is a passive feature. It contains very little 
energy and has nothing to do with the large-scale back-reaction 
mechanism that causes saturation. It is simply another signature 
of the folded structure and is present for both kinematic 
and nonlinear dynamo.

\subsection{Local Anisotropy and Steady-State Spectra}
\label{ssec_sat}

Let us now turn to the key issue of the dynamo physics: 
how does the dynamo saturate? There must exist a 
nonlinear feedback mechanism that 
(1) maintains a statistically steady magnetic-energy level 
and (2) is compatible with the strongly fluctuating velocity 
field continuously stirred by the random forcing. 
Here we outline a model for the saturated spectrum 
we proposed recently \citep[see][where it is presented 
in a somewhat different form]{SCTHMM_aniso} 
and compare its predictions with our simulations. 

\subsubsection{Dynamo Saturation via Local Anisotropization}
\label{ssec_loc_aniso}

Saturation can be achieved if stretching 
motions are nonlinearly suppressed. This does not have to 
mean complete elimination of all turbulence: only the 
$\vb\vb:\nabla\vu$ component of the rate-of-strain tensor 
leads to work being done against the Lorentz force 
and must, therefore, be suppressed. It is then natural 
that the velocity field should become 
anisotropic with respect to the local direction of the 
magnetic field. Since the magnetic field has a folded structure 
(\ssecref{ssec_fold_nlin}), such a local direction 
is defined by the tensor~$\bb^i\bb^j$, which 
varies at the scale of the flow (direction reversals cancel). 
The back-reaction term $\vB\cdot\nabla\vB$ 
in \eqref{NSEq} is quadratic in $\vB$, so it makes sense that 
the nonlinearly modified velocity should be 
indifferent to field reversals but anisotropic with respect 
to the direction of the folds. 

In order to model such an anisotropization, we 
build on the existing kinematic theory based on the 
Kazantsev velocity~\exref{KK_model}. 
Thus far, isotropic correlators $\kappa^{ij}(\vy)$ 
have been assumed. Now we let $\kappa^{ij}$ depend 
on the local direction of the folds.  
In the presence of one preferred direction defined by 
the tensor~$\bb^i\bb^j$, the correlator of an incompressible 
velocity field in $\vk$ space has the following general form
\bea
\nonumber
\kappa^{ij}(\vk) &=& \kapI(k,|\mu|)\bl(\delta^{ij} - \kk_i\kk_j\br) 
+ \kapA(k,|\mu|)\bl(\bb^i\bb^j\br.\\
&&\bl. +\,\,\mu^2\kk_i\kk_j -\mu\bb^i\kk_j - \mu\kk_i\bb^j\br),
\label{kapij_def}
\eea
where $\vkk=\vk/k$, $\mu=\vkk\cdot\vb$.
It is then possible \citep{SCTHMM_aniso} to derive an equation 
for the magnetic-energy spectrum at $k\gg\kd$ that takes into 
account the dependence of the velocity statistics on the 
local direction of the field: 
\bea
\nonumber
\d_t M &=& {1\over8}\,\gperp{\d\over\d k}
\biggl[(1+2\spar)k^2{\d M\over\d k}-(1+4\sperp+10\spar)k M\biggr]\\ 
&&+\,\,2(\sperp+\spar)\gperp M - 2\eta k^2 M,
\label{FPEq}
\eea
where 
\bea
\label{gperp_def}
\gperp &=& \int{\diff^3 k\over(2\pi)^3}\,\kperp^2\kapperp(\vk),\\ 
\label{sperp_def}
\sperp &=& {1\over\gperp}\int{\diff^3 k\over(2\pi)^3}\,\kpar^2\kapperp(\vk),\\
\label{spar_def}
\spar &=& {1\over\gperp}\int{\diff^3 k\over(2\pi)^3}\,\kpar^2\kappar(\vk),
\eea 
and we have denoted 
$\kperp = k(1-\mu^2)^{1/2}$, 
$\kpar = k\mu$, 
\bea
\nonumber
\kapperp(\vk) &=& {1\over2}\bl(\delta^{ij} - \bb^i\bb^j\br)\kappa^{ij}(\vk)\\
&=& {1\over2}\[(1+\mu^2)\,\kapI(k,|\mu|) + \mu^2(1-\mu^2)\,\kapA(k,|\mu|)\],\\
\nonumber
\kappar(\vk) &=& {1\over2}\,\bb^i\bb^j\kappa^{ij}(\vk)\\ 
&=& {1\over2}\,(1-\mu^2)\[\kapI(k,|\mu|) + (1-\mu^2)\,\kapA(k,|\mu|)\]. 
\eea
In the isotropic case, $\kapI=\kapI(k)$, $\kapA=0$, 
which gives $\sperp=2/3$, $\spar=1/6$, and $\gperp=(6/5)\gKA$ 
so \eqref{FPEq} reduces to the standard kinematic case [\eqref{M_eq}]. 

The derivation of \eqref{FPEq} depends on the assumption 
that the tensor $\bb^i\bb^j$ changes at the scale of the flow. 
This is only true in the regions where the fields 
are strong and straight. 
The regions of curved fields (bends of the folds) 
are not properly described by this formalism, but the fields 
there are weak (\ssecref{ssec_fold_nlin}) and 
occupy a small fraction of the volume compared to the strong 
straight ones (because of the steep power tail of the curvature 
pdf, \figref{fig_PK}). Since it is the stronger fields that 
dominate the back-reaction, we believe that our description 
is a reasonable one. 

With a zero-flux boundary condition imposed at some $k_*\sim\kd$ 
and in the limit $\eta\to+0$,
\eqref{FPEq} has the eigenfunction 
\bea
\label{M_efn}
M(k) &=& (\const)\, e^{\gamma t} k^s K_0(k/\kres),\\
\label{kres_def}
\kres &=& \[(1+2\spar)\gperp\over16\eta\]^{1/2},\\ 
\label{gamma_def}
\gamma &=& \gperp\biggl[2(\sperp+\spar)
-{(1+2\sperp+6\spar)^2\over8(1+2\spar)}\biggr],\\
\label{s_def}
s &=& 2\,{\sperp+2\spar\over1+2\spar}.
\eea
In the isotropic case, we get the growing 
kinematic solution: $\gamma=(3/4)\gKA$ and $s=3/2$ 
(see \ssecref{ssec_spectrum}). 
This growth can be quenched by making 
velocity anisotropic so that it varies 
more slowly along the folds than across: 
the values of~$\sperp$ and $\spar$ then drop compared 
to the isotropic case, and so does the growth rate~$\gamma$ 
until the dynamo is shut down.
Thus, saturation can be achieved by anisotropizing 
the statistics of the velocity gradients.

The parallel gradients of the 
velocity field that occur in the numerators 
$\sperp$ and $\spar$ [\eqsref{sperp_def}{spar_def}] 
represent stretching motions, while the perpendicular 
gradients that contribute to $\gperp$ [\eqref{gperp_def}] 
correspond to the two-dimensional mixing (interchange-like) motions 
that shuffle and bring the direction-reversing magnetic fields 
sufficiently close together for them to annihilate resistively. 
The decrease in the values of $\sperp$ and $\spar$ 
means that the comparative strengths of mixing and stretching 
change in favor of mixing. 
The steady solution is thus a result of a balance between 
weakened stretching and two-dimensional mixing of the folded fields 
by the partially two-dimensionalized random flow.

Note that the mixing rate $\gperp$ may or may not 
be changed by the nonlinearity, with dynamo saturation 
achievable either way, provided that the stretching rates 
weaken sufficiently. What happens to $\gperp$ is important 
because it enters the expression for $\kres$ [\eqref{kres_def}] 
and thereby determines the shape of the saturated spectrum. 
It can be seen in \ssecref{ssec_speculations} 
that whether mixing is suppressed is a key issue for the 
large-$\Re$ dynamo. For now, we assume that $\gperp$ 
is unchanged and see if we can get a reasonable model of 
saturation for our viscosity-dominated runs. 

\subsubsection{Solution of the Model} 
\label{ssec_solution}

The condition for saturation is $\gamma=0$ [see \eqref{gamma_def}]. 
This, however, does not uniquely fix $\sperp$ and $\spar$, 
but only relates them via a quadratic equation. 
Thus, to get a specific solution, we must make a further 
assumption about $\sperp$ and $\spar$. 
One plausible argument is that the anisotropization 
of the velocity gradients 
occurs in such a way that all modes with $k$ above some 
self-consistently chosen 
cutoff $\ks$ (the stretching wavenumber) are completely 
two-dimensional (have $\kpar=0$), while the motions 
with $k<\ks$ are still isotropic 
\citep[this is certainly 
a sensible assumption in the case 
of large $\Re$, see][and \ssecref{ssec_speculations}]{SCTHMM_aniso}. 
Then the reduction in the integrals in \eqsref{sperp_def}{spar_def} 
is due to the shrinking of the integration domain in 
$k$ space, but the integrands are still isotropic 
[$\kapI=\kapI(k)$, $\kapA=0$ for $k<\ks$], 
which implies that $\sperp/\spar=4$ as in the 
isotropic case. 

With this assumption, the dynamo saturates ($\gamma=0$) 
when $\sperp=4\spar\simeq0.078$. 
The spectral exponent [\eqref{s_def}] is $s\simeq0.23$, 
which, however, is clearly a model-dependent number. 

\subsubsection{Matching with Simulation Results}

The solution~\exref{M_efn} is valid in the limit 
$\kres\gg k_*\sim\kd$ ($\Pr\gg1$), but, as we are about 
to see, convergence in~$\Pr$ is only logarithmic. 
Numerical solution of \eqref{FPEq} shows that scale 
separations of several decades are required for the scaling to be 
discernible. No such scale separations are achievable in 
direct numerical simulations. Therefore, in order to make a meaningful 
comparison, we must consider model-predicted spectra for the case 
of finite separations between the resistive cutoff $\kres$ and 
the infrared cutoff $k_*$, where the boundary condition is imposed. 
Seeking steady solutions of \eqref{FPEq}, we get 
\bea
\label{M_finite}
M(k) = (\const)\, k^s K_{i\nut(\sperp,\spar)}(k/\kres),
\eea
where $\nut(\sperp,\spar)=\bl[8(\gamma/\gperp)/(1+2\spar)\br]^{1/2}$ 
and $\kres$, $\gamma$, $s$ are defined in \eqsref{kres_def}{s_def}.
The equation for $\sperp$ and $\spar$ is obtained from the 
zero-flux boundary condition: the expression in square 
brackets in \eqref{FPEq} must vanish at $k=k_*$ (i.e., no energy 
passes through the left boundary in $k$ space). This gives 
\bea
\label{ev_eq}
{k_*\over\kres}\,K'_{i\nut}\({k_*\over\kres}\) 
- {1+2\sperp+6\spar\over1+2\spar}\,K_{i\nut}\({k_*\over\kres}\) = 0.
\eea
Together with the assumption $\sperp=4\spar$, this is a transcendental 
equation for $\spar$, which can be solved numerically 
for any fixed set of parameters $k_*$, $\eta$, $\gperp$. 

Note that in the limit of $k_*\ll\kres$, \eqref{ev_eq} reduces to 
$\sin[\nut\ln(k_*/2\kres)] = 0$,
whence
\bea
\gamma = {(1+2\spar)\,\pi^2\over8\bl[\ln(k_*/2\kres)\br]^2}\,\gperp
\sim {\cal O}\biggl({1\over\bl[\ln\Pr^{1/2}\br]^2}\biggr).
\eea
As $\Pr\to\infty$, we have $\gamma\to0$, converging logarithmically. 
This is the asymptotic solution discussed 
in \ssecref{ssec_loc_aniso} and \ssecref{ssec_solution}.

To make a comparison with our numerical results, we set 
$k_*=\kf=2\pi$, the box wavenumber, and use the following 
(isotropic) expression for the mixing rate:
$\gperp = (6/5)\gKA = c\,(\epsilon/\nu)^{1/2}$,
where $\nu$ is viscosity, $\epsilon=1$ is the mean forcing power 
[see \eqref{usq_balance}], and 
$c$ is an adjustable parameter. 
We have compared the model solutions with the (normalized) spectra 
obtained in our viscosity-dominated runs. 
The adjustable parameter $c$ was the same ($c=0.3$) 
for all cases. The entire sequence is well fitted by the model 
(except at $k/2\pi=1,2$, where boundary conditions 
and discretization effects are important): 
see \figref{fig_Mk_nlin}.  

\pseudofigureone{fig_Mk_nlin}{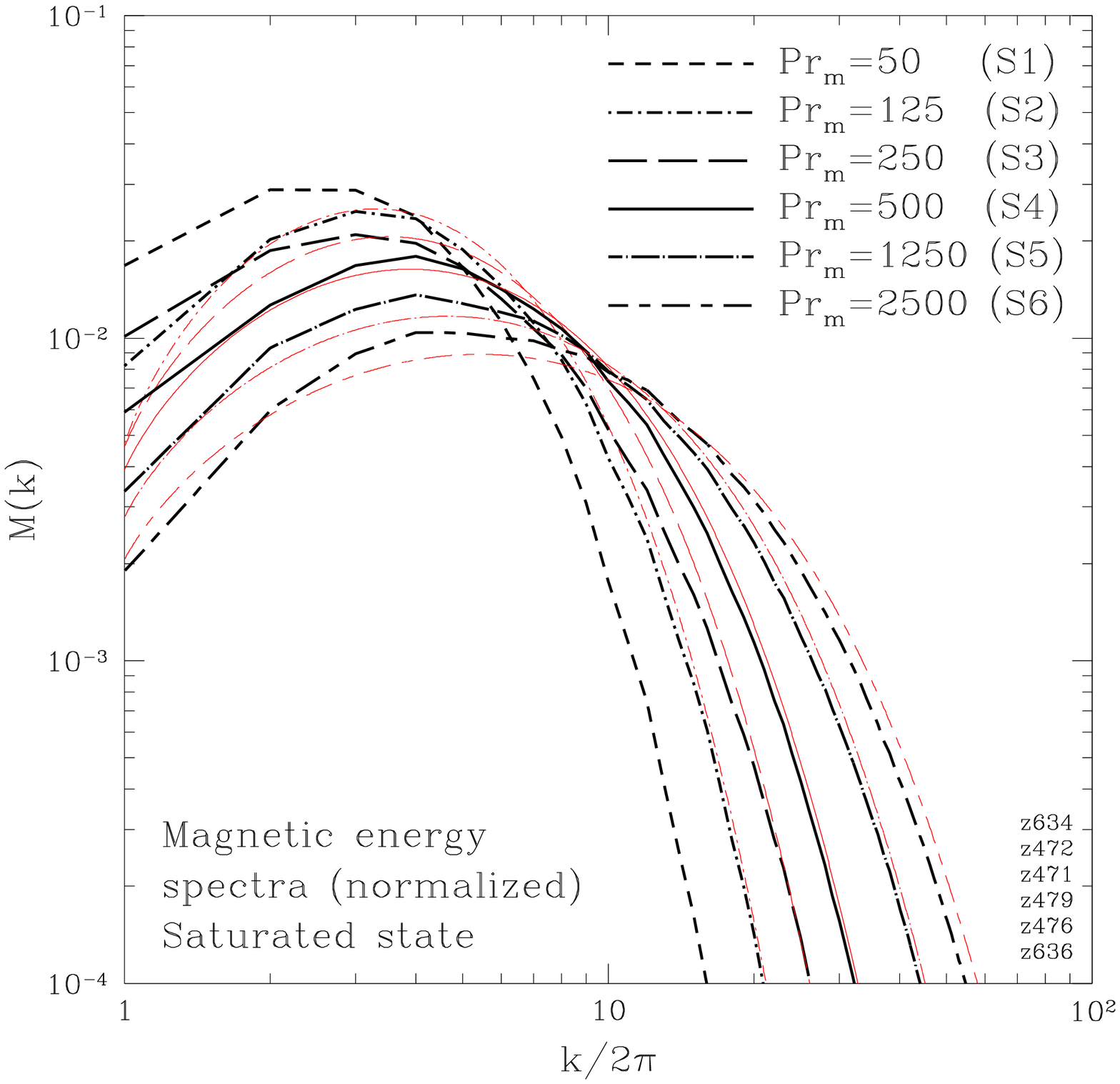}{f20.ps}{Magnetic-energy spectra 
(normalized by the total magnetic energy) in the saturated state. 
The thin red lines are spectra predicted by our model. 
For run~\run{S1}, the model does not give 
energy growth, but $\Pr$ in this case is too small 
for any asymptotic consideration to be applicable.} 

Considering the simplicity of our model, 
its ability to predict nonasymptotic 
numerical spectra so well is remarkable. 
It shows that 
numerical results are consistent with magnetic-energy 
spectra that are peaked at the resistive scale and 
would have a very flat positive spectral exponent 
in the interval $\kd\ll k\ll\kres$ if sufficiently 
large scale separations were resolved.

\subsection{Intermittency}
\label{ssec_intermittency_nlin}

In the saturated state, intermittency is reduced, implying 
tighter packing of the system domain by the magnetic field 
(\figref{fig_Bn}a). 
It is not a surprising result: nonlinear back-reaction imposes 
an upper bound 
on the field growth, and, once the strongest fields in the dominant 
substructure saturate, the weaker ones elsewhere have an opportunity 
to catch up. The pdf of the saturated field is
exponential at large $B$ \citep[\figref{fig_PB_tail}, 
cf.][]{Brandenburg_etal_structures,Cattaneo_solar}.
In the saturated state, 
the kurtosis does not depend on~$\Pr$ (\figref{fig_Bn}b). 
Furthermore, while $\Brms$ is not $\Pr$-independent 
for finite values of $\Pr$ (\figref{fig_E_eta}), 
the pdf of $B/\Brms$ is the same for all $\Pr$. 

For $B\ll\Brms$, the pdf is 
power-like, $P(B)\sim B^{1.7...1.8}$, 
and unchanged from the kinematic regime 
(see \ssecref{ssec_finite} and \figref{fig_PB_kin}b). 
This tail is made up of the weak fields in the bends of the folds. 
As we argued in 
\ssecref{ssec_fold_nlin} in the context of the curvature 
distribution, these fields remain 
quasikinematic, with no significant change in their 
statistics compared to the kinematic regime.

\subsection{Summary of the Saturated State}
\label{ssec_sum_visc}

We now summarize what we have learned about the saturated 
state of the dynamo in the viscosity-dominated regime:

\begin{enumerate}

\item The folded field structure inherited from the kinematic 
regime is preserved: magnetic fields remain organized in flux 
sheets (or ribbons) with alternating field direction and 
field strength and field-line curvature are anticorrelated. 
The scale separation between the fold length (flow scale) 
and the direction-reversal scale (resistive scale) is $\Pr^{1/2}$. 

\item The velocity field in the subviscous scale range 
($k>\kd$) develops a $k^{-4}$ spectrum, which is induced 
by the subviscous fluctuations of the Lorentz tension force 
and plays no part in the nonlinear back-reaction. 

\item The saturated magnetic-energy spectrum is flatter than 
in the kinematic case, but the bulk of the energy is still at 
the resistive scale. Accessible numerical resolutions 
do not allow us to measure the spectral 
exponent. However, the numerically obtained spectra are in an 
excellent agreement with those predicted for the same 
values of $\Pr$ by a simple saturation model, which, in the 
limit of very large $\Pr$, would give a very flat positive exponent. 
This model is based on the idea that the nonlinear saturation is 
achieved via anisotropization of the velocity-gradient 
statistics with respect to the local direction of the folds. 
The anisotropization consists in a reduction of the stretching 
component of the flow compared to its mixing component. 
The saturated spectra result from a balance 
between the weakened stretching and the mixing of the fields 
by quasi-2D interchange-like motions. 

\item The fields in the saturated state are somewhat more 
tightly packed than in the kinematic regime, but are still 
intermittent: instead of a lognormal pdf, they have an exponential 
one. Unlike in the kinematic case, the volume-filling fraction 
is independent of $\Pr$.

\end{enumerate}

\pseudofiguretwo{fig_Et_AB}{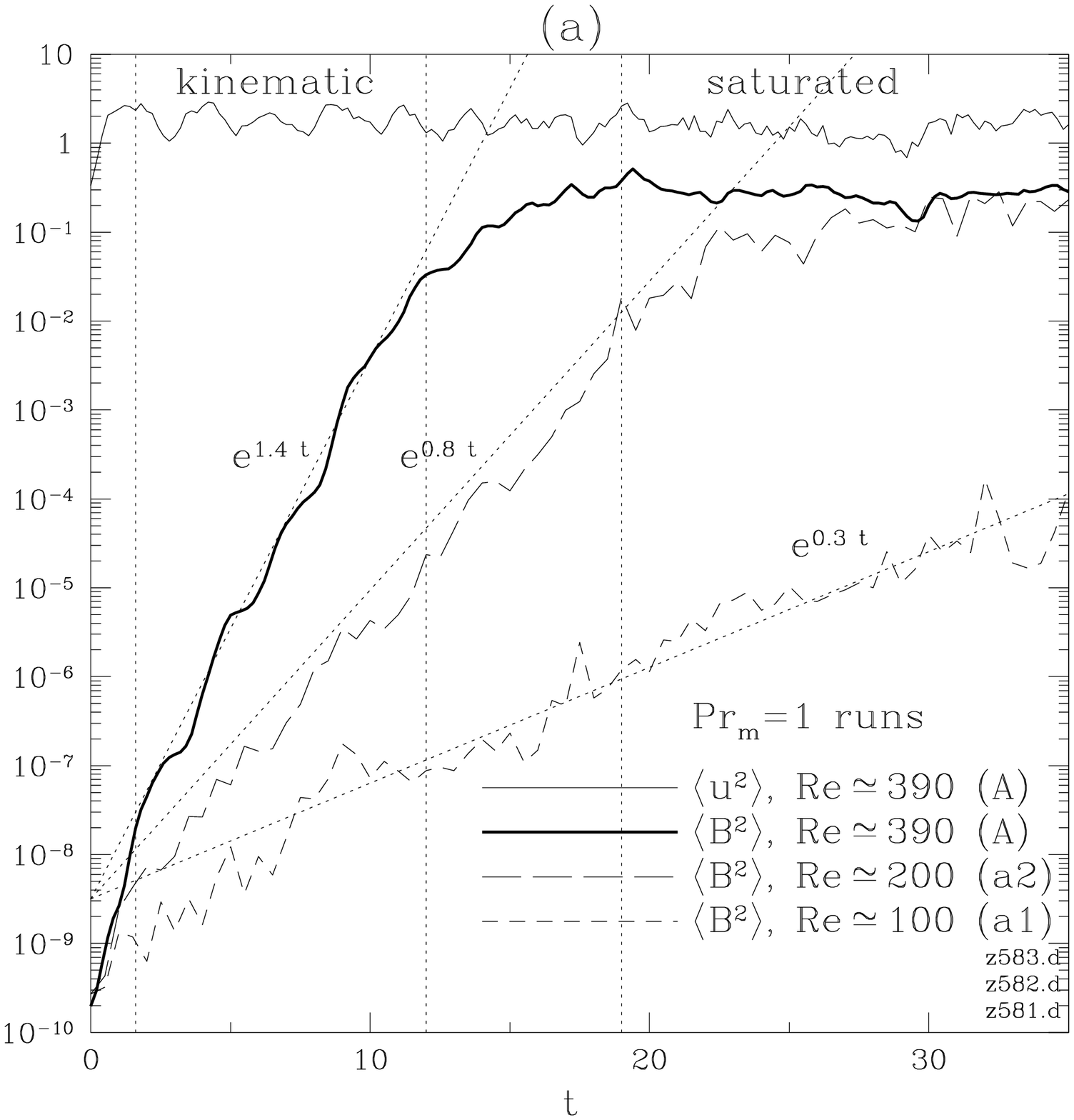}{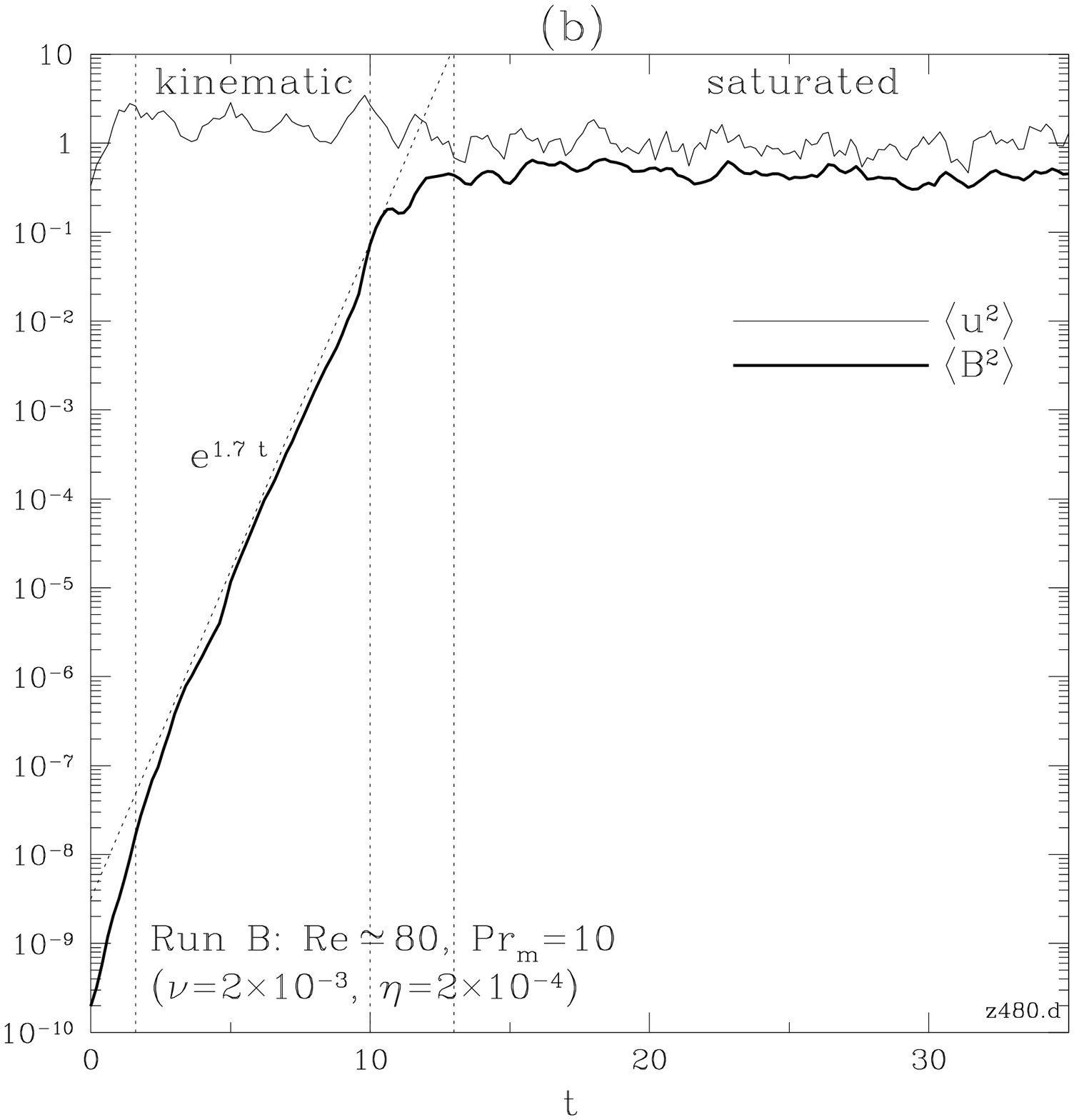}{f21a.ps}{f21b.ps}{Growth and saturation 
of the magnetic energy for (a) $\Pr=1$ runs 
(run~\run{a1} saturates around $t=50$, not shown) 
and (b) run~\run{B}.}

\section{SMALL-SCALE DYNAMO IN TURBULENCE WITH LARGE REYNOLDS NUMBERS}
\label{sec_Re}

Up to this point, we have not, in fact, attempted to study 
the turbulent dynamo in the proper sense. Because of the 
large viscosity, our velocity was spatially smooth and 
single-scale. The only feature that distinguished our model 
from synthetic dynamos in prescribed flows 
\citep[such as those reviewed in][]{STF} was that 
our velocity was random in time (because of the random 
forcing), homogeneous in space (had no fixed spatial 
features such as stagnation points), and subject to 
the physical nonlinear back-reaction associated with the 
Lorentz-tension term in \eqref{NSEq}. 
The main motivation for dwelling on this regime at such 
length was that 
a parameter scan in $\Pr$ could be afforded 
within accessible resolutions. However, if the small-scale 
dynamo theory is to make statements about physical reality, 
its ultimate goal must be to understand small-scale dynamo 
in turbulent velocity fields with large Reynolds numbers. 
In other words, we must attack the problem of {\em isotropic 
MHD turbulence}. 

This problem is essentially distinct from the more studied 
anisotropic case in which a strong mean field is imposed on the 
system. This latter case is somewhat better understood, mostly 
because we have what probably is a correct heuristic 
view of it as a turbulence of interacting Alfv\'en-wave 
packets \citep[due to][]{Kraichnan_IK}, just as our 
intuition about Kolmogorov turbulence is based on 
the idea of the constitutive fluid motions as ``eddies'' 
[note, however, that the 
more recent and increasingly widely accepted phenomenology 
of \citet{GS} with its idea of critical balance 
has blurred the distinction between Alfv\'en-wave packets 
and eddies]. 

It is the presence of the mean field that 
makes Alfv\'en waves possible. If no mean field is imposed, 
it is tempting to assume that we are still dealing 
with Alfv\'en waves, except now waves at small scales 
see the outer-scale magnetic fluctuations 
as the local mean field, 
along which these waves can propagate. However, in order 
for such a scheme to work, the outer-scale fluctuations must be 
more energetic than the small-scale ones, i.e., the bulk of 
the magnetic energy must be self-consistently kept at 
large (box) scales by some nonlinear mechanism. 
We are about to see that this does not happen (\ssecref{ssec_facts_sat}). 

Resolution constraints do not permit a meaningful parameter scan 
in the large-$\Re$ and large-$\Pr$ regime,
so we limit our discussion to two representative runs 
at the $256^3$ resolution: run~\run{A} is a traditional $\Pr=1$ run with 
$\nu=5\times10^{-4}$, while 
run~\run{B} has $\Pr=10$ and $\nu=2\times10^{-3}$.
Let us define the Reynolds number as 
\bea
\label{Re_def}
\Re={\usq^{1/2}\over\nu\kf},
\eea 
where $\kf=2\pi$ is the box wavenumber. 
Then in the saturated (kinematic) stage we have 
$\Re\simeq390\,(450)$ for run~\run{A} 
and $\Re\simeq80\,(100)$ for run~\run{B}. 
The corresponding values of the 
Taylor-microscale Reynolds number defined by 
\bea
\label{Rel_def}
\Rel={\(\usq/3\)^{1/2}\lambda\over\nu}
=\({5\over3}\)^{1/2}{\usq^{1/2}\over\nu\kl}
\eea
\citep[see][and \eqref{kl_def}]{Frisch_book} 
are $\Rel\simeq155\,(116)$ for run~\run{A} 
and $\Rel\simeq45\,(50)$ for run~\run{B}. 
In a few instances, we shall compare run~\run{A} 
with two other $\Pr=1$ runs (\run{a1} and \run{a2}; 
see \tabref{tab_index}) that have lower $\Re$. 
None of these runs are truly asymptotic in 
either $\Pr$ or $\Re$, so we do not attempt to extract 
scalings and all our conclusions are tentative.

Note that even in the case of $\Pr=1$ (\figref{fig_Et_AB}), 
the resolution constraints for isotropic MHD are more 
severe than for purely hydrodynamic simulations 
because (1) the resistive scale is 
smaller than the viscous scale (see \figref{fig_Sk_AB}a 
and \figref{fig_kt_AB}a) and (2) in order 
to achieve statistical steady state and collect adequate statistical 
information, simulations must be run for tens of box-crossing 
times, not just one or two as in the hydrodynamic case. 
The highest-resolution simulation of incompressible isotropic 
MHD to date is a $1024^3$ one 
with $\Re\simeq960$ and $\Pr=1$ by \citet{HBD_apjl}, 
who were able to run it  for a few box-crossing times. 
It appears, however, that even larger resolutions 
and $\Pr\gg1$ runs are needed to make 
statements about the truly asymptotic case.

In \ssecref{ssec_facts}, we describe the growth, saturation, and 
spectra of the magnetic energy in runs~\run{A} and~\run{B} 
on a purely factual level. A theoretical discussion is 
given in \ssecref{ssec_speculations}. 
In \ssecref{ssec_tests}, we 
return to the numerical results to test our theoretical 
arguments and to make some tentative conclusions 
(\ssecref{ssec_tent_conc}). 
Intermittency is studied in \ssecref{ssec_intermittency_AB}. 
The summary of the main results is in~\secref{sec_summary}. 

\pseudofiguretwo{fig_Sk_AB}{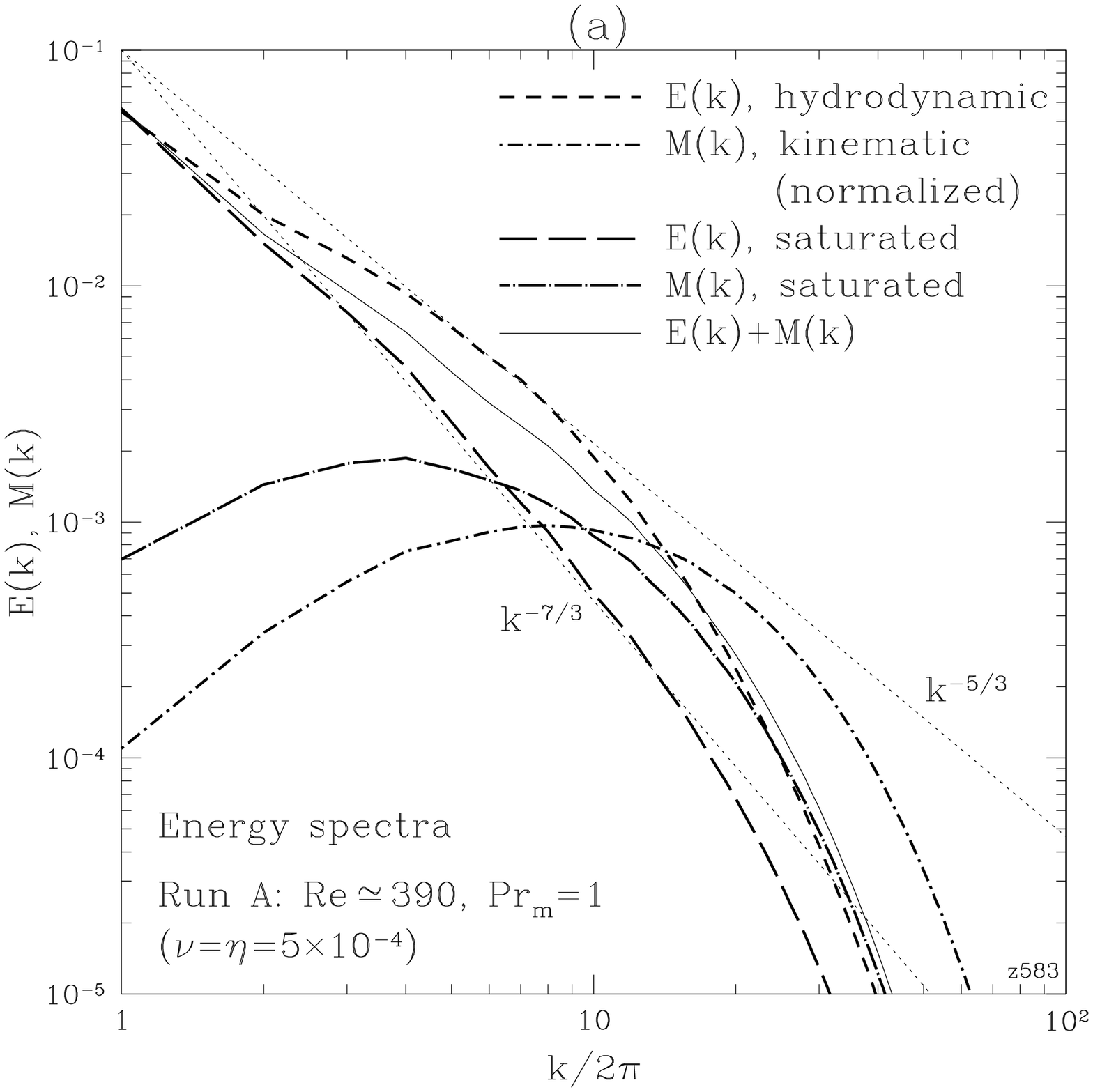}{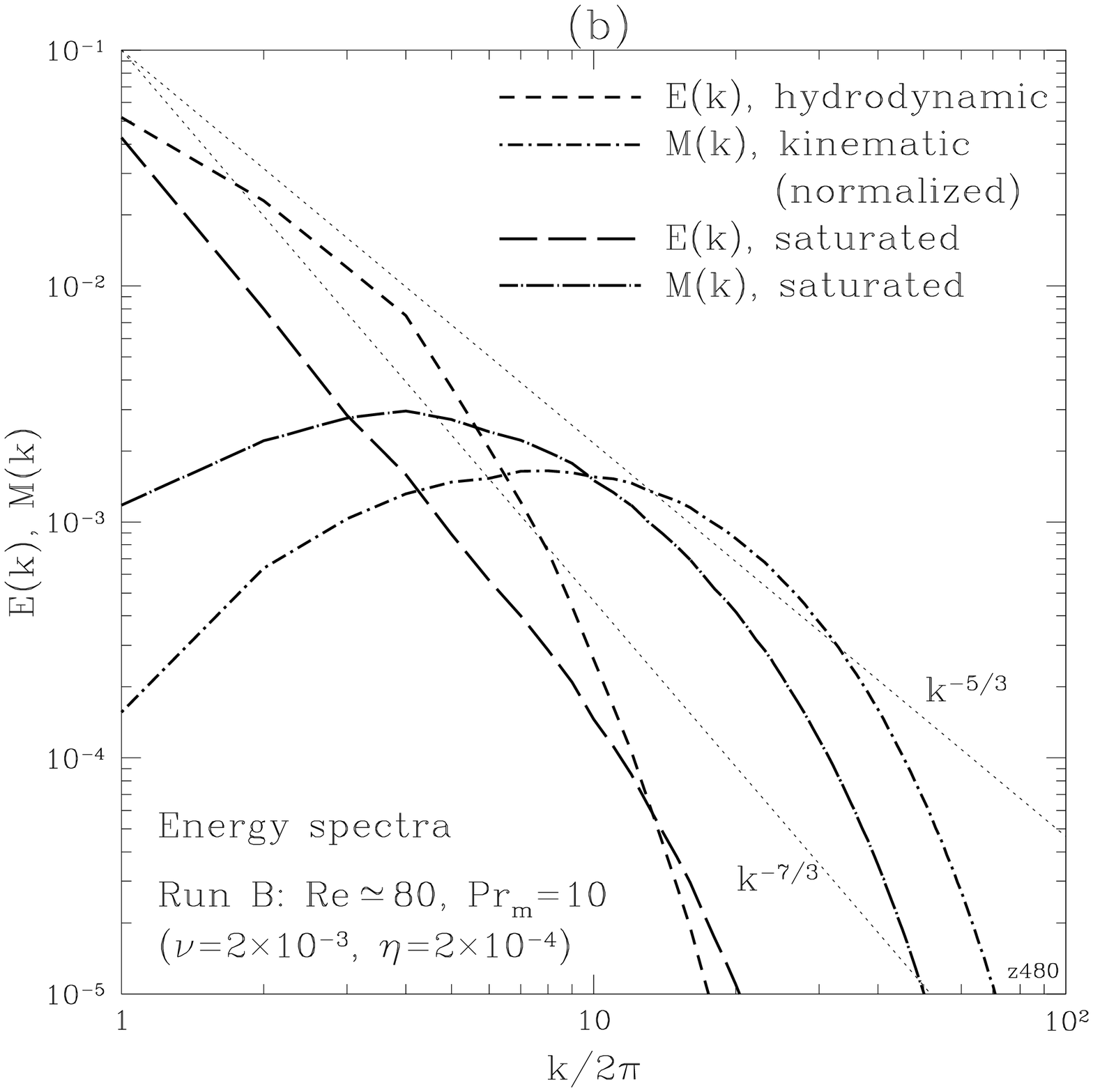}{f22a.ps}{f22b.ps}{Energy  
spectra for (a) run~\run{A} and (b) run~\run{B} in the kinematic regime 
and in the saturated state. The magnetic-energy spectra in the 
kinematic regime are normalized by the total magnetic energy, then 
averaged and rescaled by the total magnetic energy of the saturated 
state (for comparison with the saturated spectra).} 

\subsection{Growth, Saturation, and Energy Spectra: Facts} 
\label{ssec_facts} 

\subsubsection{The Kinematic Regime} 
\label{ssec_kin_AB}

The small-scale kinematic dynamo in the large-$\Pr$ limit 
depends very little on the particulars 
of the velocity field (as long as this velocity 
is three-dimensional and sufficiently chaotic). 
As seen below, all indications are that 
our runs~\run{A} and~\run{B} are nonasymptotic cases of 
the large-$\Pr$ regime; i.e., the same 
stretching and folding mechanism is responsible for the 
small-scale-field generation. 
In their kinematic stage, these runs are, therefore, very 
similar to the viscosity-dominated simulations. 

Both runs exhibit exponential magnetic-energy growth 
in the kinematic regime (\figref{fig_Et_AB}). 
In Kolmogorov turbulence, small eddies turn over faster than 
the large ones, so dominant stretching is done by the 
viscous eddies. We expect, therefore, that 
the growth rate is $\gamma\sim\Grms$ when $\Re\gg1$, where 
$\Grms = \br(\<|\nabla\vu|^2\>/15\bl)^{1/2}$ is the rms rate of strain. 
Since $\nu\<|\nabla\vu|^2\>=\epsilon=1$ [\eqref{usq_balance}], we have 
$\Grms\sim\Re^{1/2}\tbox^{-1}$, where $\tbox=\usq^{-1/2}$ is 
the box crossing time (the box size is~1). 
In principle, to check these statements, we need 
a $\Re$ parameter scan at constant large $\Pr$. 
While this is not possible at current resolutions, 
$\Pr=1$ runs can be used to establish qualitatively 
that $\gamma$ increases with $\Re$. 
In \tabref{tab_Pr1} 
we give values of $\gamma$ (least-squares fit; see also 
\figref{fig_Et_AB}a), $\Grms$, 
and $\tbox^{-1}$ for run~\run{A} 
and for two extra runs with $\Pr=1$ and smaller $\Re$. 
For comparison, we also include the data for runs~\run{B} 
and~\run{S4}. Even though the numerical values of $\gamma$ are 
substantially smaller than $\Grms$ (and might appear, 
misleadingly, to be closer to $\tbox^{-1}$), 
it is clear that both $\Grms$ and 
$\gamma$ grow with $\Re$, while $\tbox^{-1}$ stays the same. 
The smallness of the values of $\gamma$ relative 
to the rms rate of strain should probably be attributed 
to the resistive correction being order-unity for 
$\Pr=1$ and moderate values of $\Rm$ 
in our runs [see~\eqref{lambda_sln}]. 

\notetoeditor{Please put Table 2 here: it should fit in one column}

It is fair to observe that the increase of $\gamma$ with 
$\Re$ that we see has, in fact, two causes: the growth 
of the rate of strain with $\Re$ and the increasing value 
of $\Rm$. Indeed, in run~\run{a1}, $\Rm\simeq100$ is quite close to 
the critical value $\Rmc$, below which there is no dynamo: 
this is demonstrated by run~\run{a0} (\tabref{tab_index}), 
which has $\Rm\simeq52$ and decays 
[this implies $\Rmc\in(52,100)$, which is the same estimate 
as that for the viscosity-dominated runs; see \ssecref{ssec_spectrum}]. 
The effect of finite $\Rm$ on the value of $\gamma$ 
will saturate at large $\Rm$, while the large-$\Re$ 
asymptotic should be $\gamma\sim\Re^{1/2}$. 
The values of $\Re$ in our runs are not large enough for 
us to be able to make statements about scalings. 
Simulations at larger $\Re$ appear to be consistent 
with the scaling $\gamma\sim\Re^{1/2}$ 
(N.~E.~Haugen 2003, private communication). 

In both runs~\run{A} and~\run{B}, 
the magnetic energy, initially at the box scale, is quickly 
piled up the resistive scale, just as in the viscosity-dominated case. 
The evolution of the spectra is again self-similar (analogous 
to \figref{fig_Mk_kin}a). The normalized spectra averaged over 
the kinematic regime are shown in \Figref{fig_Sk_AB}. 

\subsubsection{The Saturated State}
\label{ssec_facts_sat}

The dynamo eventually saturates and a statistical steady state 
is achieved (\figref{fig_Et_AB}). 
The kinetic energy $\usq$ in this regime is smaller than 
in the purely hydrodynamic case ($\vB=0$). The magnetic 
energy $\Bsq$ is smaller than $\usq$ by a factor of 6 in run~\run{A} 
and by a factor of $\sim2.3$ in run~\run{B} (\tabref{tab_index}). 
The saturated spectra (\figref{fig_Sk_AB}) exhibit no 
scale-by-scale equipartition between the kinetic and magnetic energies 
even at $\Pr=1$. 

In both runs \run{A} and \run{B}, the kinetic-energy spectrum 
in the purely hydrodynamic regime has a 
very short segment consistent with the Kolmogorov $k^{-5/3}$ scaling 
(this is somewhat better pronounced for run~\run{A}). 
In the nonlinear saturated MHD state, the spectrum is steeper: 
in each of the runs, there is a segment best fitted by $k^{-7/3}$. 
It is not clear if this scaling (if it is really there) 
survives in the large-$\Re$ limit 
or is simply a limited-resolution effect 
\citep[spectra steeper than Kolmogorov sometimes appear 
in hydrodynamic simulations with too large viscosity; 
see theoretical discussion in][]{Frisch_book}. 
If this scaling is, in fact, asymptotic in $\Re$, 
it is a feature of the fully-developed 
isotropic MHD turbulence for which, 
at present, we have no theoretical explanation (see 
some further observations in \ssecref{ssec_speculations}). 
What is certainly true is that 
(1) the kinetic energy is dominated by the large (box) scales 
and (2) the characteristic ``eddy'' time $\sim\bl[k^3E(k)\br]^{1/2}$ 
peaks at the viscous scale. 

The magnetic-energy spectrum is 
shifted toward small (resistive) scales, with a 
substantial excess of magnetic over kinetic energy there. 
There is no discernible scaling. Note that the shapes of the 
magnetic-energy spectra in runs \run{A}, \run{B} and \run{S1--S6} 
(\figref{fig_Mk_nlin}) are not very different. 
Thus, it appears that there is no inverse cascade of magnetic 
energy and that the characteristic scale 
of the magnetic field is the resistive scale. 

The fact that numerical simulations do not support equipartition was 
first explicitly emphasized by \citet{MCM}, even though many 
prior numerical studies had found similar 
spectra \citep{Meneguzzi_Frisch_Pouquet,Kida_Yanase_Mizushima,Miller_etal,Cho_Vishniac,Chou}. 
However, there was a reluctance to accept that numerical evidence 
was at odds with the equipartition hypothesis and, therefore, 
with the standard Alfv\'enic phenomenology. 
In order to extract scalings, a popular 
device \citep[due originally to][]{Kida_Yanase_Mizushima} 
has been to consider the spectrum of the total energy, $E(k)+M(k)$. 
In our opinion, this choice of diagnostic obscures rather 
than clarifies the situation. Indeed, it only 
makes sense to add the spectra together 
if there is a symmetry between $\vu$ and $\vB$. 
For the spectra that obtain for 
$\Pr=1$ and at limited resolutions, adding $E(k)$ and $M(k)$ gives 
one what might be a false impression 
that a Kolmogorov scaling actually emerges, 
as shown in \Figref{fig_Sk_AB}a. 
However, without equipartition,
there is no obvious reason why asymmetric kinetic and 
magnetic energy spectra should add up 
to something that has a Kolmogorov scaling. 

One could argue that all current simulations are 
nonasymptotic and that, were the resolution to be increased 
dramatically, equipartition and Kolmogorov scaling 
would emerge at hitherto inaccessible small scales. 
While this cannot be ruled out, the indications so far are not 
encouraging. \citet{HBD_apjl,HBD_pre} 
performed simulations at resolutions ranging from $64^3$ 
to $1024^3$ (with $\Pr=1$ and maximum $\Re=960$). 
In none of their runs were the kinetic and magnetic energies 
in scale-by-scale equipartition, with magnetic energy always 
shifted toward the small scales. Furthermore, 
they found that the Kolmogorov scaling that seemed to 
hold at $256^3$ (cf.~\figref{fig_Sk_AB}a)
actually got worse at $1024^3$, with magnetic 
energy bulging up at small scales. 
They attributed this problem to some form of the bottleneck effect, 
i.e., essentially, to nonlocal interactions in $k$ space. 
Small-scale dynamo is, in fact, just that: 
a nonlocal interaction between velocity and magnetic fields. 

A Kolmogorov scaling for the total energy was also reported 
by \citet{Biskamp_Mueller}
who simulated {\em decaying} isotropic MHD turbulence with $\Pr=1$ 
at resolutions of $512^3$. 
It is, however, unlikely that decaying-MHD results 
apply to the forced case. Indeed, the 
kinetic and magnetic energy spectra that \citet{Biskamp_Mueller} 
found in their self-similarly decaying regime were quite 
different from those for the 
forced case: while again there was no equipartition, it was 
the magnetic energy that dominated the kinetic one 
at large scales (we presume that this is due to large-scale 
MHD equilibrium structures). 

Thus, on the balance of the available numerical results, 
while we cannot definitively claim that the equipartition theory 
is wrong, it is certainly true that there is no evidence 
in support of it and a fairly strong case against.

\subsection{The Physics of Saturation: 
Selective Decay, Alfv\'en Waves, and Mixing} 
\label{ssec_speculations}

Let us consider what are the reasonable physical scenarios 
of nonlinear transition from field growth (dynamo) to saturation 
and what they imply about the final state. 

\subsubsection{The Intermediate Nonlinear Regime}
\label{ssec_int}

The folded field structure implies the criterion 
for the onset of nonlinearity: the back-reaction 
is controlled by the Lorentz tension 
force~$\vB\cdot\nabla\vB\sim\kpar B^2$, which only 
depends on the parallel gradient of the field and does not know about 
direction reversals ($\kpar\sim\kd$). 
Balancing~$\vB\cdot\nabla\vB\sim\vu\cdot\nabla\vu$, 
we find that back-reaction is important when magnetic energy becomes 
comparable to the energy of the viscous-scale eddies. 
Eddies at scales above viscous 
are still more energetic than the magnetic field 
and continue to stretch it at their (slower) 
turnover rate. When the field energy reaches the energy of these 
eddies, they are also suppressed and the dominant remaining stretching 
is done by yet larger and slower eddies. 
Consider a time~$t$ when the total magnetic energy 
$\Bsq(t)$ is less than its final saturation value but larger than 
the energy of the viscous-scale eddies. Let 
the stretching scale $\ls(t)$ be the scale of the 
eddies whose energy is equal to $\Bsq$, i.e., 
\bea
\label{ls_def}
u_{\ls}^2\sim\Bsq.
\eea
All eddies smaller than $\ls$ have had their stretching 
component suppressed, while all the larger eddies are 
slower, so, at time $t$, the effective rate at which the field 
is being stretched is $u_{\ls}/\ls$. Using this and \eqref{ls_def}, 
we estimate 
\bea
\label{int_growth}
\Dt \Bsq \sim {u_{\ls}\over\ls}\,\Bsq \sim {u_{\ls}^3\over\ls} 
\sim \epsilon = \const,
\eea
where $\epsilon$ is the Kolmogorov energy flux. 
\Eqref{int_growth} implies $\Bsq(t)\sim\epsilon\,t$, i.e., 
there is an intermediate period of nonlinear evolution 
between the onset of nonlinearity and saturation, 
during which the magnetic energy grows proportionally 
to time rather than exponentially. 
The force balance underpinning the back-reaction 
by the folded fields on the eddies of size $\ls$ is 
\bea
\vu\cdot\nabla\vu \sim {u_{\ls}^2\over\ls} \sim 
\vB\cdot\nabla\vB \sim \kpar\Bsq,
\eea
so that $\kpar\sim\ls^{-1}$, i.e., 
the folds are elongated to the stretching scale 
\citep[for a slightly more quantitative argument in favor of fold 
elongation, see][]{SCHMM_ssim}. 

The key question is by how much $\ls$ can exceed the viscous scale~$\ld$, 
for it is the saturated value of $\ls$ that will determine 
the saturated magnetic energy and its spectral distribution. 
Saturation is achieved when the nonlinear growth according 
to \eqref{int_growth} is balanced by diffusion:
\bea
\label{ls_balance}
{u_{\ls}\over\ls} \sim {\eta\over\lres^2}.
\eea
Therefore, in order to determine $\ls$, we need to know 
what happens to the resistive scale $\lres$ in the 
nonlinear regime. There are two main possibilities 
(subject to the caveats in \ssecref{ssec_largePr}). 

\subsubsection{Saturation with Efficient Mixing} 
\label{ssec_eff}

If the velocities at $\ell<\ls$ act 
as interchange-like motions in the plane transverse 
to the local direction of the folds and, consequently, 
mix the field lines in a quasi-2D fashion, then 
the resistive scale is determined by the characteristic 
timescale of these motions {\em at the viscous scale,}
\bea
 {\eta\over\lres^2} \sim {u_{\ld}\over\ld}.
\eea
Comparing this with \eqref{ls_balance}, we see 
that $\ls$ cannot grow larger than $\ld$ by 
more than a factor of order unity. In this case, the motions 
at or just above the viscous scale are anisotropized, so 
mixing is enhanced and wins the competition 
with stretching, leading to saturation. 
The magnetic energy saturates at a value determined 
by the energy of the viscous-scale eddies. 
Thus, the signature of this type of saturation would be 
the following dependence of the saturated magnetic energy 
on~$\Re$:
\bea
\label{Bsq_visc}
\Bsq\sim u_{\ld}^2 
\sim \Re^{-1/2}\usq.
\eea

\pseudofiguretwo{fig_kt_AB}{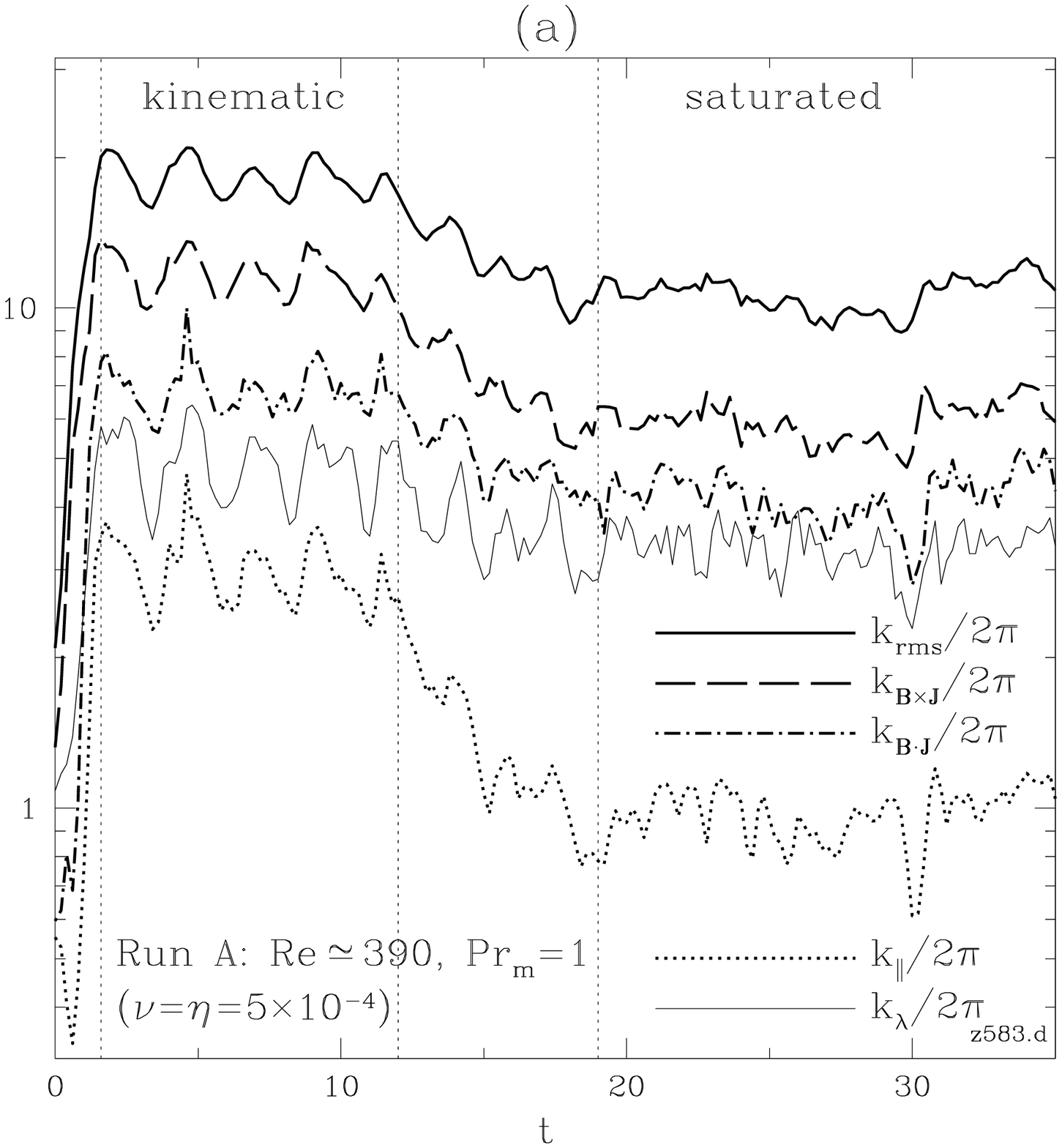}{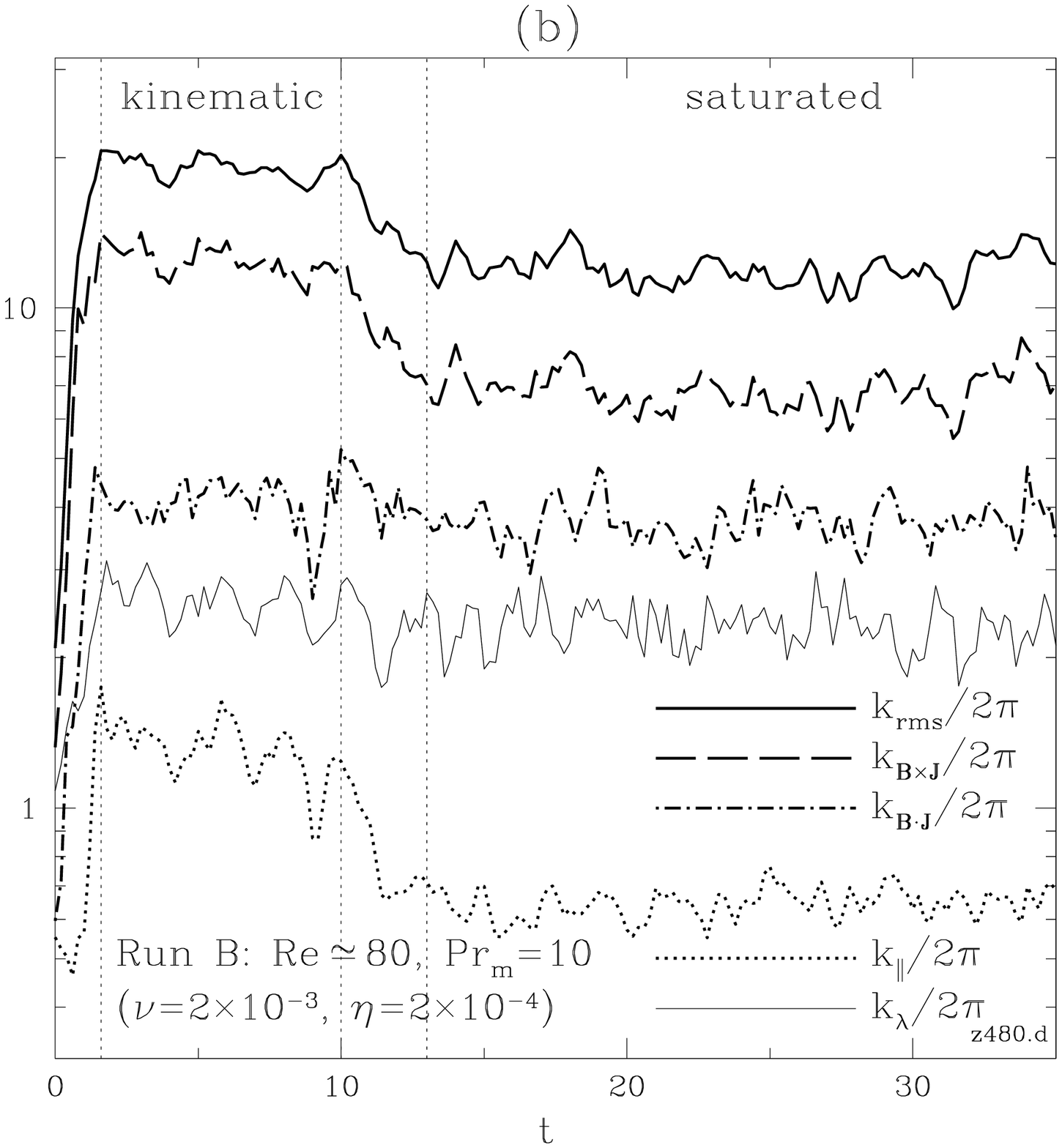}{f23a.ps}{f23b.ps}{Evolution of 
characteristic wavenumbers (defined in \ssecref{ssec_scales})
for (a) run~\run{A} and (b) run~\run{B}. 
Note that for Kolmogorov turbulence, $\kl\sim\Re^{-1/4}\kd$.}

\subsubsection{Saturation with Inefficient Mixing} 
\label{ssec_ineff}

The second possibility is that mixing by motions 
at $\ell<\ls$ is inefficient, so that 
both stretching and mixing of the field is mostly 
done by the stretching-scale eddies. 
It is then the turnover time of these eddies that determines 
the resistive scale of the magnetic fields, giving   
[from \eqref{ls_balance}, \exref{ls_def}, and \exref{int_growth}]
\bea
\lres \sim \(\eta\ls\over u_{\ls}\)^{1/2} \sim (\eta\,t)^{1/2}.
\eea
Physically, this expression describes selective decay 
of the magnetic energy at scales too small to be sustainable 
by the weakened stretching (or mixing). 
In this scenario, the magnetic energy is free 
to grow to the equipartition level, 
\bea
\label{Bsq_equip}
\Bsq\sim\usq,
\eea
at which point $\ls\sim\lf$ and the resistive scale is 
determined by the stretching rate of the outer-scale eddies:
\bea
\label{lres_nlin}
\lres\sim \(\eta\lf\over u_{\lf}\)^{1/2} \sim 
\Rm^{-1/2}\lf \sim \Re^{1/4}\Pr^{-1/2}\ld,
\eea
i.e., it is larger by a factor of $\Re^{1/4}$ than in 
the kinematic regime. Note that the condition for $\lres$ to be 
distinquishably smaller than $\ld$ in a numerical experiment is, 
therefore, 
\bea
\label{Pr_Re_condition}
\Pr \gg \Re^{1/2} \gg 1,
\eea
which is impossible to satisfy at present resolutions. 
Since the folds elongate so that $\kpar\sim\ls^{-1}\sim\lf^{-1}$ 
(\ssecref{ssec_int}), 
the separation between the direction-reversal scale and the fold 
length increases: 
\bea
\label{scale_sep_nlin}
{\kres\over\kpar} \sim {\lf\over\lres} 
\sim \Re^{1/2}\Pr^{1/2},
\eea
which is a factor of $\Re^{1/2}$ larger than in the kinematic regime.

If the velocities at $\ell<\ls$ are not
interchange-like mixing motions, what are they? 
There is another type of motion that does 
not amplify the magnetic field and can, theoretically, 
be present below the stretching scale: 
Alfv\'en waves can propagate 
along the folded direction-reversing magnetic fields 
(\figref{fig_waves}). 
These waves are damped by viscosity and make a contribution 
to both the kinetic and magnetic energy spectra. 
Their dispersion relation is 
$\omega^2 = \vk\vk:\vb\vb \Bsq$ 
for $\kd\ll k\ll\kpar\sim\ls^{-1}$ \citep{SCHMM_ssim}. 
This interval is non-empty provided that $\ls\gg\ld$. 
Therefore, mixing must be inefficient in a 
turbulence made up of such waves; otherwise, $\ls$ cannot 
grow sufficiently far above $\ld$ for them 
to exist in the first place. 

\pseudofigureone{fig_waves}{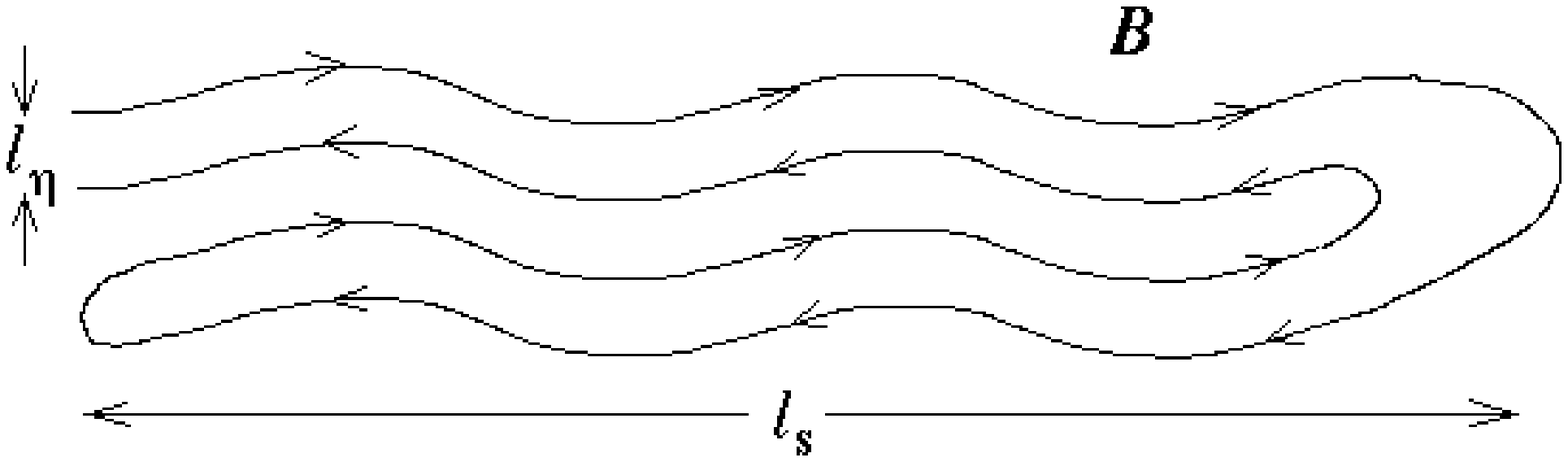}{f24.ps}{Alfv\'en waves along folded 
fields: a sketch.}

Detecting these Alfv\'en waves in numerical simulations is 
a considerable challenge both because of large scale separations 
that must be resolved and because designing the appropriate 
diagnostics is not straightforward. The simulations we present 
here are not appropriate for such a task. However, 
some of the physical possibilities we have anticipated 
can be tested. 

\subsection{Numerical Evidence: Facts and Interpretations} 
\label{ssec_tests} 

\subsubsection{Saturated Energy Levels}

The key challenge now is the resolution of the dichotomy described 
in \ssecref{ssec_eff} and \ssecref{ssec_ineff}: saturation 
with magnetic energy comparable to the viscous-eddy energy [\eqref{Bsq_visc}] 
or to the total energy of the 
turbulence [\eqref{Bsq_equip}]. The most direct 
approach would be to look at the scaling of the ratio 
$\Bsq/\usq$ with $\Re$ at constant large $\Pr$, but this 
is not yet possible at current resolutions. 
At $\Pr=1$, $\Bsq/\usq$ increases with $\Re$ (\tabref{tab_Pr1}).
In the higher-resolution $\Re$ scan with $\Pr=1$ produced by 
\citet{HBD_apjl}, $\Bsq/\usq$ appeared to saturate at 
larger~$\Re$. Similar results were reported in earlier numerical studies 
\citep{Kida_Yanase_Mizushima,Kleva_Drake,Miller_etal,Archontis_etal_nlin}.
Thus, at least for $\Pr=1$, 
numerical results do not support the scaling~\exref{Bsq_visc}, 
suggesting that it is 
the second possibility (\ssecref{ssec_ineff}) that is realized. 

\subsubsection{The Intermediate Nonlinear Regime}

We now examine the evolution of our runs \run{A} and \run{B} 
from the kinematic to nonlinear regimes to see if we can 
identify any of the features of the saturation scenarios 
proposed in \ssecref{ssec_speculations}. 

We notice immediately that, between the kinematic stage of 
exponential growth and the saturated statistically steady state, 
an intermediate nonlinear 
stage during which the magnetic energy grows slower than 
exponentially is manifest in run~\run{A} and, to a lesser extent, 
in run~\run{B} (\figref{fig_Et_AB}). 
This intermediate regime was not present in the viscosity-dominated 
runs and is clearly a large-$\Re$ effect in agreement 
with the argument in \ssecref{ssec_int}. 

\pseudofiguretwo{fig_PK_AB}{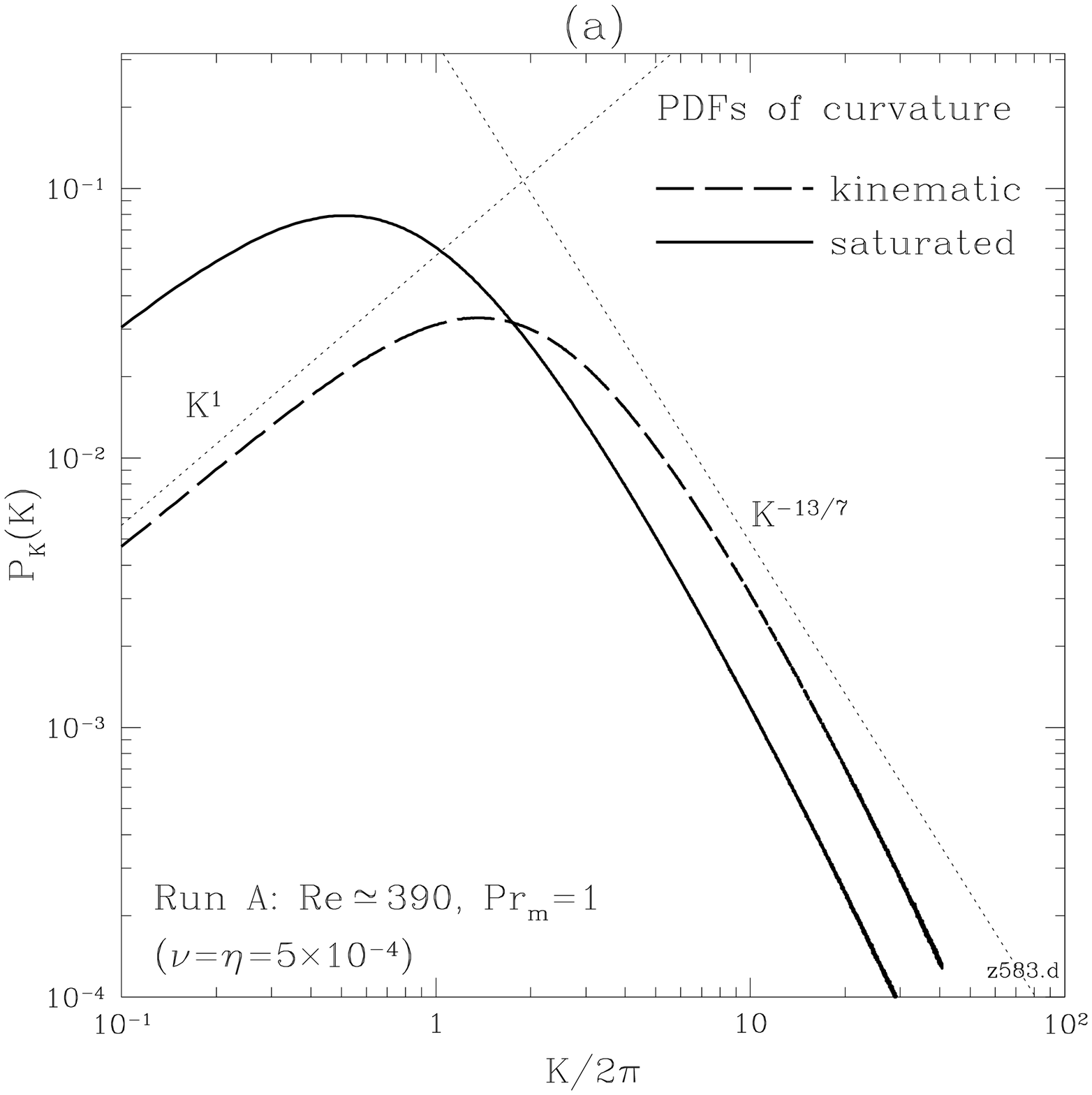}{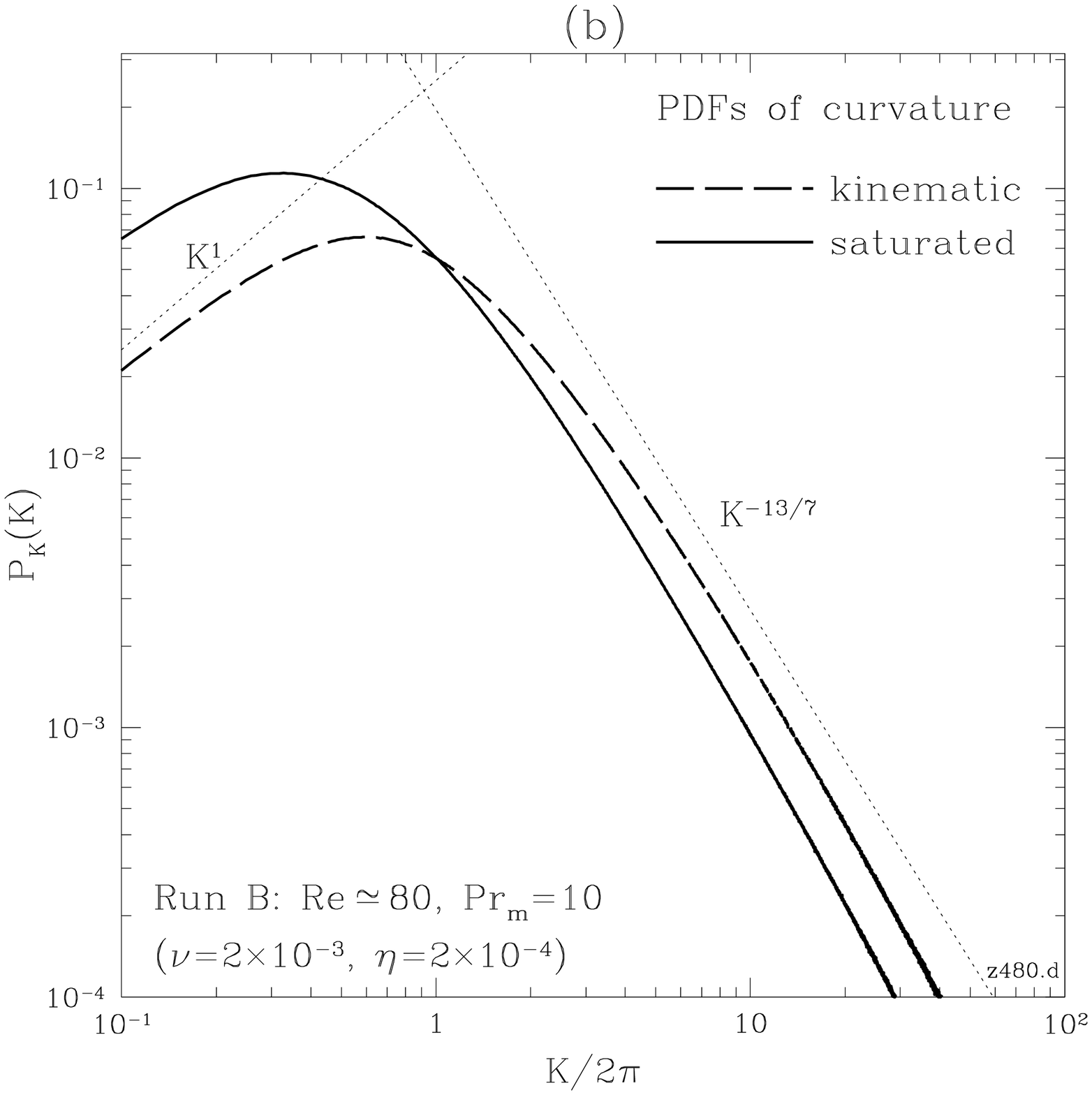}{f25a.ps}{f25b.ps}{Curvature 
pdf's for (a) run \run{A} and (b) run \run{B} during the kinematic 
regime and in the saturated state. The slopes corresponding to $K^1$ and 
$K^{-13/7}$ [see \eqref{PK_st}] are given for reference.} 

\subsubsection{The Folded Structure}
\label{ssec_fold_AB}

Employing the same diagnostics as in \secref{sec_kin}
and \secref{sec_nlin}, we find that the folded structure 
appears to survive in runs~\run{A} and \run{B} despite 
the moderate values of $\Pr$. 

The time histories of the 
characteristic wavenumbers (defined in \ssecref{ssec_scales}) 
are shown in \Figref{fig_kt_AB}. 
In the kinematic regime, $\kB\sim\kbxj\gg\kpar$.
During the intermediate nonlinear regime, 
$\kpar$ decreases further 
and becomes comparable to the box wavenumber. 
The resistive wavenumber $\kB$ also decreases, 
but by a lesser amount, which is consistent with the ideas 
of selective decay, fold elongation and increased 
scale separation in the folds in the saturated state 
set forth in \ssecref{ssec_int} and 
\ssecref{ssec_ineff} [see \eqref{scale_sep_nlin}]. 
The fold elongation is a new effect: it could not 
be seen in the viscosity-dominated runs (\secref{sec_nlin}), 
for which there was no separation between the viscous 
and box scales.  

All other features associated with the folded structure 
that we reported in \ssecref{ssec_folding} and \ssecref{ssec_fold_nlin} 
for the viscosity-dominated runs 
carry over to runs \run{A} and \run{B}: 
the magnetic-field strength and curvature are anticorrelated 
(e.g., the correlation coefficient $C_{K,B}$, \eqref{CKB_def}, 
is within 6\% of -1, see \tabref{tab_index}) 
and the pdf's of curvature (\figref{fig_PK_AB}) 
are again concentrated at velocity scales. The power tail 
appears to be somewhat steeper than $K^{-13/7}$ predicted 
for the large-$\Pr$ case (the steepening is more pronounced for 
run \run{A}). Interestingly, the scaling $P_K(K)\sim K$ 
at $K\ll\kd$ (to the left of the peak) suggested by \eqref{PK_st} 
seems to be in evidence for the kinematic pdf's. 
The elongation of the folds in the saturated state 
compared to the kinematic regime is manifested by the peak 
of the curvature pdf shifting to the left.

\subsubsection{The Tension Spectra} 
\label{ssec_tension_AB}

Let us now look at the magnetic back-reaction 
$\vB\cdot\nabla\vB$ as a function of $k$: 
the tension spectra for runs \run{A} and {B} are plotted 
in \Figref{fig_Tk_AB}. 
For run~{B}, the tension spectrum is flat up to 
the outer (box) scale; 
for run~\run{A}, it has a plateau that starts  
at $k/2\pi\sim3$, which should probably also be 
interpreted as corresponding to the outer scale. 
Based on the argument of \ssecref{ssec_4order}, 
such flat tension spectra are consistent with magnetic fields 
that are folded with the characteristic fold length at the outer 
scale. In a further confirmation of the consistency of this 
view, the relation $T(k)\simeq\kpar^2M_4(k)$ [\eqref{sln_Tk}] 
roughly holds (see \figref{fig_Tk_AB}). 

Thus, the tension spectra in runs \run{A} and \run{B} 
have similar properties to those in the 
viscosity-dominated case (\ssecref{ssec_tension}) 
except the tension force $\vB\cdot\nabla\vB$ is 
no longer passively compensated by the velocities 
as in \eqref{visc_bal_eq}: \figref{fig_Tk_AB} shows that 
the relation \exref{visc_bal} is only satisfied at very small scales. 
Instead, in the saturated state, we expect the nonlinear 
balance~\exref{ls_balance} to hold at the stretching scale~$\ls$ 
with velocities at $\ell>\ls$ remaining essentially 
hydrodynamic and velocities at $\ell<\ls$ modified 
so that they cannot significantly affect the folded structure,  
i.e., either mix or stretch the magnetic field. 
Since \eqref{ls_balance} is only an order-of-magnitude 
statement, it is not meaningful to compare the actual 
values of $\vu\cdot\nabla\vu$ and $\vB\cdot\nabla\vB$. 
However, it is still instructive to compare the behavior of $T(k)$ 
with that of $k^3E(k)^2$, a quantity that roughly 
measures the spectrum of the hydrodynamic interaction term 
$\vu\cdot\nabla\vu$. In \Figref{fig_Tk_AB}, this quantity 
is plotted in both the purely hydrodynamic case and 
the saturated MHD state [note that the absolute magnitude of 
$k^3E(k)^2$ has been rescaled in \Figref{fig_Tk_AB} and 
should not be compared to that of $T(k)$]. 
Interestingly, the intervals of flat $T(k)$ 
for both runs coincide with the intervals where there is a substantial 
suppression of $k^3E(k)^2$ in the MHD case compared to the 
hydrodynamic case. These intervals correspond to the 
steeper-than-Kolmogorov scaling of the kinetic-energy spectra 
noted in \ssecref{ssec_facts_sat}. While we reiterate the caveat 
made in \ssecref{ssec_facts_sat} that this scaling might be an artifact 
of limited resolution, it is just possible that we are 
dealing with a genuine spectral signature of the nonlinearly 
modified fluid motions in the interval $\ks<k<\kd$. 

Based on these plots, we hypothetically identify 
the stretching wavenumbers for our runs: 
for run~\run{A}, $\ks/2\pi\simeq3$; 
for run~\run{B}, $\ks/2\pi\simeq1$. 
The resistive and viscous 
cutoffs for these runs are all around 
$k/2\pi\sim10$ and hard to tell 
apart in the saturated state, which is consistent with 
the estimates in \ssecref{ssec_ineff} and $\Pr\gtrsim1$ 
[note that even run~\run{B} does not really 
satisfy the condition~\exref{Pr_Re_condition} necessary for 
$\kd$ and $\kres$ to be distinguishable]. 

\pseudofiguretwo{fig_Tk_AB}{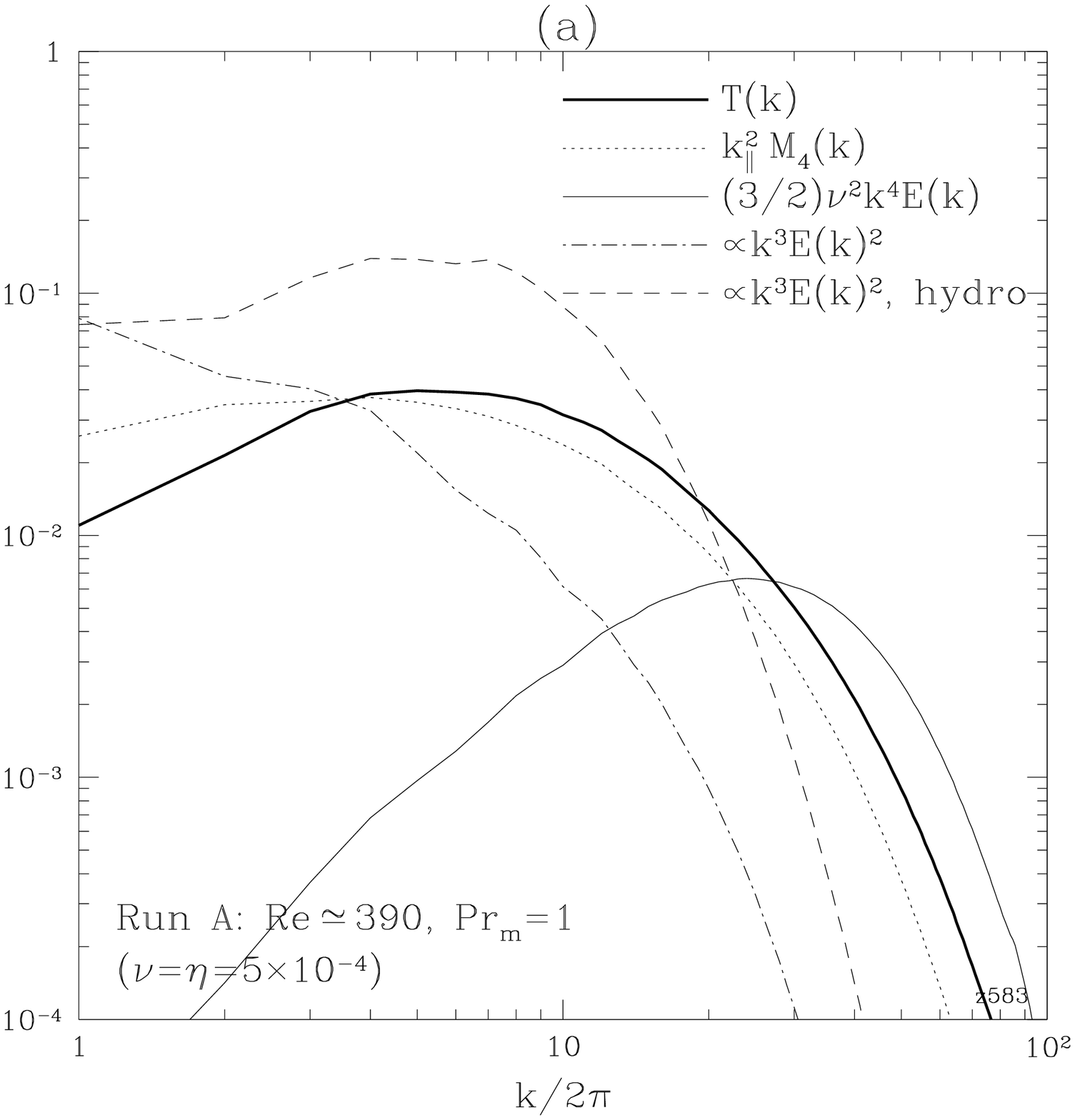}{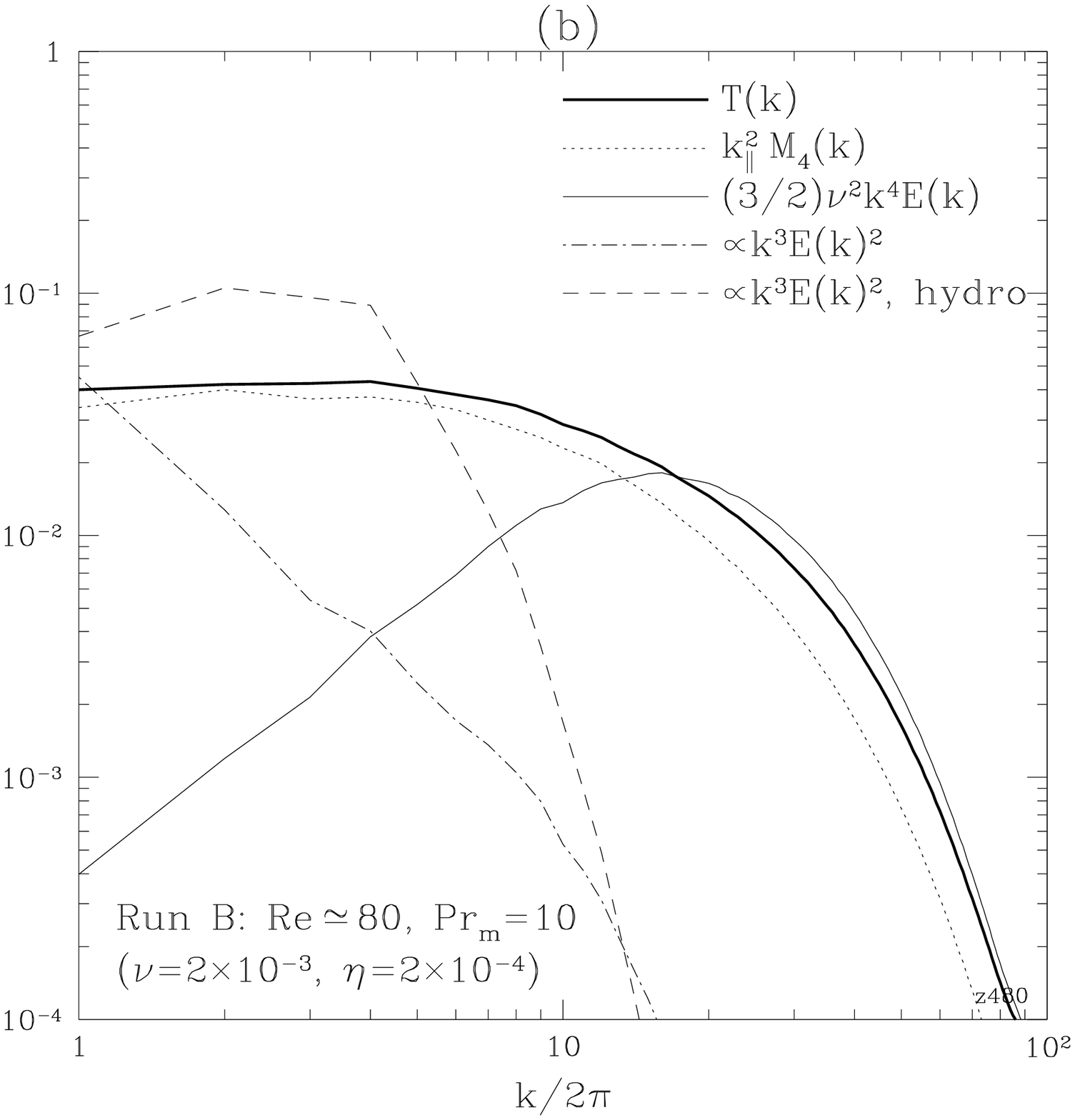}{f26a.ps}{f26b.ps}{Tension 
spectra for (a) run~\run{A} and (b) run~\run{B} in the saturated state. 
For comparison, we also plot the kinetic-energy spectrum 
$E(k)$ compensated by~$(3/2)\nu^2 k^4$,  
the spectrum of~$B^2$, multiplied by~$\kpar^2$ 
[as defined by \eqref{kpar_def}], and 
$k^3E(k)^2$ (divided by 10) 
in both hydrodynamic and nonlinear MHD regimes.}

\subsubsection{Tentative Conclusions} 
\label{ssec_tent_conc}

At the fairly low values of $\Re$ that our runs have, 
we cannot test scalings such as \eqref{Bsq_visc}, \exref{Bsq_equip}, 
\exref{lres_nlin}, or \exref{scale_sep_nlin} 
so as to make a definite distinction between the two 
scenarios of saturation (\ssecref{ssec_eff} vs.~\ssecref{ssec_ineff}). 
On the qualitative level, however, our results appear to 
favor the second scenario:

\begin{itemize}

\item increase of the ratio $\Bsq/\usq$ with $\Re$,

\item the presence of an intermediate nonlinear regime, 

\item the evolution of the characteristic wavenumbers 
(selective decay and fold elongation), 

\item the evidence of the folded structure up to the box 
scale (curvature pdf's, anticorrelation between curvature 
and field strength, flat tension spectra),

\item the nonlinear modification of the kinetic-energy spectra 
up to the box scale

\end{itemize}

\noindent are all consistent with the 
conclusion that the stretching scale in saturation is 
comparable to the outer scale, not the viscous 
scale,\footnote{In principle, it is not excluded 
that the stretching scale $\ls$ saturates at some 
$\Re$-dependent value intermediate between $\lf$ and $\ld$. 
In fact, one might consider 
the fact that $\ks$ appears to be somewhat larger than the box 
wavenumber in run~\run{A} suggestive of such an outcome. 
Much higher resolutions are necessary to test 
this possibility. However, the scale invariance 
of the inertial range makes saturation 
with $\lf\ll\ls\ll\ld$ {\em a priori} unlikely.\label{fn_ls}} 
as suggested in \ssecref{ssec_ineff}. 
In this second scenario of saturation, the velocities at scales 
below $\ls\sim\lf$, do not destroy the folded direction-reversing 
magnetic fields. The magnetic fluctuations are of two kinds: the folds 
of length $\lf$ with direction reversals at $\lres$ 
[\eqref{lres_nlin}] and (hypothetically) Alfv\'en waves propagating along 
the folds with $k$ in the inertial range (\ssecref{ssec_ineff}). 
The dynamo is saturated at an energy that is comparable 
to $\usq$. 

The mechanism of this saturation should be 
entirely controlled by what happens at the outer scale. 
Since the motions at this scale are 
directly excited by the forcing, the nonlinear 
dynamics there may not be totally dissimilar from 
the nonlinear dynamics in the viscosity-dominated 
case. In other words, we might conjecture that 
saturation results from a competition of weakened 
stretching and enhanced mixing {\em at the outer (forcing) 
scale} \citep[the hypothesis that the dynamo saturation 
is controlled by the outer timescale was first explicitly 
formulated by][]{MCM}. 

In \ssecref{ssec_sat}, we constructed a quantitatively 
successful model of saturation in the 
viscosity-dominated case by assuming that velocity gradients 
are partially anisotropized leading to weakened stretching but 
unsuppressed mixing. 
Since the viscosity-dominated runs have $\Re\sim1$, 
the stretching scale is the box scale, $\ls\sim\lf$, 
which is also the viscous scale, $\lf\sim\ld$, 
so the success of this model 
cannot be construed to favor either of the saturation 
scenarios outlined in \ssecref{ssec_speculations}. 
The fact that the mixing rate did not have to be suppressed 
in order for correct results to be obtained is consistent 
with both the scenario of efficient mixing of the field 
by the eddies at the viscous scale $\ld$ 
and the idea that the saturated state is controlled by 
the balance between mixing and stretching at the box scale~$\lf$.

The reader is referred to \ssecref{ssec_Pr_unity} for some further caveats. 

\pseudofiguretwo{fig_PB_AB}{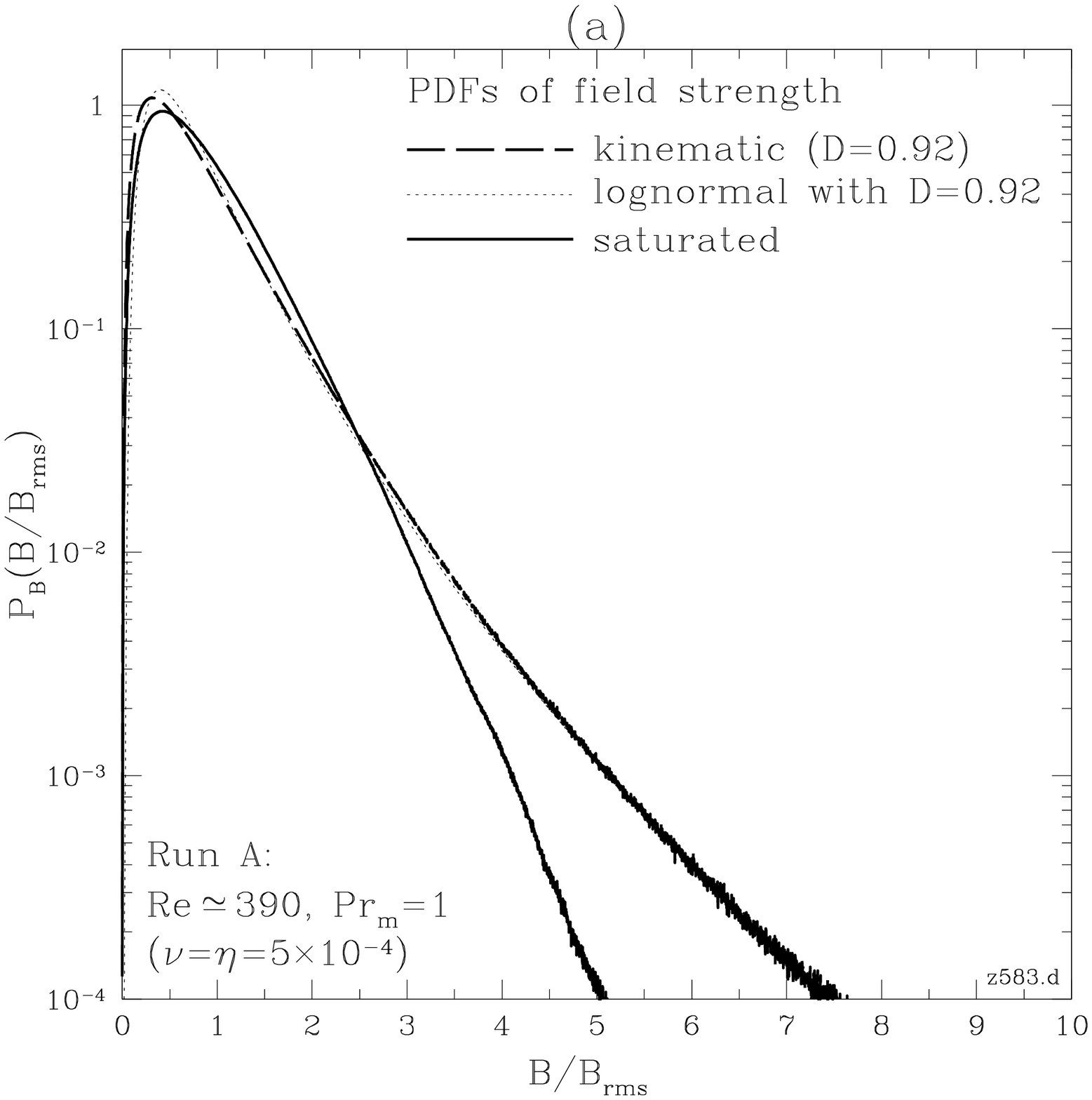}{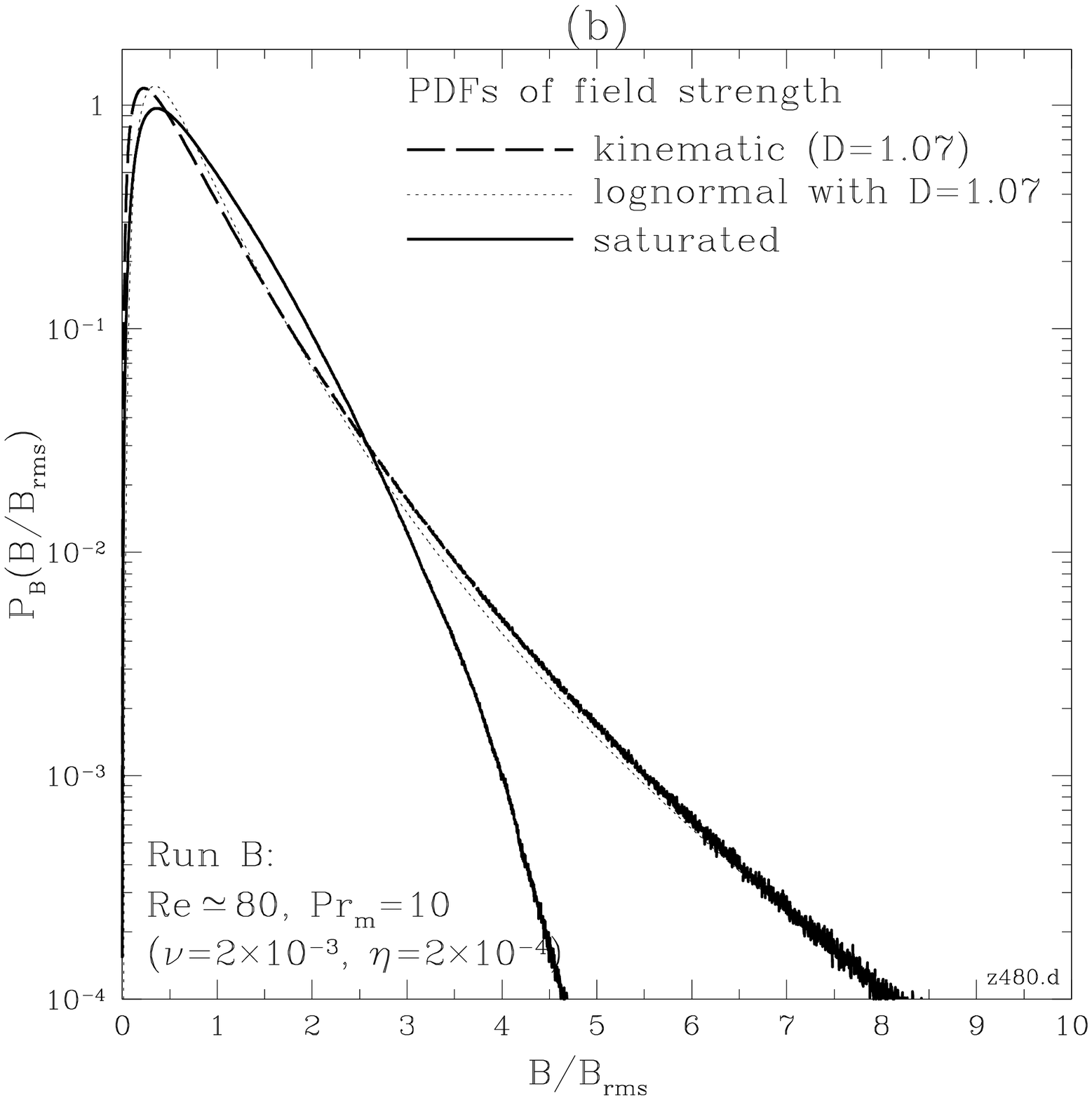}{f27a.ps}{f27b.ps}{The pdf's 
of $B/\Brms$ for (a) run~\run{A} and (b) run~\run{B} 
averaged over the kinematic stages of these runs and the corresponding 
lognormal profiles [\eqref{lognormal_fit}] with the same~$D$.
The pdf's in the saturated state are also given.}

\subsection{Intermittency}
\label{ssec_intermittency_AB}

We complete our standard set of diagnostics by looking at the 
pdf's of the magnetic-field strength in runs \run{A} and \run{B}. 

In the kinematic regime, their evolution is again self-similar 
with the pdf's of $B/\Brms$ collapsing onto stationary profiles 
(similarly to \figref{fig_PB_kin}). These profiles (averaged 
over the kinematic stages of the runs) are shown in \Figref{fig_PB_AB}. 
The lognormal fit [\eqref{lognormal_fit}] works even better for 
these runs than it did for the viscosity-dominated case in 
\ssecref{ssec_finite} (\figref{fig_PB_tail}). 

The intermittency is reduced in the saturated state: 
the evolution of the normalized moments is analogous to 
the viscosity-dominated case (\figref{fig_Bn}a) with the 
values of kurtosis in the saturated state still super-Gaussian, 
but substantially smaller than in the kinematic regime 
(see \tabref{tab_index}). 
On a semilog plot, the pdf's in the saturated state of 
both runs (\figref{fig_PB_AB}) 
have straight segments in the region of strong 
fields (exponential tails). 
However, at even larger $B$, there appears to be a knee 
followed by steeper decay. To a lesser extent, the same 
feature was also present in the viscosity-dominated case 
(\figref{fig_PB_tail}). We believe that this feature might be due 
to resolution effects and/or insufficient averaging (as these values 
of $B$ correspond to very rare events), 
so the physical pdf at large $B$ is exponential.  

Note that exponential pdf's of the magnetic-field strength
were reported previously in $\Pr\gtrsim1$ simulations of 
compressible convection with rotation \citep{Brandenburg_etal_structures} 
and of Boussinesq convection \citep{Cattaneo_solar}. 
Thus, exponential tails of the field-strength pdf 
appear to be a fairly universal feature in MHD turbulence 
\citep[at least away from the boundaries; 
see][]{Cattaneo_solar,Thelen_Cattaneo}.

\section{DISCUSSION} 
\label{sec_concerns}


\subsection{Are All Small-Scale Dynamos Large-$\Pr$ Dynamos?} 
\label{ssec_largePr}

The fundamental mechanism of magnetic-energy generation 
by three-dimensional chaotic and/or random flows 
\citep[due to][]{Zeldovich_etal_linear} is based on 
the idea that the structure of a locally uniform 
rate-of-strain tensor in three dimensions 
allows for a winning configuration of magnetic 
fields: the field aligns with the stretching direction of the 
flow but varies (reverses) along the ``null'' direction, 
so that compression cannot cancel the effect of stretching 
(see \ssecref{ssec_zeld_mech}). 
Thus, the folded structure, which has been the leitmotif 
of this paper, is an essential ingredient in our understanding 
of the very existence of the dynamo. 

The Zel'dovich et al.\ mechanism is universal in the 
sense that the particulars of the flow matter little. 
However, two properties 
are essential: (1) the flow must be spatially 
smooth and (2) the magnetic field must be able to vary at scales 
much smaller than the scale of the flow. Only if these 
conditions are satisfied, can one think of the flow as 
locally linear and make the above inferences about 
the growing field's structure. 
The turbulent dynamo with large $\Pr$ satisfies these conditions 
and so do all deterministic chaotic dynamos 
\citep[see reviews by][]{STF,Ott_review}. 
{\em The small-scale dynamo, as usually understood in the literature 
(this paper included), is the large-$\Pr$ dynamo.} 

A smooth random flow is not, strictly speaking, 
turbulence. Turbulent motions at scales above the viscous 
cutoff are spatially rough and cannot be approximated 
by a linear velocity field (locally uniform rate of strain). 
To what extent then is the Zel'dovich et al.\ 
mechanism applicable to physical 
systems and numerical simulations in which $\Pr$ is not large? 

\subsubsection{Suppression of Small-Scale Dynamo at Low $\Pr$}
\label{ssec_lowPr}

A natural way to approach this issue is to consider the 
opposite limit: $\Pr\ll1$ (but $\Re$ and $\Rm$ are still large). 
In this case, the resistive scale 
is in the hydrodynamic inertial range and the dominant 
interaction is between velocities and magnetic fields at 
that scale. Is there a dynamo in such a case? 

There currently is no physical 
understanding of the low-$\Pr$ dynamo: 
the standard heuristic argument involving 
stretching, turbulent diffusion, and resistive diffusion 
fails to produce an unambiguous prediction about the viability 
of the dynamo because all these effects are of the same order 
and the outcome of their competition is a quantitative, 
rather than a qualitative, issue. 
A number of simulations found in various MHD contexts 
that dynamo shuts down when $\Pr$ is too small 
\citep{Nordlund_etal,Brandenburg_etal_structures,Nore_etal,Christensen_Olson_Glatzmaier,MCM,SCMM_lowPr,HBD_pre}. 
In particular, in \citet{SCMM_lowPr}, we studied this 
problem using the same code and the same numerical set up 
as in this paper and found that dynamo at low $\Pr$ is shut down 
even though the magnetic Reynolds number $\Rm$ is the same as 
for successful dynamos with $\Pr\ge1$. 
We identified two possible interpretations: either a nonhelical 
small-scale dynamo with low $\Pr$ does not exist at all 
or the critical $\Rm$ for such a dynamo is much larger than 
for the case of $\Pr\ge1$ and cannot be resolved 
\citep[cf.][]{Rogachevskii_Kleeorin,Boldyrev_Cattaneo}. 

If a low-$\Pr$ dynamo does exist, what is the structure of the 
magnetic fields generated by it? The Zel'dovich et al.\ 
mechanism does not apply: we cannot have folded fields 
as there is no field-flow scale separation 
necessary to maintain them. Thus, the question 
is whether the large-$\Pr$ dynamo is the only 
kind of nonhelical small-scale dynamo or there exists 
yet another, as yet unexplored, variety. 

\subsubsection{Small-Scale Dynamo at $\Pr\sim1$} 
\label{ssec_Pr_unity}

Resolving this issue is 
important for interpreting the simulations with $\Pr\sim1$. 
It is clear that the {\em kinematic} 
dynamo in such runs is of the large-$\Pr$ kind: we have seen 
that all the main features of the viscosity-dominated runs 
carry over to the case of $\Re\sim10^2$ and $\Pr\gtrsim1$ 
\citep[the scale separations such as $\kB/\kpar$ change 
smoothly as a transition from viscosity-dominated large-$\Pr$ 
to turbulent $\Pr\sim1$ regime is made: see][]{SCMM_lowPr}. 
This dynamo is driven by the viscous-scale eddies and 
whatever the inertial-range velocities might do is guaranteed 
to happen at a slower rate. 
However, when the stretching component of the 
viscous-scale eddies is nonlinearly 
suppressed, it is crucial to know if the inertial-range 
velocities that take over 
can also generate magnetic field 
{\em at their own scale.} The heuristic saturation scenarios 
proposed in \ssecref{ssec_speculations} were based on treating 
these velocities simply as larger eddies 
acting on the magnetic field the same way as the viscous eddies  
did. This description is incomplete if there exists a second kind 
of small-scale dynamo generating magnetic fields in the inertial 
range. In the latter case, the $\Pr\sim1$ simulations are in 
a mixed regime 
containing both the folded fields resulting from large-$\Pr$-type 
dynamo and some other kind of magnetic fluctuations generated 
by the hypothetical inertial-range dynamo (plus, possibly, the Alfv\'en-like 
waves discussed in \ssecref{ssec_ineff}). 

The two alternative scenarios 
of saturation described in \ssecref{ssec_speculations} and 
interpretations of the numerical results in favor of either 
are, thus, qualified by the above discussion. 

\subsection{Large-$\Pr$ Dynamo in 3D vs.\ the 2D Case}
\label{ssec_2D}

Small-scale magnetic fields 
in the large-$\Pr$ regime were previously studied numerically 
in two dimensions by \citet{Kinney_etal}. Since the frozen-in property of the 
field lines holds equally well in two and three dimensions, 
the folded structure 
in thwo dimensions looks very similar to the folded structure in 
three dimensions. 
Such features as disparity between parallel and transverse 
scales of the field variation, power-like tails of curvature 
distributions, and anticorrelation between field strength and 
field-line curvature are present in both cases 
\citep{Kinney_etal,SCMM_folding}. The reason for such similarity 
is that the folded structure is an ideal effect: 
it is just a geometric property of the field 
lines arising from random stretching by the flow. 
Diffusion has a selection effect on the folded fields 
with opposite outcomes in three and in two dimensions: 
in three dimensions, the folds can align in such a way as to support growth 
of magnetic energy, --- in contrast to two dimensions, 
where stretching 
is always overcome by diffusion (see \ssecref{ssec_zeld_mech}). 
Indeed, the magnetic field in two dimensions 
can be written in terms of a scalar stream function, 
$\vB=(\d A/\d y,-\d A/\d x)$. It is then readily obtained that 
$A$ satisfies the advection-diffusion equation \citep{Zeldovich_antidynamo}
\bea
\d_t A + \vu\cdot\nabla A = \eta\Delta A,
\eea
implying eventual decay of the field. However, the eventuality 
of decay does not preclude transient growth and even a slowly-decaying 
nonlinear ``saturation'' phase, which was studied in some detail 
by \citet{Kinney_etal}. They found persistent folded 
structure, near-flat tension spectra, and $k^{-4}$ spectra of 
the subviscous velocities induced by the tension force
--- results that are analogous to those of 
\ssecref{ssec_fold_nlin} and \ssecref{ssec_tension}, 
confirming the geometric nature of these features.
The key difference is that the possibility of sustained 
dynamo action in three dimensions 
leads to a saturated nonlinear state 
(\ssecref{ssec_sat}), while in two dimensions, 
the nonlinear state is eroded at the resistive 
timescale and the field is destroyed. 

Another feature that arises in three dimensions, but is absent 
in two dimensions, 
is the possibility for a degree of misalignment between 
the reversing fields in the folds: this was measured tentatively 
by $\kbj$ in \ssecref{ssec_scales} and \ssecref{ssec_fold_nlin}. 
We noted in \ssecref{ssec_fold_nlin} that $\kbj$ was larger 
in the saturated state than during the kinematic growth, perhaps 
indicating a transition from flux sheets to flux ribbons. 
There is, at present, no quantitative theory of three-dimensional 
magnetic structures and field alignment. 
We have not attempted a full-fledged numerical characterization 
of these features: such a study is left for future work. 

\subsection{Diagnosing MHD Turbulence}
\label{ssec_stats}

In this paper, the set of statistical quantities that were used 
to analyze numerical data was limited to three groups: 

\begin{enumerate}

\item Energy spectra: they show energy distribution over scales 
but miss essential information about the structure of the field; 
also spectra of second-order quantities related to the Lorentz force.

\item Characteristic wavenumbers, 
pdf's of the field-line curvature, and 
correlations between the curvature and the field strength
--- these describe the structure of the field.

\item The pdf's and moments of the field strength: they 
characterize the volume-filling property of the magnetic field.

\end{enumerate}

\noindent
All diagnostics in the last two groups 
(1) are of order higher than 2 in magnetic field,
(2) contain magnetic field, but not velocity or 
its correlations with magnetic field, 
and (3) are one-point in space. 
That one has to go beyond second-order statistics 
is due to the structured and intermittent nature of 
the magnetic field and to the fact that magnetic back-reaction 
($\vB\cdot\nabla\vB$) is quadratic in $\vB$. 
Points 2 and 3 are limitations of our description of 
MHD turbulence and deserve some discussion. 

\subsubsection{Flow Statistics}

The results reported in this paper are almost exclusively 
on the statistics of the magnetic field. 
However, in our discussion of back-reaction and 
saturation in \ssecref{ssec_sat} and in \ssecref{ssec_speculations} 
we had to make conjectures about the way dynamically important 
magnetic fields modify the flow. The assumptions that helped 
us to construct a saturation model for the viscosity-dominated 
runs were fairly natural: thus, it is clear already 
from the induction equation itself that, in order to stop 
magnetic field from growing, the $\vb\vb$ component of 
the rate-of-strain tensor must be suppressed. 
Indeed, we do find that both $\langle\vb\vb:\nabla\vu\rangle$ 
and $\langle|\vb\vb:\nabla\vu|^2\rangle$ 
decrease as the transition from the kinematic to 
nonlinear regime occurs. However, in order to resolve 
the questions raised in \ssecref{ssec_speculations} about 
the mixing efficiency of the surviving motions in 
the saturated state, the existence of the Alfv\'en waves 
propagating along folds, etc., a detailed and systematic 
study of the velocity statistics and, especially, 
of the velocity--magnetic field correlations, is required. 

Very little numerical evidence on this subject is available 
in the literature. In part, this scarcity is due to the fact 
that even a qualitative physical picture of the magnetic 
back-reaction has been lacking, so it is not obvious what the 
interesting diagnostics are. Extant results include 
statements about the growth of $\langle\vu\cdot\vB\rangle$ 
correlations \citep{Pouquet_Meneguzzi_Frisch}, 
positive correlation between high current and high 
rate of strain and alignment of magnetic field parallel to 
the vorticity vector \citep{Miller_etal}, 
suppression of chaos of Lagrangian trajectories 
by the magnetic field \citep{Cattaneo_Hughes_Kim,Zienicke_Politano_Pouquet}, 
and evidence of nonlocal interaction between large-scale velocities 
and small-scale magnetic fields based on energy-transfer spectra 
\citep{Kida_Yanase_Mizushima,Cho_Vishniac} and on experiments 
with switching off velocity field \citep{MCM}. 
All these results can be made sense of on the basis of more or 
less straightforward heuristic considerations. 
While they are all reflections of some underlying back-reaction mechanism, 
none of them tell us explicitly what this mechanism is. 
Much further analysis is necessary to diagnose it 
correctly. Such an analysis is the subject of our 
ongoing work.  

\subsubsection{Two-Point Statistics: Structure Functions}

The second limitation we have imposed on ourselves 
is that all our higher-order statistics are one-point. 
We believe that most of the physics that is 
qualitatively understood is already captured in the diagnostics 
we employ. Obviously, two-point statistics contain a great wealth 
of information that cannot be accessed by the one-point approach. 
We do not, e.g., attempt to study two-point spatial intermittency 
of the magnetic field. 

In recent years, there has been an explosion of interest 
in the scaling of structure functions in MHD turbulence. 
These are defined by 
$S_n(y)=\<|\psi(\vx+\vy)-\psi(\vx)|^n\> \sim y^{\zeta_n}$,
where $\psi$ is the quantity of interest 
(e.g., the longitudinal velocity field 
$u_L=\vu\cdot\vy/y$). 
The present surge of interest followed the publication by
\citet{She_Leveque} of a phenomenological 
formula for $\zeta_n$ in hydrodynamic turbulence 
and the remarkable success it enjoyed 
in reproducing experimental and numerical scalings.  
Their model requires as input parameters 
the codimension of the most singular dissipative structures 
and the dimensional scaling of the cascade 
(eddy-turnover) time. Applications 
of their idea to the MHD turbulence followed: first, based 
on the $k^{-3/2}$ spectrum of \citet{Iroshnikov} 
and \citet{Kraichnan_IK} \citep{Grauer_Krug_Marliani,Politano_Pouquet_SL}, 
and then on the Kolmogorov $k^{-5/3}$ spectrum 
\citep[][see \citet{Biskamp_book} for a review]{Mueller_Biskamp}. 
The latter model matched numerical data for decaying MHD turbulence. 
In recent simulations, 
good fits were obtained by choosing ad hoc 
fractional values for the codimension of the dissipative structures 
\citep{Mueller_Biskamp_Grappin,HBD_pre}. 

Notably, these models seemed to work 
for structure functions not of velocity 
but of the Els\"asser fields $\vz^\pm=\vu\pm\vB$  
and it has been implicitly or explicitly assumed that 
magnetic fields and velocities are symmetric as in an Alfv\'en 
wave. However, there was no such symmetry in either 
the decaying isotropic simulations of \citet{Biskamp_Mueller} 
or the forced ones of \citet{HBD_pre}. 
Even in simulations with a mean field,  
the structure functions of $\vu$ and $\vB$, when measured 
separately, turned out to have different scalings, 
velocity being less intermittent than the magnetic field 
\citep{CLV_latest}. On a qualitative level, this property 
has, in fact, been seen in most numerical simulations 
since \citet{Meneguzzi_Frisch_Pouquet}. 

Considering Els\"asser-field scalings is made attractive 
by the exact result of \citet{Politano_Pouquet_exact2} 
stating that a mixed third-order structure function 
$\langle[\vz^\mp(\vx+\vy)-\vz^\mp(\vx)]\cdot(\vy/y)
|\vz^\pm(\vx+\vy)-\vz^\pm(\vx)|^2\rangle \propto y$ 
has a linear inertial-range scaling analogous to Kolmogorov's 
${4\over5}$ law \citep[this result only holds for $\Pr=1$; 
for arbitrary $\Pr$, linear scaling was also proved 
for two other mixed triple correlators --- not structure functions 
--- of $\vu$ and $\vB$, see][]{Politano_Pouquet_exact1}.
Thus, while the Politano--Pouquet laws provide a set of 
constraints on $\vu$--$\vB$ (or $\vz^+$--$\vz^-$) correlations, 
they do not have exact implications 
for any structure functions involving $\vu$, $\vB$, 
$\vz^+$, or $\vz^-$ alone. 

It is worthwhile to recall that the reason 
structure functions are the diagnostics of choice in 
hydrodynamic turbulence is the Galilean invariance of the 
velocity field: only velocity differences are really interesting. 
Phenomenological models of intermittency, 
the She--L\'ev\^eque model included, assume that interactions 
occur between comparable scales (locally in $k$ space), thus giving 
rise to an energy cascade. 
Neither Galilean invariance nor $k$-space locality of 
interactions holds for magnetic fields: a constant magnetic 
field cannot be transformed out, and an interaction between 
velocity and magnetic fields at disparate scales is exactly 
what the small-scale dynamo is. 
The open questions are then 
(1) whether models in the mold of She--L\'ev\^eque are justified 
for MHD and, if they work, why they do so; 
and (2) whether, in view of the asymmetry between magnetic field 
and velocity, the Els\"asser fields are meaningful. 

Thus, analyzing two-point statistics of MHD turbulence 
is complicated by the unanswered questions about the 
basic structure of this turbulence and by the lack of 
clarity about what set of diagnostics constitutes a 
physically interesting and informative description. 
We leave these matters to future work. 





\section{SUMMARY} 
\label{sec_summary}

We now itemize the main points of this paper:

\begin{enumerate}

\item Magnetic fields and turbulence in astrophysical plasmas 
exist and interact at many disparate scales, which makes 
full global description unachievable by brute-force numerical 
simulations. We concentrate on small-scale turbulent 
fields to probe universal physics 
independent of large-scale object-specific features. 
In this spirit, we simulate incompressible, isotropic, homogeneous, 
randomly forced, nonhelical MHD turbulence in a periodic box.

\item If no mean field is imposed, all magnetic fields in the system 
are generated by the small-scale dynamo, which 
consists in random stretching and folding of the field lines 
and is best revealed in the limit of large $\Pr$.  
In its kinematic stage, this dynamo is a one-time-scale 
process insensitive to the particulars of the random flow. 
In Kolmogorov turbulence, it is controlled by the viscous-scale 
turbulent eddies. It can, therefore, be studied in the numerically 
feasible limit of $\Re\sim1$ and $\Pr\gg1$ 
with random external forcing. 
In this viscosity-dominated regime, we are able to carry out 
a parameter scan in $\Pr$. 
This regime is a model for interaction between 
magnetic fields and the viscous-scale eddies, the forcing 
representing energy input from the larger-scale eddies.  

\item The salient feature of the small-scale dynamo is that 
it produces highly intermittent and structured fields. 
These fields are organized in folds whose length is comparable 
to the flow scale but that contain fields that 
reverse direction at the resistive scale. 
The scale separation between these scales is $\sim\Pr^{1/2}$. 
We diagnose the folded structure 
by measuring various average scale lengths of the field, 
the spectrum of the magnetic tension, 
the distributions of the field-strength and 
of the field-line curvature, and the (anti)correlation between them. 
The folded structure implies that the 
system becomes nonlinear (the fields start reacting back 
on the flow via the Lorentz tension force) when the energy 
of the folded field becomes comparable to the energy of 
the viscous-scale eddies. 

\item The folded structure is found to persist after 
the dynamo becomes nonlinear and saturates. 

\item Saturation of the dynamo can be achieved via partial 
two-dimensionalization of the velocity-gradient 
statistics with respect to the local direction of the folds, 
so that saturation is a result of quasi-2D mixing balancing 
weakened stretching of the field by the 
anisotropized flow. This mechanism is sufficiently 
robust for a simple semiquantitative model of saturation based 
on it to predict magnetic-energy spectra in excellent 
agreement with the numerical results for the viscosity-dominated runs. 

\item The folded structure is essential in saturation physics: 
while the magnetic field is formally small-scale (as a result 
of direction reversals), the folds themselves possess spatial coherence 
at the flow scale, which allows them to exert a back-reaction 
on the flow. Both small-scale dynamo and this nonlinear 
back-reaction exemplify the nonlocality in $k$ space of the 
interactions between the velocity and magnetic fields in MHD turbulence.

\item While the regime with both large $\Re$ and large $\Pr$ 
is not accessible 
at current resolutions, some progress is achieved by simulating 
the case of $\Re\gg1$ and $\Pr\gtrsim1$. 
The small-scale dynamo at $\Pr\sim1$ belongs to the 
same class as the large-$\Pr$ dynamo: the magnetic field 
is generated at scales somewhat smaller than viscous 
and has a structure qualitatively similar to the folded 
structure of the fields in large-$\Pr$ viscosity-dominated runs. 

\item In turbulence with large $\Re$, the nature 
of the saturated state depends on how efficiently 
the anisotropized velocity gradients can mix the 
magnetic-field lines. If mixing by the inertial-range 
motions were unsuppressed, the magnetic energy would saturate 
at a below-equipartition level comparable to the energy 
of the viscous-scale eddies. Numerical evidence, 
while not definitive, does not appear to support such an 
outcome. 

\item If both stretching and mixing are suppressed 
we predict an intermediate nonlinear regime 
with magnetic energy growing proportionally to time 
from the viscous-eddy to the outer-eddy energy 
(increase by a factor of $\Re^{1/2}$), 
the fold length increasing to the outer scale 
(a factor of $\Re^{3/4}$), 
and the resistive scale increasing via selective decay 
by a lesser factor of $\Re^{1/4}$. 
While these scalings cannot be 
verified at current resolutions, we find clear evidence 
of an intermediate nonlinear-growth regime featuring both 
fold elongation and selective decay. 

\item The fully developed, forced, isotropic MHD turbulence 
is the saturated state of the small-scale dynamo. 
While phenomenological thinking 
rooted in the Alfv\'en-wave physics of the MHD turbulence 
with a strong mean field leads one to 
expect a state of scale-by-scale equipartition between 
magnetic and kinetic energy and a Kolmogorov scaling 
of the spectra, none of the extant numerical evidence 
supports this view. Instead, the kinetic-energy spectrum is 
dominated by the outer scale and has a steeper-than-Kolmogorov 
scaling in the inertial range, while the magnetic energy 
is dominated by small scales, at which it substantially 
exceeds kinetic energy. 

\item Based on our numerical results and heuristic arguments, 
we conjecture that the saturated state of the dynamo is 
controlled by the balance of mixing and stretching by 
the anisotropized outer-scale turbulent motions. 
The magnetic fields in the saturated state are then 
a superposition of folds and Alfv\'en-like waves that 
can propagate along the folds. The existence of these 
waves in the inertial range is predicated  
on their not being efficient mixers of the magnetic fields. 
These waves are not as yet 
numerically detectable and, therefore, remain hypothetical. 

\end{enumerate}

In the context of our scenarios of saturation with efficient and 
inefficient mixing, it is interesting to recall 
the two basic alternatives put forward 
more than 50 years ago: 
magnetic-energy saturation at the viscous-eddy energy and scale 
\citep{Batchelor_dynamo} or eventual equipartition between the 
magnetic and kinetic energy at all scales 
\citep{Schlueter_Biermann,Biermann_Schlueter}. 
Biermann \& Schl\"uter's reasoning assumed 
interactions only between velocities and magnetic fields 
at the same scale. 
Batchelor's argument was based on the analogy between 
the evolution equations for the magnetic field and vorticity 
$\nabla\times\vu$. While this analogy has since been 
realized to be flawed even for $\Pr=1$ \citep[note the recent 
numerical investigations of this issue by][]{Ohkitani,Tsinober_Galanti} 
and the magnetic fields have been shown to propagate to 
the resistive scale, it is still a theoretical and numerical challenge 
to demonstrate that magnetic energy grows above the viscous-eddy 
energy and to explain why it can do so. 

Since the time these alternatives were formulated, 
theoretical predictions and numerical results have ranged from 
completely ruling out the small-scale dynamo \citep{Saffman_no_dynamo}, 
to predicting magnetic-energy pile up 
at the resistive scale \citep{Vainshtein_Cattaneo}, 
to closure-based confirmation of scale-by-scale equipartition 
\citep{PFL}, to direct numerical simulations interpreted as evidence 
that there is a tendency toward the equipartition 
or at least, Kolmogorov scaling \citep{HBD_apjl},  
to direct numerical simulations interpreted as evidence 
that such an equipartition is lacking \citep{MCM}. 
While our understanding of these alternatives 
has improved, the original dichotomy continues 
to define the terms of the debate.
 
The purpose of the present paper has been to collect 
a set of numerical results and physical considerations 
emphasizing the nonlocal (in $k$ space) nature of 
the velocity--magnetic field interaction and the 
resulting dominance of magnetic fluctuations 
at small scales, which we believe to be the defining 
properties of isotropic MHD turbulence. 

\acknowledgements

We thank S.~Boldyrev, E.~Blackman, A.~Brandenburg, 
B.~Chandran, G.~Hammett, N.~Haugen, P.~Haynes, {\AA}.~Nordlund, 
M.~Paczuski, A.~Shukurov, and J.-L.~Thiffeault for stimulating discussions 
of various aspects of this work. 
We particularly appreciate the time and effort N.~Haugen has 
invested to perform comparisons of some of our results 
with his own numerical simulations of nonhelical MHD turbulence. 
We are indebted to R.~Kulsrud for getting us interested in the physics 
and astrophysics of the small-scale dynamo.
 
Our work was supported by grants from 
PPARC (PPA/G/S/2002/00075),
EPSRC (GR/R55344/01), 
UKAEA (QS06992), 
NSF (AST~97-13241 and AST~00-98670), 
and US~DOE (DE-FG03-93ER54~224). 
A.~A.~S.~also thanks the Leverhulme Trust for support 
under the UKAFF Fellowship. 
Simulations were done at UKAFF (University of Leicester) 
and NCSA (University of Illinois, Urbana-Champaign). 
We would especially like to thank C.~Rudge of the UKAFF 
for his very competent assistance with their computer system.

\newpage 

{\footnotesize
\begin{deluxetable}{lrrrrccccccccccc}
\tablewidth{0pt}
\tablecaption{Index of runs.\tablenotemark{a}\label{tab_index}}
\tablehead{
\colhead{Run} & 
\colhead{$\nu$} &
\colhead{$\eta$} & 
\colhead{$\Pr$}  &
\colhead{Grid} & 
\colhead{$\Re$} &
\colhead{$\usq$} & 
\colhead{$\Bsq$} & 
\colhead{
\parbox{0.25in}{
$$
{\Bfr\over\Bsq^2}
$$}
} &
\colhead{
\parbox{0.25in}{
$$
{\kl\over2\pi}
$$}
} & 
\colhead{
\parbox{0.25in}{
$$
{\kB\over2\pi}
$$}
} & 
\colhead{
\parbox{0.25in}{
$$
{\kbxj\over2\pi}
$$} 
} &
\colhead{
\parbox{0.25in}{
$$
{\kbj\over2\pi}
$$} 
} & 
\colhead{
\parbox{0.25in}{
$$
{\kpar\over2\pi}
$$}
} &
\colhead{$C_{K,B}$}
} 
\startdata
\run{S0}-kin \z{587} & $5\times10^{-2}$ & $2\times10^{-3}$ &    25 &  $64^3$ &  
2.0 & 0.39 & decays & 3.9 & 1.1 & 3.4 & 2.2 & 0.77 & 0.52 & -0.948\\
\run{S1}-kin \z{586} & $5\times10^{-2}$ &        $10^{-3}$ &    50 & $128^3$ &
2.0 & 0.39 & grows  & 4.8 & 1.1 & 4.8 & 3.1 & 0.87 & 0.54 & -0.971\\
\run{S2}-kin \z{585} & $5\times10^{-2}$ & $4\times10^{-4}$ &   125 & $128^3$ &
2.0 & 0.39 & grows  & 6.2 & 1.1 & 7.1 & 4.6 & 1.0  & 0.69 & -0.984\\ 
\run{S3}-kin \z{584} & $5\times10^{-2}$ & $2\times10^{-4}$ &   250 & $128^3$ &
2.1 & 0.43 & grows  & 8.9 & 1.1 & 9.7 & 6.4 & 0.93 & 0.57 & -0.991\\ 
\run{S4}-kin \z{479} & $5\times10^{-2}$ & $       10^{-4}$ &   500 & $256^3$ &
2.0 & 0.41 & grows  & 9.4 & 1.1 & 13  & 8.8 & 1.1  & 0.55 & -0.996\\\\ 
\run{S1}-sat  \z{634} & $5\times10^{-2}$ &        $10^{-3}$ &    50 & $128^3$ &
2.0 & 0.39 & 0.13   & 3.9 & 1.1 & 4.4 & 2.8 & 0.95 & 0.53 & -0.967\\
\run{S2}-sat  \z{472} & $5\times10^{-2}$ & $4\times10^{-4}$ &   125 & $128^3$ &
1.9 & 0.36 & 0.20   & 3.7 & 1.1 & 5.8 & 3.6 & 1.2  & 0.55 & -0.977\\ 
\run{S3}-sat  \z{471} & $5\times10^{-2}$ & $2\times10^{-4}$ &   250 & $128^3$ &
1.9 & 0.34 & 0.36   & 3.6 & 1.1 & 6.9 & 4.3 & 1.5  & 0.56 & -0.981\\ 
\run{S4}-sat  \z{479} & $5\times10^{-2}$ &        $10^{-4}$ &   500 & $256^3$ &
1.9 & 0.36 & 0.33   & 4.1 & 1.1 & 8.9 & 5.5 & 1.8  & 0.57 & -0.989\\ 
\run{S5}-sat\tablenotemark{b} 
\z{476} &  $5\times10^{-2}$ & $4\times10^{-5}$ &  1250 & $256^3$ &
1.8 & 0.33 & 0.49   & 4.0 & 1.2 & 12  & 7.4 & 2.4  & 0.60 & -0.992\\
\run{S6}-sat\tablenotemark{c} 
\z{636} &  $5\times10^{-2}$ & $2\times10^{-5}$ &  2500 & $256^3$ &
1.8 & 0.32 & 0.43   & 4.1 & 1.2 & 15  & 9.6 & 3.0  & 0.77 & -0.987\\\\
\run{a0}-kin \z{709}  & $4\times10^{-3}$ & $4\times10^{-3}$ &    1 & $64^3$ &
52  & 1.7 & decays & 2.9 & 2.0 & 4.3 & 2.6 & 1.9 & 0.83 & -0.841\\
\run{a1}-kin \z{581}  & $2\times10^{-3}$ & $2\times10^{-3}$ &    1 & $128^3$ &
100 & 1.7 & grows  & 3.2 & 2.6 & 6.9 & 4.3 & 2.7 & 1.2 & -0.858\\
\run{a2}-kin\tablenotemark{c} 
\z{582}  &        $10^{-3}$ &        $10^{-3}$ &    1 & $128^3$ &
210 & 1.7 & grows  & 3.5 & 3.6 & 10 & 6.3 & 3.9 & 1.8 & -0.891\\
\run{A}-kin \z{583}  & $5\times10^{-4}$ & $5\times10^{-4}$ &    1 & $256^3$ &
450 & 2.0  & grows  & 6.3 & 5.0 & 18  & 12  & 7.0  & 3.1  & -0.941\\\\
\run{a1}-sat \z{581}  & $2\times10^{-3}$ & $2\times10^{-3}$ &    1 & $128^3$ &
100 & 1.7 & 0.062 & 2.7 & 2.4 & 6.5 & 4.0 & 2.6 & 1.1 & -0.862\\
\run{a2}-sat \z{582}  &        $10^{-3}$ &        $10^{-3}$ &    1 & $128^3$ &
200 & 1.6 & 0.20 & 2.8 & 2.7 & 7.9 & 4.4 & 3.5 & 1.0 & -0.899\\
\run{A}-sat  \z{583}  & $5\times10^{-4}$ & $5\times10^{-4}$ &    1 & $256^3$ &
390 & 1.5  & 0.25   & 3.2 & 3.3 & 11  & 6.0 & 4.2  & 1.0  & -0.972\\\\
\run{B}-kin \z{480}  & $2\times10^{-3}$ & $2\times10^{-4}$ &    10 & $256^3$ &
100 & 1.6   & grows  & 8.5 & 2.6 & 19  & 12  & 4.1  & 1.3  & -0.987\\
\run{B}-sat  \z{480}   & $2\times10^{-3}$ & $2\times10^{-4}$ &   10 & $256^3$ &
80  & 0.98  & 0.42   & 3.3 & 2.3 & 12  & 6.8 & 3.8  & 0.66 & -0.988\\\\
\enddata
\tablenotetext{a}{In the labels assigned to the runs, ``kin'' means 
kinematic regime, ``sat'' means saturated state. 
See \eqref{Re_def} for the definition of $\Re$, 
\eqsref{kpar_def}{kbj_def} for the definitions of the various $k$'s, 
\eqref{CKB_def} for the definition of $C_{K,B}$.} 
\tablenotetext{b}{In the kinematic regime, this run requires higher resolution.}  
\tablenotetext{c}{These runs are slightly underresolved.}  
\end{deluxetable}}

\begin{deluxetable}{cccccc}
\tablewidth{0pt}
\tablecaption{Timescales and energy ratios.\label{tab_Pr1}}
\tablehead{
\colhead{Run} & 
\colhead{$\Re$} &
\colhead{$\tbox^{-1}$} & 
\colhead{$\Grms$} & 
\colhead{$\gamma$} &
\colhead{
\parbox{0.25in}{
$$
{\Bsq\over\usq}
$$}}
} 
\startdata
\run{a1} & 110 & 1.3  & 5.6 & 0.30 & 0.037\\
\run{a2} & 210 & 1.3  & 7.7 & 0.79 & 0.12\\
\run{A}  & 450 & 1.4  & 11  & 1.4  & 0.16\\\\
\run{B}  & 100 & 1.2  & 5.3 & 1.7  & 0.43 \\
\run{S4} & 2.0 & 0.64 & 1.1 & 0.53 & 0.92 \\
\enddata
\end{deluxetable}

\end{document}